\documentclass[12pt]{article}
\usepackage[margin=1in]{geometry}
\usepackage[round,compress, sort, authoryear]{natbib}
\usepackage{setspace}
\usepackage{xcolor}
\usepackage{hyperref}
\usepackage{amsmath,amsthm,amssymb}
\usepackage{enumitem}

\usepackage{rotating,multirow,makecell,array}
\usepackage {tikz}
\usepackage{tikz}
\usetikzlibrary{arrows.meta}

\usepackage{float}
\usepackage{pgfplots}
\usepackage{mathtools}
\usepgfplotslibrary{groupplots}
\usepackage{textcomp}
\usetikzlibrary{calc} 
\usepackage{subcaption}
\usepackage{algorithm}
\usepackage{algpseudocode}

\usetikzlibrary {positioning}
  \newtheorem{cor}{Corollary}
  \newtheorem{thm}{Theorem}
\theoremstyle{definition}
 \newtheorem{defn}{Definition}
  \newtheorem{ass}{Assumption}
 \newtheorem{setting}{Setting}
  \usepackage{multibib}
\usepackage[utf8]{inputenc}   
\usepackage[T1]{fontenc}      
\usepackage{graphicx}         
\usepackage{booktabs}         
\usepackage{epstopdf}  
\usepackage{siunitx}
\usepackage{booktabs}   
\usepackage{multirow}   
\usepackage{booktabs}
\usepackage{multirow}      
\usepackage{makecell}      
 \newtheorem{exmp}{Example}
 \newtheorem{rem}{Remark}
 
\usepackage{booktabs}   
\usepackage{amsmath}    

\usepackage{caption}              
\ifxetex
  \usepackage{mathspec}
\fi

\title{Learning What to Learn: Experimental Design when Combining Experimental with Observational Evidence\footnote{We benefited from comments from several participants at the UChicago Causal Inference workshop, Cornell Decision Theory workshop, Harvard seminar, NBER Summer Institute 2026 (Frontiers in Econometrics), SEA conference, SITE Development (Stanford),  Munich econometric conference.  We are especially grateful to Arun Chandrasekhar, Guido Imbens, Gabriel Kreindler, Jesse Shapiro, and Elie Tamer for their encouragement and extensive feedback 
 and Giacomo Opocher, Nabin Poudel and Sasha Stasovskyi for exceptional research assistance. Davide Viviano acknowledges support by the Harvard Griffin Fund in Economics and NSF Grant SES 2447088. All mistakes are our own. An R package for its implementation is available at \url{https://github.com/dviviano/robustExpDesign}.}
}
\author{Aristotelis Epanomeritakis%
\footnote{\raggedright Department of Economics, Harvard University. Email address: \texttt{aristotle\_epanomeritakis@fas.harvard.edu}.}
\and Davide Viviano%
\footnote{Department of Economics, Harvard University.  Email address: \texttt{dviviano@fas.harvard.edu}.}
}
\date{First version: October, 2025\\ 
This version: August, 2026}

\begin{document}

\maketitle

\vspace{-24pt}

\begin{abstract}

Experiments deliver credible treatment-effect estimates but, because they are costly, are often restricted to specific sites, small populations, or particular mechanisms. A common practice across several fields is therefore to combine experimental estimates with reduced-form or structural external (observational) evidence to answer broader policy questions, such as those involving general equilibrium effects or external validity. We develop a unified framework for the design of experiments when combined with external evidence, i.e.,   choosing which experiment(s) to run and how to allocate sample size under arbitrary budget constraints.  Because observational evidence may suffer bias unknown ex-ante, we evaluate designs using a robust regret criterion that compares any candidate design to an oracle with knowledge about the observational study bias bound that jointly chooses the design and estimator. This yields a transparent bias-variance trade-off that does not require the researcher to specify a bias bound and relies only on information already needed for conventional power calculations. We illustrate the framework for studying general equilibrium effects of cash transfer programs.

\end{abstract}

\newpage 

\onehalfspacing 

\section{Introduction}

Randomized controlled trials (RCTs) have improved economics by providing internally valid estimates. However, in practice, feasibility constraints often confine trials to localized effects such as the effect in a specific site or subpopulation, or of a particular mechanism.\footnote{Under tabulations in \cite{muralidharan2017experimentation} only $31\%$ of experiments  in top economics journals sample units from a larger population. Of these, experiments are in median ``representative of a population of 10,885 units [...] and randomized in clusters of 26 units per cluster''.} These effects, although useful, are often not sufficient to answer relevant policy questions about external validity and general equilibrium (GE) effects of at-scale interventions. 

In response, a large literature in development \citep[e.g.,][]{ToddWolpin2006AER, attanasio2012education, de2025decoupling}, as well as labor economics and industrial organization  \citep[e.g.,][]{chetty2016effects, katz2025digital, allcott2025sources} combine experiments with external evidence—observational and/or results from experiments in other settings—to extrapolate policy-relevant counterfactuals that no single experiment in the population of interest can identify. In our review of AEA journals over the past decade, about one-third of experimental papers complement experimental with observational inputs, either reduced-form or structural (Figure~\ref{fig:AEA_journals}).\footnote{This practice is particularly common in development economics. For instance, close to \textit{$50\%$} of papers with an experimental component presented at the NBER Summer Institute in Development in 2024--2026 combine experiments with external/observational evidence to study mechanisms or counterfactuals (Figure~\ref{fig:NBER_SI_dev}).} Since budget constraints often bind, this practice raises the question of how to choose and design the most informative experiments from a menu of feasible experiments \citep[see discussion in][]{ravallion2012fighting, niederle2025experiments}.


Our main contribution is a framework for choosing which experiment to run and how to design it in this common setting of  combining experimental with external evidence.
By external evidence, we mean reduced-form or structural estimates based on observational data, and/or results from experiments in other contexts. External evidence may be biased in ways not known ex ante (e.g., due to confounding or lack of external validity). 
We provide a procedure that (1) helps researchers disentangle the role of each experiment, (2)
selects which experiments to run subject to a budget (e.g., which treatment arms and/or sub-populations), (3) optimally allocates sample size.
We accommodate a broad class of user-specified feasibility constraints on the design and choice of the estimator.

As an illustrative example, consider a government piloting a cash-transfer program in a small set of districts to increase school attendance. The trial delivers a local average effect, but policy decisions often require counterfactuals at scale, allowing prices and wages to adjust \citep[e.g.][]{ToddWolpin2006AER, egger2022general}.  Researchers therefore combine experimental evidence with a supply-demand model that leverages external data to map local impacts into economy-wide outcomes. The design question is (i) which mechanisms are the most valuable to learn experimentally given constraints on the size of the experiment
and (ii) how to allocate sample size across treatment arms (and locations). 

The first step is to specify the estimand and the underlying parameters. Define $\tau(\theta)$ the (policy-relevant) object of interest, where $\tau$ is a known smooth function and $\theta$ is a vector of unknown parameters; $\tau$ may be scalar or vector-valued. Researchers also observe external baseline estimates for $\theta$ (or relevant moment estimates)--henceforth \emph{observational estimates}--which may be biased. Each feasible experiment provides a set of unbiased moments, e.g., by identifying a subset of parameters without bias. This parameterization makes explicit which components of $\theta$ are learned experimentally and which may rely on (biased) observational evidence; not all components need or can be learned experimentally.

Returning to the cash-transfer example, $\tau(\theta)$ is the effect on schooling when all poor households in rural Kenya receive the subsidy. The entries of 
$\theta$ consist of (i) the direct schooling response to a conditional cash transfer (CCT), (ii) the income effect of transfers in partial equilibrium, and (iii) wage adjustments that shift the returns to schooling. Due to cost constraints, researchers may only run small, partial-equilibrium experiments and then combine these with prior evidence to recover $\tau(\theta)$. For instance, they may choose between a CCT arm to estimate the direct effect, or an unconditional-cash arm (UCT) to measure the income elasticity, and extrapolate the GE effects using evidence from a previous study in a different context, such as the Progresa program in Mexico \citep{ToddWolpin2006AER}. When site selection is also part of the design, $\theta$  additionally collects site-specific effects.

The next step is to choose which experimental evidence to generate and how precisely to estimate it. We consider two ways in which this evidence may subsequently be used. Researchers may commit ex ante to an (asymptotically) linear rule that combines experimental and observational moments, as with methods of moments and minimum-distance estimators. Alternatively, they may report the full evidence and allow audiences to adapt their beliefs about observational bias to form their posterior, as when communicating results to stakeholders, such as researchers, policy-makers, or founding agencies with different priors.

The first challenge is that the bias may be hard to elicit for all relevant parameters when designing the experiment. A worst-case approach would be overly-conservative and disregard information about the variance.  Motivated by a large decision-theory literature \citep[e.g.][]{stoye2009minimax, montiel2026decision, armstrong2024adaptingmisspecification}, we measure the performance relative to the regret, defined as the ratio of the researcher's worst-case mean-squared error relative to an oracle that knows the worst-case bound on the observational bias and chooses the \textit{design} and estimator.\footnote{See Remark \ref{rem:alternative_objectives} for a discussion about minimax and minimax additive regret.} Taking its supremum over the worst-case bias yields a robust procedure without requiring a prior about the bias. While previous references take the design as fixed, here we optimize over the design choice, which changes the target objective.

Section \ref{sec:linear}
provides an explicit and novel characterization of proportional regret for (asymptotically) linear estimators in experimental and observational moments. 
These include generalized methods of moments where we may also optimize over the weighting matrix. For any design and weighting matrix that combine the experimental and observational point estimates, the regret is the maximum of two normalized components. The first is a \textit{variance regret}: the estimator’s sampling variance under the chosen design and weights, divided by the smallest attainable variance. The second is a \textit{bias regret}: the worst-case squared bias induced by using external evidence, divided by the smallest feasible worst-case bias over the class of designs. 
The variance regret depends on the (expected) variance-covariance matrix of the observational and experimental estimates implied by the design, the weighting matrix, and the sensitivity of the policy parameter, i.e.,  the gradient of $\tau$
with respect to the parameters evaluated, under mild conditions,  at the observational estimates.\footnote{For  this result to hold, we assume that $\tau(\theta)$ is twice differentiable with bounded derivatives. For non-linear $\tau(\theta)$ in $\theta$, we also require that the observational estimates bias is local to zero in the spirit of 
\cite{andrews2020transparency, armstrong2021sensitivity, BonhommeWeidner2021arXiv}, where its rate of convergence can be the same, faster, or even slower than the estimators' standard error. Section \ref{sec:gmm-extension} provides details.} The bias regret depends on the chosen experiment, sensitivity vector and weighting matrix, known ex-ante.

This result makes the bias-variance trade-off intuitive. At the optimum, the two normalized components tend to be equalized: when the bias dominates, the design prioritizes high-sensitivity parameters; when variance dominates, the procedure invests sample where it most reduces variance, trading off precision with sensitivity. It finds the point on the variance-sensitivity frontier closest to the (infeasible) variance and sensitivity minima.

From a theoretical (and computational) perspective, our key insight is that, as we restrict the class of estimators to be linear in the moments, the worst-case proportional regret is quasi-convex in the radius of the bias, for any norm characterizing the bias' uncertainty set.  This structure is particularly useful in the extrapolation environments that
motivate our analysis, where feasibility constraints prevent any experiment
from identifying the full counterfactual $\tau(\theta)$, making the bias regret finite. When an unbiased estimator of the full target is
instead available, non-linear estimators are more natural; a case we discuss in Section~\ref{sec:adaptive_weights}.

Section~\ref{subsec:fully_adaptive_benchmark} turns to settings in which the
researcher does not commit ex ante to a particular estimator; we allow different readers to heterogeneously adapt their beliefs, motivated by recent work on scientific communication \citep[e.g.][]{banerjee2020theory, andrews2021model, abadie2020statistical}. We complement this literature by studying which experiments to conduct and their
precision. We evaluate each
design using the expected loss of a Bayesian audience that observes all the moments reported by the researcher; researchers face a potentially large class of priors, with the bias magnitude unknown to researchers. 

Our next main result shows that, under mild restrictions on the class of priors, the solution to the researcher's problem admits a
variational representation with the same variance-bias structure as the
linear-estimator problem, where 
the estimator is now chosen ex-post by the audience (possibly as a non-linear estimator). The solution is intuitive and corresponds to the
best uniform approximation to the Pareto frontier that trades-off sensitivity and precision.

 These results provide us with tractable optimization solved off-the-shelf, i.e., a convex program conditional on the choice of the treatment arms, and mixed integer over a (finite) menu of feasible experiments.  The design only requires knowledge of the expected variance-covariance matrix and observational estimates and can be pre-specified.
In our leading applications with independent experimental and observational estimates, it requires the same variance inputs as for standard power calculations \citep[e.g.,][]{GerberGreen2012}.


In Section \ref{sec:extensions}, we then provide several extensions, including results for minimizing confidence interval length, partial identification, and settings with a known bias bounds.

We illustrate the framework in a cash‐transfer application in Kenya, where the researcher combines new experimental evidence with external evidence from Mexico’s Progresa program \citep{ToddWolpin2006AER, attanasio2012education}; these preliminary estimates may lack external validity. We compare three designs and their combinations: (i) a CCT arm to estimate direct schooling effects; (ii) a UCT arm to estimate income effects; and (iii) an employment program that identifies individual 
schooling response to
wages. We show that the method equalizes bias and variance regret, and achieves the same level of mean-squared error of the standard variance-optimal (Neyman) allocations with a significantly smaller sample size.



\vspace{-4mm} 
\paragraph{Related literature} This paper links the experimental-design literature—which mostly focuses on settings where all parameters are identified within the experiment and leaves aside questions of data combination—to recent work that integrates experimental and (reduced-form or structural) observational evidence to extrapolate effects in complex scenarios.

Recent advances for experimental design include balancing and variance-minimizing allocations \citep[e.g.][]{cytrynbaum2021optimal, 
tabord2023stratification, dette1997designing}, adaptive designs for policy choice \citep[e.g.][]{kasy2019adaptive}, and experimental design under correct model specification \citep[e.g.][]{higbee2024experimental, kiefer1959optimum, kasy2016experimenters}. Traditional work on experimental design for robust-model estimation either focused on testing competing models   \citep{atkinson1975design, lopez2007optimal}, or on using a-priori knowledge of (worst-case) bias
\citep{box1959basis, wiens1998minimax, sacks1984some}. All these references leave aside questions about observational data combination. When minimizing the experimental variance, \cite{rosenman2021designing} use observational studies to construct confidence bounds on the experimental variance, and then focus on variance-optimal  stratification. This problem differs from ours, where instead we study the use of observational studies with experiments for extrapolation. 

In summary, we complement this literature by studying which experiment to run (and how) when not all parameters are learned experimentally, a common problem in economics.

Our minimax regret criterion connects to a long-standing decision-theoretic tradition for experimental design. Recent work include \cite{manski2016sufficient}, \cite{banerjee2020theory}, \cite{manski2004}, \cite{dominitz2017more}, \cite{olea2024externally}, \cite{hu2024minimax}, \cite{breza2025generalizability} among others. These references focus on settings where researchers only have access to experimental variation, instead of combining it with external (observational) evidence, motivating different designs and objective functions. We also complement literature that leverages correctly-specified models for site-selection  \citep{gechter2024selecting, abadie2021synthetic} by allowing for misspecification in observational estimates.

Related work analyzes estimation under misspecification  \citep[e.g.][]{armstrong2024adaptingmisspecification, armstrong2021sensitivity, kaido2024information, BonhommeWeidner2021arXiv, kwon2025estimating, sacks1978linear,sacks1984some}, and combining existing experiments with observational studies \citep[e.g.][]{gechter2022combining, athey2025experimental, de2020empirical}. Unlike the references above, where the design is fixed, our main contribution is to optimize
the design itself. This changes the structure of the problem, including the optimization, 
solution and definition of regret, computed against the best design-estimator.

Specifically, in the linear estimator framework, we characterize the proportional regret as a quasi-convex function in the bias bound, which simplifies optimization; it differs from solutions under the fixed-design optimal estimators \citep[e.g.][]{armstrong2024adaptingmisspecification, tsybakov1998pointwise},  where, due to non-linearity of the estimator, their solution corresponds to a  minimax program over worst-case priors.
Here we show how misspecification is captured through the first-order sensitivity of the estimand to each parameter, an important measure for sensitivity analysis \citep[e.g.][]{andrews2020transparency}; we provide a foundation for simultaneously protecting against a correctly specified and misspecified model  \citep[e.g.][]{bickel1984parametric, kempthorne1988controlling}, formalized via a bias and variance regret. From the audience perspective, our main characterization shows how our solution corresponds to a novel Pareto frontier approximation along this same sensitivity and variance objective. It complements previous literature where the choice of the experiment is fixed \citep[e.g.][]{andrews2021model, andrews2020transparency} and therefore no variance-sensitivity trade-off arises on the choice of the experiment.


\section{Setup}
\label{sec:setup}

Consider a researcher interested in an arbitrary target estimand
\(\tau(\theta)\in\mathbb R\), indexed by a low-dimensional unknown parameter vector
\(\theta\in\mathbb R^p\) and a known mapping \(\tau\) (see Section~\ref{sec:multivalued_estimands} for extensions to multivariate estimands).
The goal is to design an experiment that is the most informative about
\(\tau(\theta)\) when combining existing evidence (e.g., observational studies) with
experimental variation, under  feasibility or cost constraints on the experiments.  

\subsection{Observational and experimental estimates}

We assume that researchers have access to a set of observational estimates
\(S^{\mathrm{obs}}\in\mathbb R^{p_o}\). We impose no restrictions on how
\(S^{\mathrm{obs}}\) is formed: it can be based on arbitrary exclusion
restrictions implied by an economic or statistical model. However,
\(S^{\mathrm{obs}}\) may have
biases collected in a vector \(b\in\mathbb R^{p_o}\) unknown at the experimental design
stage. Examples include an estimate from an
instrumental variable regression that may fail the exclusion restriction, or an experimental
estimate from a country different from the one of interest that may lack external validity.
Researchers can then collect additional experimental estimates from a set of
feasible experiments $\mathcal{E}$, for example by learning a \emph{subset} of the relevant parameters.

The reported statistics may or may not be direct estimates of individual components of
\(\theta\). In particular, their population counterparts may correspond to known
linear combinations of \(\theta\), captured through a known matrix $\Lambda$ below, or asymptotically linear as in Example \ref{exmp:gmm}.

\begin{ass}[Reported estimates]
\label{ass:1a}

Given a maximum number of possible experiments $p_e$, for an experimental choice \(\mathcal E \subseteq \{1,\ldots,p_e\}\), and for a given level of precision \(\Sigma\) associated with the experimental choice \(\mathcal E\), denote by
\(S_{\mathcal E,\Sigma}^{\mathrm{exp}}\in\mathbb R^{|\mathcal{E}|}\) the set of experimental estimates collected by the researcher. Researchers observe
$ 
\tilde{S}_{\mathcal E,\Sigma}
\equiv
\begin{pmatrix}
S^{\mathrm{obs}}\\
S^{\mathrm{exp}}_{\mathcal E,\Sigma}
\end{pmatrix}
\in\mathbb R^{p_o+|\mathcal{E}|},
$ 
such that
\begin{equation}
\small 
\begin{aligned} 
\label{eqn:linear_representation}
\mathbb E[\tilde{S}_{\mathcal E,\Sigma}]
-
\Lambda_{\mathcal E}\theta
=
\begin{pmatrix}
b\\
0_{|\mathcal{E}|}
\end{pmatrix},
\qquad
b\in\mathbb R^{p_o},
\qquad
\mathbb V(\tilde{S}_{\mathcal E,\Sigma})=\Sigma,
\end{aligned} 
\end{equation}
for known matrices
\[
\Lambda_{\mathcal E}\in\mathbb R^{(p_o+|\mathcal{E}|)\times p},
\qquad
\Sigma\in\mathbb R^{(p_o+|\mathcal{E}|)\times(p_o+|\mathcal{E}|)}.
\]
Here \(b\) denotes the vector of observational biases, potentially \textit{unknown} to researchers.
\end{ass}

Here, \((\mathcal E,\Sigma)\) characterizes the experimental design, encoding which
statistics are learned experimentally and with what precision. The matrix
\(\Lambda_{\mathcal E}\) determines how the reported estimates identify the primitive
parameter \(\theta\). The key distinction between the observational and experimental estimates is that observational estimates may have a bias \(b\), potentially unknown. The number of statistics $p_o + |\mathcal{E}|$ may exceed the number of parameters \(p\).

\vspace{2mm} 

A few remarks about $\Sigma$, which we assume to be known: 
\begin{itemize} 
\item First,  our framework can accommodate arbitrary positive-definite $\Sigma$ with potentially correlated observational and experimental estimates. In practice, the leading example is the case where observational estimates are independent of the experimental estimates, as in our application where estimates are computed on independent samples.
In this case, 
$ 
\Sigma=\operatorname{diag}(\Sigma_{\mathrm{obs}},\Sigma_{\mathrm{exp}}), 
$ 
with the first block corresponding to observational estimates, not subject to design choices. 

The component of $\Sigma$ corresponding to the observational estimates can often be estimated consistently, in which case our characterizations in subsequent sections will hold up-to an asymptotically negligible error. 

On the other hand, the experimental component $\Sigma_{\mathrm{exp}}$ does not have to be pre-determined as a function $\mathcal{E}$; for a given experimental choice $\mathcal{E}$, these components may be chosen optimally by researchers by optimizing sample size allocation described below.

\item Second, assuming that the experimental components $\Sigma_{\mathrm{exp}}$ are known (for given choice of the sample size) is standard in experimental planning  for the candidate experiments \citep{GerberGreen2012}. 
We can also interpret $\Sigma_{\mathrm{exp}}$ as a researcher's  prior expectation over the true variances, indexed by the sample size allocated to each experiment.\footnote{In this case we interpret the mean-squared error in the subsequent section as integrating over the corresponding prior for $\Sigma$ given $\theta$.} 

In some applications, $\Sigma_{\mathrm{exp}}$ can be estimated using the first and second moments from the preliminary observational estimates.  Section  \ref{sec:sigma} shows that misspecification and estimation error in this case is negligible for estimating $\tau(\theta)$ under standard local misspecification; that is, its error is asymptotically negligible in the limit experiment, whereas the bias $b$ remains of first order.
 \end{itemize} 

\begin{exmp}[Direct parameter estimation]
\label{exmp:direct}
Suppose that each $S_j^{\mathrm{obs}}$ is a direct estimate of $\theta_j$, with \(p_o=p\), and that the experiment may identify one or some of the components of
\(\theta\): 
\[
\mathbb E[S^{\mathrm{obs}}_j]
=
\theta_j+b_j,
\qquad j=1,\ldots,p, \qquad \mathbb E[S^{\mathrm{exp}}_j]
=
\theta_j,
\qquad j\in\mathcal E. 
\]
In this example, \(\Lambda_{\mathcal E}\) is the \((p+ |\mathcal{E}|)\times p\) matrix whose first
\(p\) rows form the identity matrix \(\mathbb{I}_p\), and whose remaining rows are the rows of
\(\mathbb{I}_p\) corresponding to the coordinates in \(\mathcal E\).  Whenever the observational
and experimental estimates are from independent samples and the
experimental estimates corresponds to mutually independent treatment arms, 
$
\Sigma
=
\begin{pmatrix}
\Sigma_{\mathrm{obs}}
&
0_{p\times|\mathcal E|}
\\[3pt]
0_{|\mathcal E|\times p}
&
\operatorname{diag}
\left(
\left\{
\frac{v_j^2}{n_j}
\right\}_{j\in\mathcal E}
\right)
\end{pmatrix},
$
where \(\Sigma_{\mathrm{obs}}\) denotes the covariance matrix of the
observational estimates, \(v_j^2\) denotes the per-unit variance of
experimental estimate \(j\), and \(n_j\) denotes the sample size allocated to
that experiment, potentially chosen by researchers. \qed 
\end{exmp}

Our framework also encompasses method-of-moments estimators: exactly for linear moments,
and up to a negligible remainder under a standard local misspecification framework
\citep[e.g.][]{andrews2020transparency,armstrong2021sensitivity}.
For non-linear moments, we will think of the linear relationship as a first-order (asymptotic) approximation of moments
around a preliminary estimate. For this case, we 
define $\theta_0^{\mathrm{obs}} \in \mathbb{R}^p$ as the probability limit of a preliminary (possibly biased) estimate $\widehat{\theta}_0^{\mathrm{obs}}(S^{\mathrm{obs}}) \rightarrow_p \theta_0^{\mathrm{obs}}$ of $\theta$.

\begin{exmp}[Method-of-moments estimators]
\label{exmp:gmm}
Let \(\theta \in\mathbb R^p\) denote the primitive
structural parameter, and let \(\theta_0^{\mathrm{obs}}\) denote the probability limit of a
preliminary observational estimate for $\theta$. For
experiments \(\mathcal E\), let
$ 
g^{\mathrm{obs}}(\theta)\in\mathbb R^{p_o},
g^{\mathrm{exp}}_{\mathcal E}(\theta)\in\mathbb R^{|\mathcal E|}
$ 
denote population moments, and let
\(\bar G^{\mathrm{obs}}\) and
\(\bar G^{\mathrm{exp}}_{\mathcal E,\Sigma}\) denote their sample analogues, with 
$ 
\mathbb E[\bar G^{\mathrm{obs}}]
-
g^{\mathrm{obs}}(\theta) = b,
\mathbb E[\bar G^{\mathrm{exp}}_{\mathcal E,\Sigma}]
- 
g^{\mathrm{exp}}_{\mathcal E}(\theta) = 0
$, for an unknown vector $b$. Define
\begin{equation} \label{eqn:tilde_theta}
\small 
\begin{aligned} 
\tilde{S}_{\mathcal E,\Sigma}
=
\begin{pmatrix}
S^{\mathrm{obs}}\\
S^{\mathrm{exp}}_{\mathcal E,\Sigma}
\end{pmatrix}
=
\begin{pmatrix}
\bar G^{\mathrm{obs}}
-
g^{\mathrm{obs}}(\theta_0^{\mathrm{obs}})
+
\Lambda^{\mathrm{obs}}\theta_0^{\mathrm{obs}}\\[3pt]
\bar G^{\mathrm{exp}}_{\mathcal E,\Sigma}
-
g^{\mathrm{exp}}_{\mathcal E}(\theta_0^{\mathrm{obs}})
+
\Lambda^{\mathrm{exp}}_{\mathcal E}\theta_0^{\mathrm{obs}}
\end{pmatrix}, \qquad \Lambda_{\mathcal E}
=
\begin{pmatrix}
\Lambda^{\mathrm{obs}}\\
\Lambda^{\mathrm{exp}}_{\mathcal E}
\end{pmatrix}
=
\begin{pmatrix}
\frac{\partial g^{\mathrm{obs}}(\theta)}
{\partial\theta^\top}
\big|_{\theta=\theta_0^{\mathrm{obs}}}\\[4pt]
\frac{\partial g^{\mathrm{exp}}_{\mathcal E}(\theta)}
{\partial\theta^\top}
\big|_{\theta=\theta_0^{\mathrm{obs}}}
\end{pmatrix}.
\end{aligned} 
\end{equation} 
The affine normalization is without loss and simply removes the deterministic intercept in the local expansion. From a Taylor expansion, under sufficient smoothness conditions
\[
\small 
\begin{aligned} 
\mathbb E[\tilde{S}_{\mathcal E,\Sigma}]
-
\Lambda_{\mathcal E}\theta
=
\begin{pmatrix}
b\\
0_{|\mathcal E|}
\end{pmatrix}
+
O(\|\theta-\theta_0^{\mathrm{obs}}\|^2).
\end{aligned} 
\]
The linear representation in Setting~\ref{setting:1},
Equation~\eqref{eqn:linear_representation}, therefore holds up to a negligible remainder
under the local asymptotic regime formalized in Section~\ref{sec:gmm-extension}, where $n^{1/4} ||\theta - \theta_0^{\mathrm{obs}}|| = o(1)$, with $n$ denoting the sample size used to estimate $\theta$. This encompasses cases where the bias may be of the same or different order of the standard error, but local to zero. 
The same expansion applies asymptotically if \(\theta_0^{\mathrm{obs}}\) and \(\Lambda_{\mathcal E}\) are replaced by
consistent estimates. 
\qed
\end{exmp}

The class of experiments is assumed to be in a user-specific set
\(\mathcal D\).

\begin{ass}[Feasible experiments]
\label{ass:constraint_set}
We write
$
(\mathcal E,\Sigma)\in\mathcal D
$
indicating that \(\mathcal D\) is the set of feasible designs, i.e., choices of
\(\mathcal E\) and \(\Sigma\). The set \(\mathcal D\) is such
that: (i) each feasible \(\Sigma\) has uniformly bounded entries and is strictly positive
definite, with $\bar{\lambda} I \succ \Sigma \succ \underline{\lambda} I$ for positive constants $\bar{\lambda}, \underline{\lambda} \in (0,\infty)$; (ii) for each feasible \(\mathcal E\), \(\Lambda_{\mathcal E}\) has full column rank and it is uniformly bounded, so that its operator norm $||\Lambda_{\mathcal{E}}||_{\mathrm{op}} \le U$ for a constant $U < \infty$.
\end{ass}

We let \((\mathcal E,\Sigma)\) be within arbitrary constraint sets that may encode
feasibility or budget constraints. In our framework, restrictions on the experiments
researchers can run and on their precision \(\Sigma\) can take any desired form. In most
applications, we think of such constraints as arising from fixed costs of running an
experiment and/or constraints on their power
\citep[e.g][]{duflo2007using, athey2017econometrics}. We require that
\(\Sigma\) is uniformly bounded: once we commit to learn a set of
experimental estimates, their variance is finite, and strictly positive definite; this latter condition simplifies exposition but it is not necessary.\footnote{For this latter condition to
hold in an asymptotic framework with growing sample size, \(\theta\) and
\(\tilde{S}_{\mathcal E,\Sigma}\) can be defined without loss as parameters and statistics
rescaled by the square-root of the sample size; in this case \(\Sigma\) denotes the asymptotic
variance. See Section \ref{sec:sigma} for details. Also, although omitted for expositional convenience, from an inspection of the proof of Theorem \ref{thm:1}, it is possible to allow for degenerate $\Sigma$ as long as the overall variance of the estimator for $\tau(\theta)$ is non-degenerate. }  
\(\Lambda_{\mathcal E}\) is full rank (and uniformly bounded to avoid degenerate solutions), which guarantees identification in
the absence of observational bias (i.e., when $b = 0$).

\begin{rem}[Prior information about $b$ and unbiased observational estimates or biased experiments] \label{rem:adaptation_B}
We will consider settings where researchers  have limited knowledge about the (local) bias $b$ before running an experiment. This is motivated by the fact that learning $b$ would require running pilots for \textit{each} design in the menu of feasible experiments, which can be difficult in practice. Section \ref{sec:bound_known} show however how prior knowledge of $b$ can be incorporated within our framework by restricting (priors on) $b$ in a given set. 

Also, in some applications, researchers may assume that some observational estimates are not
biased, or conversely that some experimental estimates may be biased. This can be
accommodated by redefining which entries of $\tilde{S}_{\mathcal{E},\Sigma}$ are potentially biased.\qed 
\end{rem}

\begin{rem}[Why observational evidence remains informative even under misspecification]
\label{rem:informative_observational_evidence}
The linear representation in
Equation~\eqref{eqn:linear_representation} can be interpreted asymptotically as in Example \ref{exmp:gmm}, under local misspecification. With $n^{-1/2}$ denoting the convergence rate of $\tilde{S}_{\mathcal{E},\Sigma}$
the observational bias is represented by a sequence \(b_n\)
satisfying
$ 
b_n\to 0,
\sqrt n\,\|b_n\|^2=o(1).
$ 
The bias can be at a rate that can be the same, slower or faster than the standard error. This feature of our framework distinguishes informative (but potentially biased) estimates from any random guess for $\theta$ that would instead not satisfy the local asymptotic framework.
\qed
\end{rem}

\subsection{Properties of \(\tau\)}

A key feature of our framework is that researchers explicitly parametrize \(\tau(\cdot)\),
thereby pre-specifying and making transparent which biases drive the design problem; it clarifies the role of the experiment in complementing existing evidence, often a desideratum \citep[e.g.][]{christensen2018transparency}; when unknown
Section~\ref{sec:CI} studies partially identified $\tau$.

The
sensitivity of \(\tau\) to \(\theta\), captured by its gradient,
\(\omega(\theta)=\partial\tau(\theta)/\partial\theta\), determines how, to first order, bias in
the estimates propagates to the final estimand.

\begin{ass}[First-order approximation] \label{ass:linearity} For a preliminary value \(\theta_0^{\mathrm{obs}} \in \mathbb{R}^p\), let $\tau$ be differentiable and define 
$ 
\omega= \frac{\partial \tau(\theta)}{\partial \theta}\Big|_{\theta = \theta_0^{\mathrm{obs}}},
\omega_0=\tau(\theta_0^{\mathrm{obs}})-\omega^\top\theta_0^{\mathrm{obs}}. 
$
Define the first-order approximation
\[
\tau_L(\theta)
=
\tau(\theta_0^{\mathrm{obs}})+\omega^\top(\theta-\theta_0^{\mathrm{obs}})
=
\omega_0+\omega^\top\theta.
\]
for known weights \(\omega\in\mathbb R^p, \omega \neq \mathbf{0}\) and uniformly bounded.
\end{ass}

The analysis below is conducted for \(\tau_L(\theta)\). When \(\tau\) is linear, this approximation is
exact; when \(\tau\) is smooth, it is the first-order approximation. As for Example \ref{exmp:gmm}, Section~\ref{sec:gmm-extension} shows how focusing on \(\tau_L(\theta)\) is
asymptotically equivalent to studying \(\tau(\theta)\) under standard local-misspecification, and similarly when replacing $\theta_0^{\mathrm{obs}}$ with its consistent estimate for estimating $\omega$. Thus, building on standard local misspecification frameworks, \(\omega\)
captures the first-order effect of perturbations in \(\theta\) 
on $\tau$.

\begin{exmp}[Choosing the site for an experiment for external validity]
\label{exmp:site_selection}
Consider the problem of choosing where to conduct an experiment
\citep{gechter2024selecting,olea2024externally}.

In our notation, consider two sites \(j\in\{1,2\}\) with
site-specific average treatment effects \(\theta_j\). The target is the cross-site
average
$ 
\tau(\theta)=\omega_1\theta_1+\omega_2\theta_2,
$ 
where \(\omega_j\) denotes the population share in site \(j\). Let
$ 
S^{\mathrm{obs}}
=
(S^{\mathrm{obs}}_1,S^{\mathrm{obs}}_2)^\top
$ 
denote observational estimates of the two site-specific effects. These estimates
may be biased, for example because of confounding. Write
$ 
\mathbb E[S^{\mathrm{obs}}]-\theta
=
b
=
(b_1,b_2)^\top .
$ 
Suppose a budget constraint allows an experiment in only one site. If site \(j\) is chosen, the researcher obtains
an experimentally valid estimate \(S^{\mathrm{exp}}_j\) of \(\theta_j\);
the other site \(k\neq j\) remains informed only by the observational estimate.  
\qed
\end{exmp}

\begin{exmp}[Choosing which survey to conduct]
\label{exmp:survey_design}
\citet{egger2022general} study the efficacy of cash-transfer programs on the marginal
propensity to consume (MPC). Measuring the MPC requires capturing both short- and
long-run effects. Because survey rounds are limited, the authors complement experimental
data that lacks short-run effects with auxiliary information from prior studies that use
short-run surveys collected in other regions; see \cite{egger2022general}. This raises the
question of which survey to conduct and with which frequency.

In stylized form, suppose researchers can observe, for \(s\in\{1,2\}\), potential outcomes
\(Y_s(t)\) denoting consumption in period \(s\) when measured \(t\) periods after the
intervention. Auxiliary estimates from previous studies identify 
$ 
\theta_1
=
\mathbb E\!\left[Y_1(t=1)-Y_1(t=\infty)\right],
\theta_2
=
\mathbb E\!\left[Y_2(t=1)-Y_2(t=\infty)\right],
$
possibly with external-validity bias. The target is
$ 
\tau(\theta)=\theta_1+\theta_2.
$ 
Researchers consider one of two possible survey designs: an early survey design that experimentally learns
\(\theta_1\), and a later survey design that experimentally learns \(\theta_2\). 
\qed
\end{exmp}

\begin{exmp}[Experiment with nonlinear target and equilibrium model]
\label{exmp:3}
\cite{bergquist2020competition} conduct demand and supply experiments in food markets in
Kenya. Suppose here we are interested in similar applications in Uganda. For exposition,
consider a stylized linear demand and supply model (similar reasoning applies to more
complex models)
\[
\small 
\begin{aligned} 
Q^D=a-\theta_D P+u_D,
\qquad
Q^S=c+\theta_S P+u_S,
\end{aligned} 
\]
with \(\theta_D\neq-\theta_S\). Several estimands may be of interest; one such estimand is the
effect of a tariff \(t\) on prices,
$ 
\tau(\theta) \equiv \frac{\theta_S}{\theta_D+\theta_S}\,t.
$ 
Researchers observe baseline
estimates from Kenya, which may lack external validity. Due to fixed costs of each experiment,
$ 
\mathcal E\in\{\{D\},\{S\}\},
$ 
i.e., we can learn either demand, estimating \(\theta_D\) with
randomized price discounts, or supply, estimating \(\theta_S\) by
introducing regulatory cost shocks on firms. 
Let
\(\theta_0^{\mathrm{obs}}=(\theta_{D0},\theta_{S0})^\top\) denote the probability limit of a
preliminary estimator in Kenya. From a first-order approximation in
Assumption~\ref{ass:linearity} (see Section~\ref{sec:gmm-extension}),  
$ 
\tau_L(\theta)
=
\tau(\theta_0^{\mathrm{obs}})
+
\omega^\top(\theta-\theta_0^{\mathrm{obs}}), \omega
=
\frac{\partial \tau(\theta)}{\partial \theta}
\Big|_{\theta=\theta_0^{\mathrm{obs}}}$.  \qed 
\end{exmp}

\subsection{Design problem}
\label{sec:bias_variance_geometry}

Our design problem has three components:
\begin{enumerate}
\item \textit{Use of the reported evidence.} For each feasible
$(\mathcal E,\Sigma)$, specify how the report
$\widetilde S_{\mathcal E,\Sigma}$ is used to learn the target estimand
$\tau(\theta)$. We consider two scenarios:
\begin{itemize}
\item[(A)] Section~\ref{sec:linear} studies researchers who report an
(asymptotically) linear estimator, as with the methods-of-moments estimators
in Example~\ref{exmp:gmm}. The weighting matrix may be fixed ex ante or
optimized jointly with the design.
\item[(B)] Section~\ref{subsec:fully_adaptive_benchmark} studies researchers
who report the available evidence to an audience that forms its own
posterior, without restrictions on the class of estimators.
\end{itemize}

\item \textit{Choice of precision.} For each feasible set of designs, choose $\Sigma$, which determines the sample allocation and
precision of the experimental estimates.

\item \textit{Choice of experiments.} Choose a feasible set of experiments
$\mathcal E$.
\end{enumerate}

Choosing how to combine existing evidence for a fixed design
follows a long-standing tradition
\citep[e.g.][]{armstrong2024adaptingmisspecification,
andrews2021model,de2020empirical,BonhommeWeidner2021arXiv}.
Here we study the different question of how to generate experimental data which makes optimization more challenging and changes the decision problem and notion of oracle. Even in settings where researchers may not need to commit to a single estimator/statistic to report, and instead report the full evidence (Case (B)), choosing the most informative design entails significant trade-offs due to cost constraints on which moments to generate.

\vspace{3mm} 

The natural starting point is to choose the design that minimizes the MSE
(or, as in Section~\ref{sec:CI}, confidence-interval length). The difficulty
is that the MSE depends on the observational bias $b$, whose magnitude is
unknown when the experiment is designed. 

\begin{defn}[Uncertainty set]
\label{defn:uncertainty_set}
Let $\mathcal C\subseteq\mathbb R^{p_o}$ be an arbitrary user-specified compact and
convex set containing zero. For $B\geq0$, define
$ 
\mathcal B(B)
\equiv
\left\{
Bu:u\in\mathcal C
\right\}.
$ 
The scalar $B$ indexes the worst-case magnitude of potential observational bias,
whereas $\mathcal C$ describes additional  restrictions. $B$ is \textit{unknown} ex-ante to the researcher designing the experiment. 
\end{defn}

The set $\mathcal C$ can incorporate prior restrictions on the shape or
sign of the observational bias and can be specified before the experiment
is conducted. For example, a natural choice, after standardizing parameters measured on
different scales, is the $\ell_\infty$-ball,
\[
\mathcal C
=
\left\{
u:\lVert u\rVert_\infty\leq1
\right\},
\qquad
\mathcal B(B)
=
\left\{
b:\lVert b\rVert_\infty\leq B
\right\}.
\]
The $\ell_{\infty}$ has the advantage that does not force a trade-off across coordinates, as biases may be
simultaneously large in several components. Weighted versions can
accommodate parameters measured on different scales. 
The main challenge however is that we do not assume that
$B$ is known ex ante as it may be difficult to elicit for all feasible experiments \textit{before} running the experiment. (When a conservative bound $\bar{B}$ is known Section~\ref{sec:bound_known} considers 
$B\leq \bar{B}$).

The structure  in Definition \ref{defn:uncertainty_set} provides us with the following variance-sensitivity trade-off. 

\begin{defn}[Variance, sensitivity, and risk curve]
\label{defn:variance_regret}
For a vector $v \in \mathbb{R}^{p_o}$ define $\lVert v\rVert_*^2
\equiv
\sup_{u\in\mathcal C}(v^\top u)^2$. 
For any $a\in\mathbb R^{p_o+|\mathcal E|}$ satisfying
$\Lambda_{\mathcal E}^{\top}a=\omega$, and for  $B \ge 0$, define
\begin{equation}
\label{eqn:alpha_beta}
\small
\begin{aligned}
\alpha(\mathcal E,\Sigma,a)
&\equiv
a^\top\Sigma a,
&
\beta(\mathcal E,a)
&\equiv
\lVert a_{1:p_o}\rVert_*^2,
& R_B(\mathcal E,\Sigma,a)
\equiv
\alpha(\mathcal E,\Sigma,a)
+
B^2\beta(\mathcal E,a)
\end{aligned}
\end{equation}
\end{defn}

The component $\alpha$ measures sampling uncertainty, whereas $\beta$
measures sensitivity to observational misspecification, which is closely related to \cite{andrews2020transparency}'s sensitivity definition. To see this, for Case~(A), it follows immediately that 
\begin{equation}
\label{eqn:linear_risk_curve}
\sup_{b\in\mathcal B(B)}
\mathbb E_{\mathcal E,\Sigma,b}
\left[
\Big(
\widehat\tau_a-\tau_L(\theta)
\Big)^2
\right]
=
R_B(\mathcal E,\Sigma,a), \qquad \widehat\tau_a
\equiv
\omega_0+a^\top\widetilde S_{\mathcal E,\Sigma}.
\end{equation}

In Case~(B), we show that for a class priors $\pi \in \Pi_B$ that bounds the second moment of $b$ within $\mathcal{B}(B)$, 
\begin{equation}
\label{eqn:audience_risk_curve_intro}
\small 
\begin{aligned} 
\sup_{\pi \in \Pi_B} \mathbb{E}_\pi\left[\Big(\tau_L(\theta) - \delta_\pi(\tilde{S}_{\mathcal{E},\Sigma})\Big)^2\right]
=
\min_{a:\Lambda_{\mathcal E}^{\top}a=\omega}
R_B(\mathcal E,\Sigma,a), \qquad \delta_\pi(\tilde{S}) =\mathbb{E}_\pi[\tau_L(\theta) | \tilde{S}], \qquad (\theta,b) \sim \pi 
\end{aligned} 
\end{equation}
Thus, both formulations are characterized by the same variance and
sensitivity objects, where, in Case~(B), the auxiliary weights in
\eqref{eqn:audience_risk_curve_intro} are profiled out separately for each
$B$ even when the audience's posterior estimator may be nonlinear. The following sections  use these structures as starting points to study optimal design, by letting $B$ be unknown.

\section{Experimental design for linear estimators}
\label{sec:linear}

In this section we study the design of experiments in settings where researchers consider linear estimators, as for typical weighting/GMM or minimum distance estimators.

We introduce below a class of exactly linear estimators where the weighting matrix does not depend on the estimates $\tilde{S}$. Our results continue to hold asymptotically if we replace the weighting matrix with a consistent estimator that admits a fixed probability limit. 

\begin{setting}[Linear estimators] \label{setting:1} For a given \((\mathcal E,\Sigma)\), let \(W\) be a symmetric positive semidefinite
weighting matrix of dimension \(p_o+|\mathcal E|\), so that \(\Lambda_{\mathcal E}^\top W\Lambda_{\mathcal E}\) is nonsingular. Consider linear estimators of the parameters $\theta$ and corresponding (linearized) estimator for $\tau(\theta)$
\[
\small 
\begin{aligned} 
\hat\theta_W
\in
\arg\min_{\theta'\in\mathbb R^p}
\left(
\tilde{S}_{\mathcal E,\Sigma}
-
\Lambda_{\mathcal E}\theta'
\right)^\top
W
\left(
\tilde{S}_{\mathcal E,\Sigma}
-
\Lambda_{\mathcal E}\theta'
\right), \qquad \widehat{\tau}_{a_W} \equiv \tau_L(\hat\theta_W)
=
\omega_0 + a_W^\top \tilde{S}_{\mathcal{E},\Sigma} 
\end{aligned}
\]
where $a_W^\top = \omega^\top \left(
\Lambda_{\mathcal E}^\top W\Lambda_{\mathcal E}
\right)^{-1}
\Lambda_{\mathcal E}^\top W$. 
It follows that $a_W$ satisfies $\Lambda_{\mathcal{E}}^\top a_W = \omega$. 


The matrix \(W\) can be chosen as a function of
\(\Sigma\), \(\Lambda_{\mathcal E}\), \(\mathcal E\) and/or may belong to some pre-specified constraint set that may incorporate a particular choice or preference of the researcher.
To allow for all such cases, we will refer, for a given choice of design $(\mathcal{E},\Sigma)$ broadly to $W \in \mathcal{W}(\mathcal{E},\Sigma)$ where the space $\mathcal{W}$ may incorporate any user-specific additional constraint imposed on $W$. The matrix $W$ is not a function of $\tilde{S}_{\mathcal{E},\Sigma}$.

\end{setting}

Although, for simplicity, we assume $W$ does not depend on realized estimates $\tilde{S}$, in an asymptotic regime, researchers may replace \(W\) by an estimated matrix \(\hat W\), possibly constructed also using \(\tilde{S}_{\mathcal E,\Sigma}\). Our results below will apply in the limit provided
$ 
\hat W \to_p W
$, 
for some \textit{non-random} matrix \(W\) such that \(\Lambda_{\mathcal E}^\top W\Lambda_{\mathcal E}\) is nonsingular. In this case
$
a_{\hat{W}} = a_W+o_p(1),
$ 
under which the feasible estimator is first-order equivalent to the linear estimator indexed by \(W\). To choose ex-ante the design, the estimated weighting matrix $\hat{W}$ can be evaluated at locally misspecified preliminary estimates $\widehat{\theta}_0^{\mathrm{obs}}$ under local misspecification in Section \ref{sec:gmm-extension}.\footnote{Under the local asymptotics, to also allow for cases where $b$ grows faster than $1/\sqrt{n}$, the rate of convergence of $\hat{W}$ to $W$ needs to be faster than $n^{-1/4}$ rate in the experimental sample size, which is standard.} 

Similarly, all the calculations are for the first order expansion \(\tau_L\). If the researcher
instead reports the \textit{nonlinear} plug-in estimator \(\tau(\hat\theta_W)\), then under the same local
misspecification argument, \(\tau(\hat\theta)\) and
\(\tau_L(\hat\theta)\) are first-order equivalent and our results hold asymptotically.  

The focus on this class is to accommodate estimators common in applications, including minimum-distance and GMM estimators. A special case is also the practice of relying solely on experimental evidence for the parameters for which an experiment is available, and use observational estimates for the parameters not identified by the experiment.

This class excludes more general weighting rules whose first-order limit remains random ($\hat{W} \not \rightarrow_p W$ for some fixed $W$), as for adaptive non-linear estimators in \cite{armstrong2024adaptingmisspecification} and \cite{de2020empirical}. See Section~\ref{sec:adaptive_weights} for a discussion.

\paragraph{Example \ref{exmp:direct} continued (direct parameter estimation)} 
Returning to Example~\ref{exmp:direct}, suppose \(W\) is diagonal for simplicity. With appropriate normalization, consider the estimator
\begin{equation} \label{eqn:weight_estimator}
\hat{\theta}_{W,j} = \begin{cases} S_j^{\mathrm{exp}} (1 - W_{jj})  + W_{jj} S_j^{\mathrm{obs}},  & j \in \mathcal{E} \\ 
 S_j^{\mathrm{obs}}  & j \not \in \mathcal{E}
\end{cases}, \qquad W_{jj} \in [0,1], \qquad  j \in \{1, \cdots, p\}, 
\end{equation} 
for a user choice of $W_{jj}$. 
The linear class includes the limiting empirical rule that uses
experimental evidence for the coordinates learned experimentally and relies on
observational evidence only for the components of the target not covered by the experiment ($W_{jj} = 0$). 

\paragraph{Example \ref{exmp:gmm} continued (methods of moments)} The use of methods of moments is common in the class of applications we consider. For instance, \cite{meghir2022migration} propose a structural model that leverage observational information combined with experimental variation for estimating welfare effects of subsidies for migration. They estimate the structural model via GMM fixing the weighting matrix to be the identity matrix. In our framework this corresponds to asymptotically linear estimator with $W = \mathbb{I}$. \cite{attanasio2012education} uses a method of moments estimator (corresponding to the score of the likelihood function) for a schol choice model. \cite{de2025decoupling} study the effect of a randomized digital campaign on voting behavior combined with a structural model estimated via GMM. \qed 
\begin{exmp}[Additional examples]
\citet{athey2025experimental} propose combining experimental and observational data to correct
confounding in observational estimates. In their linear model of confounding bias, their selection correction estimator is an exactly linear estimator, while it is asymptotically linear under mild smooth moment restrictions. \cite{ishihara2021evidence} also study useful linear estimators for aggregating existing evidence. \qed 
\end{exmp}

\subsection{Objective function} \label{sec:robust_design}

Following the identity between MSE and risk-curve in Equation \eqref{eqn:linear_risk_curve}, if an oracle knew $B$, a natural choice would be to pick the minimax design
\[
\small 
\begin{aligned} 
R^*(B)\;\equiv\;
\inf_{(\mathcal{E},\Sigma)\in\mathcal{D}, W \in \mathcal{W}(\mathcal{E},\Sigma)}
\ R_B(\mathcal{E},\Sigma,a_W).
\end{aligned} 
\]
For finite $B$ this would lead to natural bias-variance trade-offs. 
However, having to specify $B$ can pose a large burden on the researchers and make the choice of the experiment sensitive to $B$. Therefore, we seek designs that perform as close as possible to the oracle that knows $B$, uniformly over the values of $B$, following a long-standing tradition in decision theory \citep[e.g.][]{manski2007admissible, montiel2026decision, stoye2011axioms}.

\begin{defn}[Proportional regret] We minimize the worst-case proportional regret 
\begin{equation} \label{eqn:regreta}
\mathcal{R}^{\mathrm{MSE}}(\mathcal{E},\Sigma,W)
\;\equiv\;
\sup_{B\ge 0}\ 
\frac{\displaystyle R_B(\mathcal{E},\Sigma,a_W)}
{\displaystyle R^*(B)}. 
\end{equation}
\end{defn}
The use of proportional regret follows a long-standing tradition in
optimization and robust decision-making
\citep[e.g., also common in standard textbooks][]{vazirani2001approximation}. 
Related ratio criteria have also
been used in several contexts \citep[e.g.][]{ tsybakov1998pointwise,armstrong2024adaptingmisspecification}, with the difference that all such references keep the design fixed. 

Standardized ratios were used as efficiency measures in classical optimal design literature under a correct linear model
specification \citep[][]{dette1997designing}, where, each
coefficient-specific variance is normalized by its coefficient-specific
minimum variance. Here instead, regret depends on unknown biases due to the different extrapolation problem with data combination.

\begin{rem}[Alternative objectives]
\label{rem:alternative_objectives}
Two natural alternatives to proportional regret are minimax MSE and minimax additive
regret. These objectives are useful in many settings \citep[see][for relevant examples]{manski2007admissible, manski2004, stoye2009minimax}. In our context, however, if one takes a
worst case over all \(B\ge 0\), minimax MSE and minimax additive regret  would either be uninformative or place
lexicographic priority on minimizing bias, selecting designs with minimal
bias, at the expense of potentially very large variance. This also occurs in our leading applications when not all parameters can be learned experimentally.\footnote{For minimax solutions this follows directly due to additivity of MSE in $B^2$. For minimax additive regret, defined as the difference between the researcher and the oracle MSE, if the bias is larger than the smallest achievable bias ($\beta > \beta^\star$ with $\beta^\star$ its minimum), additive regret can also be made arbitrary large.} By contrast, proportional regret preserves a bias-variance trade-off that may better reflect researcher's preferences towards interior (non-lexicographic) solutions in these same leading applications. (Also, if a conservative upper bound $\bar{B}$ is available, the same criterion can be defined worst-case over $B \le \bar{B}$, see Section \ref{sec:bound_known}).
\qed
\end{rem}




\subsection{Optimal design as an intuititive geometric solution} \label{sec:optimal_3}

Our first result shows that the proportional regret provides natural trade-offs between the variance and the bias. To do so, it will be useful to characterize the ideal minima. 

\begin{defn} \label{defn:minima} Let
$ 
\alpha^\star \;\equiv\; \inf_{(\mathcal E, \Sigma) \in \mathcal{D}, W \in \mathcal{W}(\mathcal{E},\Sigma)} \alpha(\mathcal E,\Sigma,a_W),   \beta^\star \equiv \inf_{(\mathcal E, \Sigma) \in \mathcal{D}, W \in \mathcal{W}(\mathcal{E},\Sigma)} \beta(\mathcal{E},a_W).
$ 
Denote $\alpha/\alpha^\star, \beta/\beta^\star$ the variance and bias regret, respectively.  
\end{defn}

The variance regret denotes the ratio between the variance of a given design and estimator relative to the smallest achievable variance; $\beta^\star$ denotes the smallest achievable sensitivity. 

\paragraph{Example \ref{exmp:direct} continued} Continue with the estimator in Equation \eqref{eqn:weight_estimator},
 and consider worst-case bias deviations of the form $||b||_{\infty} \le B$, so that $\mathcal{C} = \{u: ||u||_{\infty} \le 1\}$. Then
$$
\small 
\begin{aligned} 
\sqrt{\beta(\mathcal{E}, W)} = \underbrace{\sum_{j \not \in \mathcal{E}} |\omega_j|}_{\text{sensitivity to observational estimates}} + \underbrace{\sum_{j' \in \mathcal{E}} W_{j'j'} |\omega_{j'}|}_{\text{sensitivity due to shrinkage towards observational}};  
\end{aligned} 
$$ 
in the special case where $W_{jj} = 0$ for all $j \in \mathcal{E}$, so that those experimental estimates collected by researchers receive all the weight, $\sqrt{\beta(\mathcal{E},W)} = ||\omega||_1 - ||\omega_{\mathcal{E}}||_1$ denote the absolute sum of the sensitivity of $\tau(\cdot)$ to those parameters that haven't been learned experimentally. 

In our leading applications, we expect $\beta^\star \neq 0$: it is often infeasible to run an experiment able to identify all the parameters of interest. 
\qed

\vspace{2mm} 

An ideal design would set the variance equal to $\alpha^\star$ and bias equal to $\beta^\star$. Unfortunately, this may be infeasible as it might require different choices of experiments and estimators to achieve one or the other. This raises the question of how to trade-off these two components. 

This trade-off is formally characterized in our theorem below. We will use the convention throughout that 0/0 = 1.

\begin{thm}\label{thm:1}
Consider Setting \ref{setting:1} and let Assumptions~ \ref{ass:1a}, \ref{ass:constraint_set}, \ref{ass:linearity} hold. Then, for any $(\mathcal E,\Sigma) \in \mathcal{D}, W \in \mathcal{W}(\mathcal{E},\Sigma)$,
\[
\small 
\begin{aligned} 
\mathcal{R}^{\mathrm{MSE}}(\mathcal E,\Sigma,W)
\;=\;
\max\!\left\{
\frac{\alpha(\mathcal E,\Sigma,a_W)}{\alpha^\star},\ \ 
\frac{\beta(\mathcal E,a_W)}{\beta^\star}
\right\}.
\end{aligned} 
\]
\end{thm}

\begin{proof}
See Appendix~\ref{proof:thm:regret_gmm}.
\end{proof}

Theorem \ref{thm:1} offers a simple and novel expression of the regret which does not require researchers to specify $B$.  
As discussed in Remark \ref{rem:opta}, Theorem \ref{thm:1} will significantly simplify the optimization program. 
The key insight is that the proportional regret for linear estimators and arbitrary designs leads to a quasi-convex objective function, whose worst-case solution is the maximum between the variance and the bias regret. This differs from 
the characterization of non-linear estimators as in \cite{tsybakov1998pointwise} and \cite{armstrong2024adaptingmisspecification}, where optimization is instead over worst-case priors due to lack of quasi-convexity, which makes optimization more challenging (see Appendix \ref{app:armstrong}).  A key property here is that provided that the class of designs $\mathcal{D}$ is sufficiently flexible (outside boundary solutions), it follows that at the optimum we would expect to equalize
$$
\small 
\begin{aligned} 
\frac{\alpha}{\alpha^\star}
\;=\;
\frac{\beta}{\beta^\star}. 
\end{aligned} 
$$ 
That is, we allocate sample size to noisier treatment arms when the variance dominates, and select treatment arms with largest sensitivity when the bias dominates. 

\begin{figure}[H]
\centering
\begin{tikzpicture}[scale=1.0]
\draw[->] (0,0) -- (5.2,0) node[below] {$\alpha/\alpha^\star$};
\draw[->] (0,0) -- (0,5.2) node[left] {$\beta/\beta^\star$};

\draw[dashed] (0,0) -- (5,5) node[above right] {$y=x$};

\draw[thick,blue]
(0.6,4.5) .. controls (1.2,2.6) and (2.2,1.8) .. (3.8,1.1);

\filldraw[black] (1,1) circle (2pt)
node[below right] {\footnotesize ideal point $(1,1)$};

\filldraw[red] (2.12,2.12) circle (2pt)
node[above right,align=left]
{$(\mathcal E^\star,\Sigma^\star,W^\star)$\\[-1pt]
 \footnotesize closest frontier point};

\draw[dotted] (0,2.2) -- (2.2,2.2) -- (2.2,0);
\end{tikzpicture}

\caption{\footnotesize
Each feasible design $(\mathcal E,\Sigma,W)$ maps to a point
$\big(\alpha/\alpha^\star,\beta/\beta^\star\big)$: the $x$-axis is the
variance ratio and the $y$-axis is the worst-case bias ratio. The blue
curve depicts the attainable frontier as we vary shrinkage $W$ and
precision $\Sigma$. The point $(1,1)$ is the ideal point, whose two
coordinates are individually attainable but need not be attainable
simultaneously. Level sets of
$\mathcal R=\max\{\alpha/\alpha^\star,\beta/\beta^\star\}$ are
axis-aligned squares. Since every feasible point
lies weakly northeast of $(1,1)$, the red point is the point on the
frontier closest to the ideal point under the normalized
$\ell_\infty$ distance. }
\label{fig:equalization}
\end{figure}

Theorem~\ref{thm:1} also provides a geometric justification for the
optimal design that does not rely on interpreting the objective as
proportional regret over $B$. Map each feasible choice
$(\mathcal E,\Sigma,a)$ into its variance-sensitivity pair
$ 
\left(
\alpha(\mathcal E,\Sigma,a),
\beta(\mathcal E,a)
\right).
$ 
The point $(\alpha^\star,\beta^\star)$ is the ideal point, but infeasible in practice.
Theorem~\ref{thm:1} selects the smallest $R\geq1$ such that
\[
\small 
\begin{aligned} 
\alpha(\mathcal E,\Sigma,a)\leq R\alpha^\star,
\qquad
\beta(\mathcal E,a)\leq R\beta^\star
\end{aligned}
\]
Interestingly, this solution is not specific to Setting \ref{setting:1}; Section \ref{sec:CI} shows how this same solution arises for minimizing confidence interval length when an audience has prior belief about $B$, unknown to the designer.

\begin{rem}[Optimization] \label{rem:opta} Appendix \ref{sec:optimization} provides an explicit optimization routine. In our leading applications of selecting arms and choosing sample size across them, with unrestricted weighting matrix $W$, optimization is convex in the choice of the sample size and estimator. It is mixed-integer over the choice of the experiment $\mathcal{E}$, typically from a small set. This makes optimization easy to solve off-the-shelf using mixed integer convex program solvers.

In addition, the solution in Theorem \ref{thm:1} has the desirable feature that, \emph{conditional} on the chosen experiment and $W$, it takes the form of a Neyman (variance-optimal) allocation. It is therefore natural to refer to our design as a (novel) “bias-aware” Neyman allocation, where the choice of the experiment and  weights trade-off bias and variance, and optimal sample size minimizes the variance conditional on these choices. \qed  
\end{rem}

\begin{exmp}\label{subsec:two_param}
To build intuition, consider a two-parameter model with
\(\theta=(\theta_1,\theta_2)^\top\), and $\mathcal{B}(B) = \{b: ||b||_{\infty} \le B\}$. For each \(j\in\{1,2\}\), suppose that we have an observational estimator \(S_j^{\mathrm{obs}}\) and, if experiment \(j\) is run, an experimental estimator \(S_j^{\exp}\), with mutually independent errors
$ 
S_j^{\mathrm{obs}}-\theta_j\sim \mathcal N(b_j,\sigma_j^2),
S_j^{\exp}-\theta_j\sim \mathcal N(0,v_j^2),
j=1,2.
$ 
For simplicity, \(\sigma_j^2\) and \(v_j^2\) are fixed. Because of budget constraints, the researcher can run only one experiment,
$
\mathcal E\in\big\{\{1\},\{2\}\big\}.
$
If experiment \(j\) is run, write with an abuse of notation \(W_j=W_{jj}\in[0,1]\) for the weight placed on the observational estimator for component \(j\). We write the $j^{th}$-entry estimator as
$\hat\theta_j(W_j)
=
(1-W_j)S_j^{\exp}
+
W_j S_j^{\mathrm{obs}},
\hat\theta_{-j}
=
S_{-j}^{\mathrm{obs}}. 
$ 
To first order,
$
\tau(\theta)-\tau(\hat\theta)
\approx
\omega_1(\theta_1-\hat\theta_1)
+
\omega_2(\theta_2-\hat\theta_2),
$
where we assume $\omega_1, \omega_2 \neq 0$. 

For a fixed experiment \(j\) and weight \(W_j\), the variance-minimizing weight is
$
\frac{v_j^2}{\sigma_j^2+v_j^2},
$
and
\[
\small 
\begin{aligned} 
\alpha^\star
=
\min_{j\in\{1,2\}}
\left\{
\omega_{-j}^2\sigma_{-j}^2
+
\omega_j^2 \sigma_j^2 \frac{v_j^2}{\sigma_j^2+v_j^2}
\right\}, 
 \qquad \beta(j,W_j)
=
\left(
|\omega_{-j}|+W_j|\omega_j|
\right)^2, \qquad \beta^\star
=
\min\{\omega_1^2,\omega_2^2\}
\end{aligned} 
\]
Since \(\alpha(j,W_j)\) is minimized at \(\frac{v_j^2}{\sigma_j^2+v_j^2}\), while \(\beta(j,W_j)\) is increasing in \(W_j\), no value \(W_j>\frac{v_j^2}{\sigma_j^2+v_j^2}\) can be optimal. Therefore, \(W_j^\star\in[0,\frac{v_j^2}{\sigma_j^2+v_j^2}]\), and (see Appendix \ref{proof:cor:gamma_2_param} for details), outside the two boundary cases, we have 
$$
\small 
\begin{aligned} 
W_j^\star = W_j\in(0,\frac{v_j^2}{\sigma_j^2+v_j^2})
\text{ such that }
\dfrac{\alpha(j,W_j)}{\alpha^\star}
=
\dfrac{\beta(j,W_j)}{\beta^\star}
\end{aligned} 
$$ 
assuming such $W_j$ exist.  
The optimal experiment is
$
j^\star
\in
\arg\min_{j\in\{1,2\}}
\max\left\{
\frac{\alpha(j,W_j^\star)}{\alpha^\star},
\frac{\beta(j,W_j^\star)}{\beta^\star}
\right\}.
$ 
\end{exmp}

 \subsection{Main applicability and non-linear estimators}
\label{sec:adaptive_weights} 

 We pause here to discuss benefits and limitations of the solution in Theorem \ref{thm:1}, and comparisons to settings with non-linear estimators studied for fixed-design problems in the literature. 

First, note that the characterization in Theorem~\ref{thm:1} is most useful in the regular case in which
$
\alpha^\star>0,
\beta^\star>0.
$
The first condition is a standard variance non-degeneracy condition. The second condition defines the main class of applications. It says that, given the budget and feasibility constraints encoded in \(\mathcal D\), no feasible design and estimator can eliminate worst-case bias in \(\tau(\theta)\). This is the extrapolation problem that motivates our framework: experiments can identify some policy-relevant parameters or moments, but not the full counterfactual.

The boundary case \(\beta^\star=0\) is outside our main scope of application; it implies that the feasible set contains a design and estimator that identify the target without any worst-case bias. It arises naturally, however, in the classical problem of combining two estimators for a single target parameter, when one estimator is unbiased for that same target \citep[e.g.][]{gerber2004illusion}. When $\beta^\star = 0$, the proportional-regret criterion over fixed linear rules collapses to a lexicographic preference as other minimax criteria in Remark \ref{rem:alternative_objectives}. In this case non-linear procedures, such as those in \cite{armstrong2024adaptingmisspecification,de2020empirical}, avoid such lexicographic preferences. Although useful for estimation, these can be difficult to compute for the different problem of choosing the  \textit{design}, see Appendix \ref{sec:gaussian_priors}.  

In the next section we study design for an audience that may ex-post adapt their belief about $b$ potentially non-linearly. In addition, in Section \ref{subsec:minimax_adaptive} we show that optimal designs under linear rules provide surrogate upper bounds for arbitrary non-linear rules.

\section{Experimental design for general reporting rules}
\label{subsec:fully_adaptive_benchmark}

In some applications, researchers may not want to restrict attention to a particular estimator
for combining experimental and observational evidence. Instead, they may report the available
statistics and allow an audience to update according to their own prior beliefs about the
observational bias. This subsection studies such a formulation.

\begin{setting} \label{setting:2}
    Let \(\pi\) denote an audience prior on \((\theta,b)\), so that
\[
\small 
\begin{aligned} 
(\theta,b)\sim \pi, \qquad \pi \in \Pi_B, 
\end{aligned} 
\]
for a class $\Pi_B$ indexed by a parameter $B \ge 0$. 
The prior \(\pi\) captures the audience's beliefs about both the target parameter and the
possible bias in the observational evidence. The prior \(\pi\) and parameter $B$ indexing $\Pi_B$ are unknown to the researcher.  For a design \((\mathcal E,\Sigma)\), researchers report for known $\Lambda_{\mathcal{E}}, \Sigma$
\begin{equation} \label{eqn:guassian}
\small 
\begin{aligned}
\tilde{S}_{\mathcal E,\Sigma} | (\theta,b) \sim \mathcal{N}\left(\Lambda_{\mathcal{E}} \theta + \begin{pmatrix}
b\\
0_{|\mathcal{E}|}
\end{pmatrix}, \Sigma\right), \qquad b \in \mathbb{R}^{p_o}.
\end{aligned}
\end{equation} 
Given the report
\(\tilde{S}_{\mathcal E,\Sigma}\), an audience with prior \(\pi\) forms the posterior
distribution of \(\theta\), and of the (linearized) estimand $\tau_L(\theta)$. See Figure \ref{fig:communication_model} for an illustration.\qed 
\end{setting}

Setting \ref{setting:2} captures scenarios where researchers may report the vector of statistics, and allow the audience to form their posterior belief akin to models of scientific communication \citep[e.g.][]{andrews2021model, banerjee2020theory}, here studied for choosing which experiments to conduct to report (and their precision).

We capture ambiguity through the unknown prior $\pi$ and restriction class $\Pi_B$ (while the map $B \mapsto \Pi_B$ is assumed to be known, $B$ is unknown). 
The Gaussian assumption is motivated by standard asymptotic approximations; similarly, following Assumption \ref{ass:linearity}, we focus on the linearized estimand $\tau_L$, which is interpreted as a local approximation.

Under squared-error loss, the  expected posterior risk of the audience updating their own beliefs under prior $\pi$, and its worst-case formulation over $\pi \in \Pi_B$ are
\begin{equation} \label{eqn:L_B_audience}
\small 
\begin{aligned} 
L_\pi(\mathcal E,\Sigma) \equiv \mathbb{E}_\pi\left[\Big(\tau_L(\theta) - \mathbb{E}_\pi[\tau_L(\theta) | \tilde{S}_{\mathcal{E},\Sigma}]\Big)^2\right], \qquad \bar{L}_B(\mathcal{E},\Sigma) \equiv \sup_{\pi \in \Pi_B} L_\pi(\mathcal E,\Sigma)
\end{aligned} 
\end{equation} 
where the expectation is taken over the prior predictive distribution of
\(\tilde{S}_{\mathcal E,\Sigma}\) induced by \(\pi\) and the sampling model in
Setting~\ref{setting:2}. 
These criteria separate the design problem from the choice of the estimator: the
researcher chooses what evidence to generate and report without committing to a given estimator (either linear or non-linear), while the audience chooses the
posterior estimator implied by the prior. Because  $\tilde{S}_{\mathcal{E},\Sigma}$ is not observed ex-ante, we evaluate designs by the expected
Bayes risk, where take the expectation over $\tilde{S}_{\mathcal{E},\Sigma}$.

Similarly to our linear formulation, $B$ controls the bias-variance trade-off. In practice, the researcher's knowledge may be limited, and may lack information about relevant restrictions on the bias captured through $B$. Therefore, consistent with the previous section, we evaluate designs using
proportional regret relative to an oracle that knows the prior class $\Pi_B$. 

\begin{defn}[Audience proportional regret] \label{defn:audience_regret} For a given set of priors $\pi \in \Pi_B$, indexed by $B \ge 0$, denote the audience proportional regret relative to an oracle with knowledge of $\Pi_B$ as  
\[
\small 
\begin{aligned} 
\mathcal R^{\mathrm{Bayes}}(\mathcal E,\Sigma)
=
\sup_{B \ge 0} 
\frac{ \bar{L}_B(\mathcal{E},\Sigma)
}{
\inf_{(\mathcal{E}',\Sigma')\in \mathcal{D}}\bar{L}_B(\mathcal{E}',\Sigma')
}.
\end{aligned} 
\]
\end{defn} 
The numerator is the worst-case expected posterior risk induced by the chosen design, while the
denominator is the smallest worst-case expected posterior risk attainable by an oracle that knows
\(\pi \in \Pi_B\) for a given feasible subset of priors $\Pi_B$. The objective asks for a design that performs well uniformly over the
audience priors in \(\Pi_B, B \ge 0\), relative to an oracle that knows $B$, without the researcher committing to a single estimator.  

The proportional normalization plays the same role here as in the previous section. To see this, suppose that \(\Pi\) contains priors which are highly diffuse for $b$. In this case, an audience places little weight on observational evidence, and the posterior risk of any design is driven by the components of the target that can be learned experimentally. A minimax posterior-risk criterion, \(\sup_{\pi\in\Pi}L_\pi(\mathcal E,\Sigma)\), would therefore be dominated by the most skeptical priors in \(\Pi\). Similarly, an additive-regret criterion may give lexicographic priority to designs that perform best in this large-bias-prior region. The ratio allows the researcher to remain agnostic over the audience's prior beliefs without letting the design be dictated only by the most pessimistic beliefs about observational bias. The constraints in $\mathcal{D}$ control the oracle risk to be bounded away from zero.

\begin{figure}[t]
\centering
\begin{tikzpicture}[
    >=latex,
    timeline/.style={thick},
    event/.style={
      rectangle, draw, rounded corners,
      align=left, inner sep=4pt, font=\footnotesize,
      text width=3.35cm
    },
    note/.style={
      align=center, font=\scriptsize,
      text width=3.35cm
    }
]

\draw[timeline,->] (0,0) -- (10.7,0);

\foreach \x/\lab in {0/$t=0$,5/$t=1$,10/$t=2$} {
  \draw[timeline] (\x,0.1) -- (\x,-0.1);
  \node[below=0.22cm] at (\x,0) {\lab};
}

\node[event, above=0.55cm] (design) at (0,0) {%
    \textbf{Design}\\
    Researcher chooses\\
    experiment set and precision\\
    $(\mathcal E,\Sigma)\in\mathcal D$
};

\node[event, above=0.55cm] (report) at (5,0) {%
    \textbf{Communication}\\
    Researcher reports\\
    $\tilde{S}_{\mathcal E,\Sigma}$ and $(\mathcal E,\Sigma)$
};

\node[event, above=0.55cm] (audience) at (10,0) {%
    \textbf{Audience updating}\\
    Audience has prior $\pi\in\Pi$\\
    forms posterior for $\tau(\theta)$\\
    and incurs Bayes risk
};

\draw[->,timeline] (design.east) -- (report.west);
\draw[->,timeline] (report.east) -- (audience.west);

\end{tikzpicture}
\caption{\footnotesize Timeline of the communication model. The researcher chooses what evidence to generate and report, while heterogeneous audiences update according to their own priors.}
\label{fig:communication_model}
\end{figure}

In the following lines we provide an exact characterization under the class of priors below.

\begin{ass}[Audience priors]
\label{defn:class_priors}
For each \(B\geq 0\), let \(\Pi_B\) denote the class of priors \(\pi\)
satisfying
$ 
\mathbb E_\pi
\left[
\left(q^\top b\right)^2
\mid\theta
\right]
\leq
\sup_{\widetilde b\in\mathcal B(B)}
\left(q^\top\widetilde b\right)^2
\qquad
\text{for every }q\in\mathbb R^{p_o},
\quad \pi\text{-almost surely}.
$ 
\end{ass}

Here the scalar $B^2$ indexes the magnitude of the bias under the audience's prior. 
An example include settings where $\mathcal{C} = \{v: ||v||_{\infty} \le 1\}$,  and  $\Pi_B = \{\pi: \mathbb{E}_\pi[b_j^2|\theta] \le B^2, \text{ for all $j$}\}$, or weighted versions of it ($\mathbb{E}_\pi[b_j^2|\theta] \le \kappa_j B^2$ for weights $\kappa$). This restriction is also intuitive, as it imposes that the second moment is bounded by $B^2$, while the  correlations between bias components, which are often hard to specify, remain unrestricted. The oracle may know the prior variance scale \(B^2\), in the same way that the oracle in
Section~\ref{sec:linear} may know the bias radius. See Figure \ref{fig:directional_second_moment_prior}. 

\begin{figure}[!ht]
\centering
\begin{tikzpicture}[scale=0.95,>=Stealth]
\tikzset{
  every node/.style={font=\scriptsize},
  axis/.style={->,gray!70},
  priorA/.style={thick},
  priorB/.style={thick,densely dashed},
  title/.style={font=\small}
}

\begin{scope}[xshift=0cm]
\node[title,align=center] at (0,2.25)
{(a) Direction is unrestricted};

\draw[axis] (-1.75,0) -- (1.85,0) node[right] {$b_1$};
\draw[axis] (0,-1.45) -- (0,1.65) node[above] {$b_2$};

\draw[priorA,rotate around={30:(0,0)}]
  (0,0) ellipse [x radius=1.15,y radius=0.12];

\draw[priorB,rotate around={118:(0,0)}]
  (0,0) ellipse [x radius=1.05,y radius=0.24];

\fill (0,0) circle (1.2pt);

\draw[priorA] (-1.60,-1.82) -- (-1.18,-1.82);
\node[anchor=west] at (-1.08,-1.82)
  {$b$ concentrated in one direction};

\draw[priorB] (-1.60,-2.12) -- (-1.18,-2.12);
\node[anchor=west] at (-1.08,-2.12)
  {$b$ concentrated in another direction};

\node[gray!70,align=center] at (0,-2.52)
{both are possible under the same \(\Pi_B\)};
\end{scope}

\begin{scope}[xshift=6.8cm]
\node[title,align=center] at (0,2.25)
{(b) \(B\) bounds magnitude};

\draw[axis] (-1.75,0) -- (1.85,0) node[right] {$b_1$};
\draw[axis] (0,-1.45) -- (0,1.65) node[above] {$b_2$};

\fill (0,0) circle (1.2pt);

\draw[priorA,rotate around={30:(0,0)}]
  (0,0) ellipse [x radius=0.62,y radius=0.18];

\draw[priorB,rotate around={30:(0,0)}]
  (0,0) ellipse [x radius=1.22,y radius=0.35];

\draw[priorA] (-1.48,-1.82) -- (-1.06,-1.82);
\node[anchor=west] at (-0.96,-1.82) {$B_1$};

\draw[priorB] (-1.48,-2.12) -- (-1.06,-2.12);
\node[anchor=west] at (-0.96,-2.12)
  {$B_2$, \(B_2>B_1\)};

\node[gray!70,align=center] at (0,-2.52)
{same direction, different second-moment bounds};
\end{scope}

\end{tikzpicture}

\caption{\footnotesize
Geometry of the second-moment prior class for \(p_o=2\).
Panel (a) fixes \(B\) and illustrates two priors in \(\Pi_B\) for
which uncertainty about \(b\) is concentrated in different directions.
The direction is not specified by the designer.
Panel (b) fixes an illustrative directional pattern and increases
\(B\). The ellipses are only illustrative: priors in \(\Pi_B\)
need not be Gaussian or conditionally mean zero.}
\label{fig:directional_second_moment_prior}
\end{figure}

\subsection{Solution as a Pareto frontier characterization}

The solution of the design problem will take an intuitive Pareto frontier characterization. 

\begin{defn}[Pareto frontier] \label{defn:pareto}
For each design $(\mathcal{E},\Sigma) \in \mathcal{D}$, define its supported bias-variance frontier as
\begin{equation}
\label{eq:design_frontier}
\small 
\begin{aligned} 
\mathcal F(\mathcal{E},\Sigma)
\equiv
\bigcup_{\lambda \in[0,1]}
\left\{
\bigl(\alpha(\mathcal{E},\Sigma, a_\lambda^\star),\beta(\mathcal{E}, a_\lambda^\star)\bigr):
a_\lambda^\star\in
\arg\min_{a: \Lambda_{\mathcal{E}}^\top a = \omega}
\left\{
(1 - \lambda) \alpha(\mathcal{E},\Sigma,a)+\lambda \beta(\mathcal{E},a)
\right\}
\right\}.
\end{aligned} 
\end{equation}
Similarly, define the oracle frontier as 
\begin{equation}
\small 
\begin{aligned} 
\label{eq:oracle_frontier}
\mathcal{F}^\star \equiv 
\bigcup_{\lambda \in [0,1]}
\left\{
\bigl(\alpha(\mathcal{E}_\lambda^\star,\Sigma_\lambda^\star, a_\lambda^\star),\beta(\mathcal{E}_\lambda^\star, a_\lambda^\star)\bigr):
(\mathcal{E}_\lambda^\star,\Sigma_\lambda^\star,a_\lambda^\star) \in
\arg\min_{(\mathcal{E},\Sigma) \in \mathcal{D}, a: \Lambda_{\mathcal{E}}^\top a = \omega}
\left\{
(1-\lambda) \alpha(\mathcal{E},\Sigma,a)+\lambda \beta(\mathcal{E},a)
\right\}
\right\}, 
\end{aligned} 
\end{equation}
where we assume throughout that such $\mathcal{E}_\lambda^\star,\Sigma_\lambda^\star, a_\lambda^\star$ exist.\footnote{Existence is mild and typically holds in our leading examples in Section \ref{sec:optimization}. If the solution does not exist, the oracle
frontier may instead be defined to include limits of approximately optimal $\alpha$ and $\beta$.}
\end{defn}

The Pareto frontier is defined using the same notions of variance,
\(\alpha\), and sensitivity to observational bias, \(\beta\) studied in previous sections. For each \(\lambda\in[0,1]\), it records the
pair generated by a linear estimator \(a_\lambda^\star\) that minimizes
$ 
(1-\lambda)\alpha(\mathcal E,\Sigma,a)
+\lambda\beta(\mathcal E,a).
$ 
Thus, \(\lambda\) indexes the relative importance assigned to sensitivity. For \(\lambda<1\), this problem is equivalent to
minimizing \(\alpha+B^2\beta\), with
\(B^2=\lambda/(1-\lambda)\); the endpoint \(\lambda=1\) corresponds to
the limiting case \(B\to\infty\). The oracle frontier allows both the
design and the estimator to be chosen separately for each 
\(\lambda\).

\begin{thm}[Solution characterization]
\label{thm:audience_frontier}
Let Assumptions \ref{ass:1a}, \ref{ass:constraint_set}, \ref{ass:linearity}, \ref{defn:class_priors} hold and consider Setting \ref{setting:2}. 
For every
$(\mathcal{E},\Sigma) \in \mathcal{D}$, $\bar{L}_B(\mathcal{E},\Sigma) = \min_{a: \Lambda_{\mathcal{E}}^\top a = \omega} R_B(\mathcal{E},\Sigma,a)$ as in Definition \ref{defn:variance_regret} and
\begin{equation}
\label{eq:frontier_domination}
\mathcal R^{\mathrm{Bayes}}(\mathcal E,\Sigma)
=
\inf\left\{
R\geq1:
\begin{array}{l}
\text{for every }(\alpha,\beta)\in\mathcal F^\star,\text{ there exists}\\
(\widetilde\alpha,\widetilde\beta)
\in\mathcal F(\mathcal E,\Sigma)
\text{ such that }
\widetilde\alpha\leq R\alpha,\quad
\widetilde\beta\leq R\beta
\end{array}
\right\}.
\end{equation}
\end{thm}

\begin{proof} See Appendix \ref{proof:audience_frontier}.
\end{proof}

Theorem \ref{thm:audience_frontier} provides a novel experimental design through a Pareto frontier approximation. 
Figure \ref{fig:audience_frontier} illustrates the intuition of the solution in Theorem \ref{thm:audience_frontier}. The audience-regret optimal design corresponds to finding the smallest point $R$ to make the frontier in terms of bias and sensitivity the closest to the smallest frontier we can achieve.

The characterization of the worst-case Bayes risk for fixed $B$ is closely related to the worst-case MSE studied in Section \ref{sec:optimal_3}: the variance takes a quadratic form as a function of $\Sigma$ and the bias is equivalent to $B^2 \beta$ in Definition \ref{defn:variance_regret}.

The proportional regret has however an important distinction: in the Bayesian audience formulation, the audience forms posterior
beliefs using a prior on the observational bias. As a result, the posterior mean
of the target, and in turn the variational representation through \(a\), depends on \(B\) (i.e., $\lambda$). Thus, when we evaluate
the posterior risk, the implicit worst-case estimator may vary with \(B\). This
differs from Theorem~\ref{thm:1}, where the researcher commits ex ante to a
single linear rule for which $B$ remains unknown.

\begin{figure}[t]
\centering
\begin{tikzpicture}[
    x=1.35cm,
    y=1.05cm,
    >=Latex,
    every node/.style={font=\small}
]

\draw[->,thick] (0,0) -- (7.55,0)
    node[below right] {variance $\alpha$};
\draw[->,thick] (0,0) -- (0,6.15)
    node[above left,align=right]
    {sensitivity\\to bias $\beta$};
\node[below left] at (0,0) {$0$};

\path[fill=red!8]
    (0.90,5.00)
      .. controls (1.02,2.70) and (1.65,1.80)
      .. (2.816,1.60)
      .. controls (3.80,1.05) and (5.00,0.72)
      .. (6.55,0.62)
    -- (7.40,0.62)
    -- (7.40,6.05)
    -- (0.90,6.05)
    -- cycle;

\begin{scope}
  \clip (0,0) rectangle (7.40,6.05);
  \begin{scope}[scale=1.28]
    \path[fill=green!12]
      (0.75,4.30)
        .. controls (0.95,2.80) and (1.30,1.65)
        .. (2.20,1.25)
        .. controls (3.00,0.85) and (4.00,0.65)
        .. (5.40,0.55)
      -- (5.80,0.55)
      -- (5.80,4.75)
      -- (0.75,4.75)
      -- cycle;

    \draw[green!45!black,thick,dashed]
      (0.75,4.30)
        .. controls (0.95,2.80) and (1.30,1.65)
        .. (2.20,1.25)
        .. controls (3.00,0.85) and (4.00,0.65)
        .. (5.40,0.55);
  \end{scope}
\end{scope}

\draw[red!75!black,very thick]
    (0.90,5.00)
      .. controls (1.02,2.70) and (1.65,1.80)
      .. (2.816,1.60)
      .. controls (3.80,1.05) and (5.00,0.72)
      .. (6.55,0.62);

\draw[blue!75!black,very thick]
    (0.75,4.30)
      .. controls (0.95,2.80) and (1.30,1.65)
      .. (2.20,1.25)
      .. controls (3.00,0.85) and (4.00,0.65)
      .. (5.40,0.55);

\draw[gray!65,densely dotted]
    (0,0) -- (3.30,1.875);

\draw[->,black!75,thick]
    (2.20,1.25) -- (2.816,1.60)
    node[midway,above left,inner sep=1pt]
    {$\times R^\star$};

\fill[blue!75!black]
    (2.20,1.25) circle (1.5pt)
    node[below left,text=blue!75!black]
    {$z^\star$};

\fill[red!75!black]
    (2.816,1.60) circle (1.7pt)
    node[above right,text=red!75!black]
    {$R^\star z^\star$};

\node[text=blue!75!black,anchor=north east]
    at (1.88,1.52)
    {oracle frontier $\mathcal F^\star$};

\node[text=red!75!black,anchor=north west]
    at (3.65,0.78)
    {design frontier
     $\mathcal F(\mathcal E,\Sigma)$};

\node[text=green!40!black,anchor=south west,align=left]
    at (4.65,1.28)
    {dilated oracle frontier\\
     $R^\star\mathcal F^\star$};

\end{tikzpicture}

\caption{\footnotesize 
Geometric interpretation of audience regret.
The blue curve is the oracle bias--variance frontier, while the red
curve is the frontier attainable under the fixed design
$(\mathcal E,\Sigma)$. Both coordinates of the oracle upper image are
inflated by a common factor. Audience regret is the smallest factor
$R^\star$ for which the resulting upper image is contained in the
upper image attainable under the fixed design. 
}
\label{fig:audience_frontier}
\end{figure}

The frontier characterization also facilitates computation. Replacing the continuum of directions $\lambda$ (by design in a compact space) with a
sufficiently fine finite grid produces a finite-dimensional
approximation. The next corollary provides an approximation
guarantee.

\begin{cor}[Optimization via finite grid approximation]
\label{cor:finite_frontier_approximation}
Maintain the conditions of Theorem \ref{thm:audience_frontier}. Define
$ 
F^\star(\lambda)
\equiv
\min_{(\mathcal{E},\Sigma) \in \mathcal{D}, a:\Lambda_{\mathcal E}^{\top}a=\omega}
\left\{
(1-\lambda)\alpha(\mathcal E,\Sigma,a)
+\lambda\beta(\mathcal E,a)
\right\}. 
$ 
Let 
$ 
0=\lambda_0<\lambda_1<\cdots<\lambda_K=1
$ 
be a finite grid, with $\lambda_{k+1} - \lambda_k \le \varepsilon \min\{F^\star(1), F^\star(0)\}$ for all $k$ and arbitrary small $\varepsilon > 0$. Suppose that $F^\star(0), F^\star(1) > 0$. Consider a grid-approximation to the audience-regret problem 
\begin{equation}
\label{eq:finite_lambda_program}
\begin{aligned}
& \underline R_K
=
\min_{\substack{(\mathcal E,\Sigma)\in\mathcal D,\ R\geq1\\
                a_0,\ldots,a_K}}
\quad R\\
\textnormal{s.t.}\quad
&\Lambda_{\mathcal E}^{\top}a_k=\omega,
\qquad  (1-\lambda_k)\alpha(\mathcal E,\Sigma,a_k)
+\lambda_k\beta(\mathcal E,a_k)
\leq
R F^\star(\lambda_k),
&&k=0,\ldots,K.
\end{aligned}
\end{equation}
Then for a finite constant  $M<\infty$, $\underline{R}_K \le  \min_{(\mathcal{E},\Sigma) \in \mathcal{D}} \mathcal{R}^{\mathrm{Bayes}}(\mathcal{E},\Sigma) \le (1 + \varepsilon M) \underline{R}_K$.
\end{cor}

\begin{proof} See Appendix \ref{proof:cor1}.  
\end{proof}

Corollary \ref{cor:finite_frontier_approximation} shows that we can solve the optimization problem via a simple finite grid approximation, with weights within $\lambda \in [0,1]$. The approximation error is controlled by the size of the grid $\varepsilon$ times a finite constant $M$, where therefore we can make $\varepsilon M$ arbitrary small. 
Appendix \ref{sec:optimization_profiled_posterior} derives the complete optimization program. The program is similar in spirit to the one under a linear regret, with the additional grid choices over a uni-dimensional $\lambda$. In our leading applications, this corresponds to a mixed-integer convex program (and a convex program for fixed experiment choice $\mathcal{E}$) which can be solved numerically using standard mixed-integer solvers. When $B \le \bar{B}$ this can be directly incorporated as part of the grid.

\section{Extensions} \label{sec:extensions}

\subsection{Confidence intervals and partial identification}
\label{sec:CI}

Next, we generalize our framework to minimizing confidence interval length and partially identified $\tau(\theta)$.  
For convenience, we focus on one-sided hypothesis testing, which matches the most common scenarios of development policy interventions where policy makers are interested in studying positive effects. Two sided testing follows similarly with appropriate adjustments that incorporate information about worst-case lower and upper bounds.

Researchers may not know $\tau(\cdot)$ exactly and only know a lower bound 
\begin{equation} \label{eqn:lower_bound1}
\small 
\begin{aligned} 
\underline{\tau}_L(\theta) = \underline{\omega}_0 + \underline{\omega}^\top \theta, \quad \tau(\theta) \ge \underline{\tau}_L(\theta)
\end{aligned} 
\end{equation} 
where linearity can be justified as a first-order linear approximation under local asymptotics, or in finite samples by constructing a strict linear minorant. 

We use $\underline{\omega}$ to indicate the weights associated with the \textit{lower bound} for $\tau(\theta)$. We implicitly maintain throughout the assumption that $\underline{\omega}$ is bounded and has at least one non-zero entry.

\begin{setting} \label{setting:3}
Consider linear estimators in Setting \ref{setting:1}, with $\mathcal{C} = -\mathcal{C}$ being a symmetric set. Suppose that $\tilde{S}_{\mathcal{E},\Sigma}$ is normally distributed as in Equation \eqref{eqn:guassian}. For a candidate design and estimator \((\mathcal E,\Sigma,W)\), and a given observational bias \(b\in\mathbb R^{p_o}\), define the one-sided \((1-\eta)\) worst-case lower-bound for \(\tau_L(\theta)\) as
$$ 
\small
\begin{aligned} 
\iota_B(\mathcal E,\Sigma,W)
\equiv
&\inf_{b \in \mathcal{B}(B)} \ell_b(\mathcal{E},\Sigma,W), \qquad \ell_b = \underline{\tau}_L(\hat\theta_W)
-
\underline{a}_{W, 1:p_o} b
-
z_{1-\eta}
\sqrt{\alpha(\mathcal E,\Sigma,\underline{a}_W)}
\end{aligned}
$$ 
where \(z_{1-\eta}\) denotes the \((1-\eta)\)-quantile $(\eta < 0.5)$ of the standard normal distribution and 
$ 
\underline{a}_W^\top \equiv 
\underline{\omega}^\top\left(
\Lambda_{\mathcal E}^\top W\Lambda_{\mathcal E}
\right)^{-1}
\Lambda_{\mathcal E}^\top W
$ 
subtracts the bias from the estimated lower bound. An audience indexed by a worst-case bias radius \(B\ge 0\) forms a conservative lower-bound $\iota_B$ by taking the most conservative endpoint over
$ 
b \in \mathcal B(B)
$.\footnote{This connects to the notion of bias-aware confidence intervals \citep{armstrong2021sensitivity}, but differs in one important aspect: an audience knows $B$ and adjust confidence intervals accordingly, while $B$ remains unknown to the researcher. Instead of reporting the single best case confidence set as a function of a particular $B$, here we use a proportional regret criterion which is worst-case over the audience action. The resulting characterization of the proportional regret as the end-point solution (and its quasi-convexity) is an insight we provide here, which significantly simplifies optimization even when $B$ is unknown to the designer.}

The researcher does not need to know the audience's choice of \(B\). Researchers may report the point estimate, its standard error, and the sensitivity measures for the lower bound
\(\sqrt{\beta(\mathcal E,\underline{a}_W)}\), allowing audiences with different values of \(B\) to form their preferred bias-aware confidence intervals. See Figure \ref{fig:ci} for an illustration.
Denote 
$$
\small 
\begin{aligned}
\mathrm{Len}_B(\mathcal{E},\Sigma, W) \equiv
\sup_{b \in \mathcal{B}(B)} \mathbb{E}_{\mathcal{E},\Sigma,b}[|\underline{\tau}_L(\theta) - \iota_B(\mathcal{E},\Sigma,W)|]
\end{aligned}
$$ 
the worst-case excess length for a bias $b \in \mathcal{B}(B)$. 
\qed 
\end{setting} 
\begin{figure}[!ht]
\centering
\begin{tikzpicture}[
    >=latex,
    timeline/.style={thick},
    event/.style={
      rectangle, draw, rounded corners,
      align=left, inner sep=2pt, font=\footnotesize,
      text width=3.0cm
    }
]

\draw[timeline,->] (0,0) -- (12.5,0) node[right]{};

\foreach \x/\lab in {0/$t=0$,4/$t=1$,8/$t=2$,11.5/$t=3$} {
  \draw[timeline] (\x,0.1) -- (\x,-0.1);
  \node[below=0.2cm] at (\x,0) {\lab};
}

\node[event, above=0.4cm] (design) at (0,0) {%
    \textbf{Design}\\
    Choose $(\mathcal E,\Sigma,W)$
};

\node[event, above=0.4cm] (report) at (4,0) {%
    \textbf{Reporting}\\
    Compute $ \underline{\tau}_L(\hat\theta_W)$\\
    Report $\alpha$\\
    Report $\beta$
};

\node[event, above=0.4cm] (audience) at (8,0) {%
    \textbf{Audience}\\
    Observe reported statistics\\
    Choose $B\ge 0$\\
    Form $\iota_B(\mathcal E,\Sigma,W)$
};

\node[event, above=0.4cm] (loss) at (11.5,0) {%
    \textbf{Ex-post loss}\\
    Excess length\\
    $\left|\underline{\tau}(\theta) - \iota_B(\mathcal E,\Sigma,W)\right|$
};

\draw[->,timeline] (design.east) -- (report.west);
\draw[->,timeline] (report.east) -- (audience.west);
\draw[->,timeline] (audience.east) -- (loss.west);

\end{tikzpicture}
\caption{\footnotesize Setting when minimizing confidence intervals. Timeline of experimental design, reporting, audience choice of the bias radius, and confidence interval length.}
\label{fig:ci}
\end{figure}

As in the MSE analysis, define proportional regret relative to the oracle that minimizes this excess length with knowledge of $B$ (but not $b$) as
\[
\small 
\begin{aligned} 
\mathcal R^{\mathrm{CI}}(\mathcal E,\Sigma,W)
& \equiv
\sup_{B\ge 0}
\frac{\mathrm{\mathrm{Len}_B(\mathcal{E},\Sigma,W)}
}{
\inf_{(\mathcal{E}',\Sigma')\in \mathcal{D}, W' \in \mathcal{W}(\mathcal{E}',\Sigma')} \mathrm{Len}_B(\mathcal{E}',\Sigma',W')
}. 
\end{aligned}
\]
The oracle therefore is aware of the audience worst-case action (indexed by $B$), and minimizes the excess length, worst-case over the bias set.

\begin{thm}[Interval set]
\label{thm:CI_equivalence}
Consider Setting~\ref{setting:3} and let Assumptions~\ref{ass:1a},
\ref{ass:constraint_set}, and \ref{ass:linearity} hold. 
\begin{itemize} 
\item For a point identified model, i.e., for $\underline{\tau}_L(\theta) = \tau(\theta)$,   
$ 
\arg\min_{\substack{
(\mathcal E,\Sigma)\in\mathcal D,\\
W\in\mathcal W(\mathcal E,\Sigma)
}}
\mathcal R^{\mathrm{CI}}(\mathcal E,\Sigma,W)
=
\arg\min_{\substack{
(\mathcal E,\Sigma)\in\mathcal D,\\
W\in\mathcal W(\mathcal E,\Sigma)
}}
\mathcal R^{\mathrm{MSE}}(\mathcal E,\Sigma,W).
$ 
\item For a partially identified model where potentially $\underline{\tau}_L(\theta) \neq \tau(\theta)$, $\underline{\omega} \neq 0$ and uniformly bounded,
$
\mathcal R^{\mathrm{CI}}(\mathcal E,\Sigma,W)^2 = \max\Big\{ \frac{\alpha(\mathcal{E},\Sigma, \underline{a}_W)}{\alpha_{\underline{\omega}}^\star}, \frac{\beta(\mathcal{E},\underline{a}_W)}{\beta_{\underline{\omega}}^\star} \Big\}, 
$
where $\alpha_{\underline{\omega}}^\star$ and $\beta_{\underline{\omega}}^\star$ are the minima of $\alpha(\mathcal{E},\Sigma, \underline{a}_W)$ and $\beta(\mathcal{E}, \underline{a}_W)$ over the set of feasible designs and weights $W$. 
\end{itemize} 
\end{thm}

\begin{proof} See Appendix \ref{proof:ci}.  
\end{proof}

 \subsection{Partial knowledge of the bias bound}
\label{sec:bound_known}

We now extend the baseline analysis to settings in which researchers have prior information about the magnitude of the observational bias. We focus on Setting \ref{setting:1}, since this information can be directly incorporated for Setting \ref{setting:2} in the choice of the grid in Corollary \ref{cor:finite_frontier_approximation}. 

Suppose, in particular, that researchers know that
$ 
B\in[0,\bar B]
$ 
for a known upper bound \(\bar B<\infty\). All other elements of the problem remain as in Section~\ref{sec:robust_design}. For a feasible design and estimator \((\mathcal E,\Sigma,W)\), define proportional regret over the restricted range of bias magnitudes as
\begin{equation}
\label{eqn:restricted_B_regret}
\small 
\begin{aligned}
\mathcal R_{\bar B}^{\mathrm{MSE}}(\mathcal E,\Sigma,W)
\equiv
\sup_{B\in[0,\bar B]}
\frac{
R_B(\mathcal{E},\Sigma,a_W)
}{
R^*(B)
},
\end{aligned}
\end{equation}
where  \(R^*(B)\) is the oracle MSE defined in Section~\ref{sec:robust_design}. The next theorem provides the corresponding regret characterization.

\begin{thm}[Regret with a bounded bias radius]
\label{thm:restricted_B}
Consider Setting~\ref{setting:1} and let Assumptions~\ref{ass:1a},
\ref{ass:constraint_set}, and \ref{ass:linearity} hold. Then, for any feasible
\((\mathcal E,\Sigma,W)\),
\[
\small 
\begin{aligned}
\mathcal R_{\bar B}^{\mathrm{MSE}}(\mathcal E,\Sigma,W)
=
\max\Big\{
\frac{\alpha(\mathcal E,\Sigma,a_W)}{\alpha^\star},
\frac{
\alpha(\mathcal E,\Sigma,a_W)
+
\bar B^2\beta(\mathcal E,a_W)
}{
\displaystyle
\inf_{\substack{
(\mathcal E',\Sigma')\in\mathcal D,\\
W'\in\mathcal W(\mathcal E',\Sigma')
}}
\left\{
\alpha(\mathcal E',\Sigma',a_{W'})
+
\bar B^2\beta(\mathcal E',a_{W'})
\right\}
}
\Big\}.
\end{aligned}
\]
\end{thm}

\begin{proof}
See Appendix \ref{proof:thm:regret_gmm}.
\end{proof}

Theorem~\ref{thm:restricted_B} shows that, when researchers know an upper bound \(\bar B\) on the magnitude of the observational bias, the worst-case proportional regret over \(B\in[0,\bar B]\) is again determined by two terms. The first is the variance regret,
which corresponds to the case \(B=0\). The second compares the worst-case MSE of the chosen design and estimator at the largest admissible bias radius \(\bar B\) with the smallest worst-case MSE attainable by an oracle that knows \(B=\bar B\). The result therefore has the same structure as Theorem~\ref{thm:1}, but the large-bias limit is replaced by performance at the finite upper bound \(\bar B\). 

Compared to minimax solutions over $\bar{B}$, proportional regret avoids over-conservative solutions through the max operator even if $\bar{B}$ is particularly large. 

\begin{rem}[Partial knowledge about relative biases] In some applications, researchers may have partial knowledge about the \textit{relative} bias of the estimates or sign restrictions on the bias. This can be readily incorporated within our framework as it can be captured by the specific set $\mathcal{C}$ chosen by the researcher, such as imposing $|b_j| \le \sqrt{\kappa_j} B$ for a user specified choice of $\kappa_j$, or imposing sign restrictions for instance $B \ge b_j \ge 0$. \qed 
\end{rem}

\subsection{Multivalued estimands}
\label{sec:multivalued_estimands}

We now extend the framework to a multivalued target
$ 
\tau(\theta)\in\mathbb R^q.
$ 
Specifically, as in Assumption \ref{ass:linearity}, let \(\tau\) be differentiable at the preliminary value \(\theta_0^{\mathrm{obs}}\), and define the multivariate linear approximation
\[
\small 
\begin{aligned} 
\tau_L(\theta)
=
\tau(\theta_0^{\mathrm{obs}})+\Omega(\theta-\theta_0^{\mathrm{obs}})
=
\Omega_0+\Omega\theta, \qquad \Omega
=
\frac{\partial \tau(\theta)}
{\partial\theta^\top}
\Big|_{\theta=\theta_0^{\mathrm{obs}}}
\in\mathbb R^{q\times p},
\qquad
\Omega_0
=
\tau(\theta_0^{\mathrm{obs}})-\Omega\theta_0^{\mathrm{obs}}. 
\end{aligned} 
\]
When \(\tau\) is linear, this representation is exact. When \(\tau\) is nonlinear, the approximation is first-order equivalent to \(\tau(\theta)\) under the local asymptotic conditions in Section~\ref{sec:gmm-extension}.

Following Setting \ref{setting:1}, consider a linear estimator of the form
$ 
\tau_L(\hat\theta_W)
=
\Omega_0+\Omega\hat\theta_W
=
\Omega_0+
\Omega\Gamma_{\mathcal E}(W)
\tilde{S}_{\mathcal E,\Sigma}.
$ 
For simplicity, 
we evaluate this estimator using the sum of the mean-squared errors of the \(q\) target components. For a given observational bias \(b\) as defined in Assumption \ref{ass:1a}, define
\begin{equation}
\small 
\begin{aligned} 
\label{eqn:MSE_multivalued}
\mathrm{MSE}_b(\mathcal E,\Sigma,W)
=
\mathbb E_{\mathcal E,\Sigma,b}
\left[
\left\|
\tau_L(\hat\theta_W)-\tau_L(\theta)
\right\|_2^2
\right].
\end{aligned} 
\end{equation}

We retain the ambiguity set introduced in Definition \ref{defn:uncertainty_set} for 
$ 
\mathcal B(B) = B \mathcal{C}$. 
With a slight abuse of notation, redefine the variance and sensitivity component, 
\begin{equation}
\label{eqn:alpha_multivalued}
\small 
\begin{aligned} 
\alpha(\mathcal E,\Sigma,W)
\equiv
\operatorname{Trace}\left(
\Omega\Gamma_{\mathcal E}(W)
\Sigma
\Gamma_{\mathcal E}(W)^\top
\Omega^\top
\right), \qquad \beta(\mathcal{E},W) \equiv \sup_{u \in \mathcal{C}}
\left\|
\left[
\Omega\Gamma_{\mathcal E}(W)
\right]_{.,1:p_o}u
\right\|_2^2. 
\end{aligned} 
\end{equation}
The quantity \(\beta(\mathcal E,W)\) measures the largest joint perturbation in the multivalued target induced by an observational bias of unit norm. Define the corresponding benchmarks
\begin{equation} \label{eqn:alpha_star_multivalued} 
\small 
\begin{aligned}
\alpha^\star
&\equiv
\inf_{\substack{
(\mathcal E,\Sigma)\in\mathcal D,\\
W\in\mathcal W(\mathcal E,\Sigma)
}}
\alpha(\mathcal E,\Sigma,W), \qquad 
\beta^\star
&\equiv
\inf_{\substack{
(\mathcal E,\Sigma)\in\mathcal D,\\
W\in\mathcal W(\mathcal E,\Sigma)
}}
\beta(\mathcal E,W).
\end{aligned}
\end{equation} 
The oracle MSE and proportional regret are defined as in Section~\ref{sec:robust_design}, replacing the scalar squared-error loss with the loss in Equation~\eqref{eqn:MSE_multivalued}. 

\begin{thm}[Multivalued estimands]
\label{prop:multivalued_estimands}
Consider Setting~\ref{setting:1} and let Assumptions~\ref{ass:1a},
\ref{ass:constraint_set} hold, alongside the multivariate extension of Assumption \ref{ass:linearity} described in this section, with $\tau_L(\theta) = \Omega_0 + \Omega\theta \in \mathbb{R}^q$ and $\Omega \neq 0$.
Let the MSE, variance, and bias sensitivity be defined in
Equations~\eqref{eqn:MSE_multivalued},
\eqref{eqn:alpha_multivalued}. Then, for any feasible
\((\mathcal E,\Sigma,W)\),  $ 
\mathcal R^{\mathrm{MSE}}(\mathcal E,\Sigma,W)
=
\max\left\{
\frac{\alpha(\mathcal E,\Sigma,W)}{\alpha^\star},
\frac{\beta(\mathcal E,W)}{\beta^\star}
\right\}.
$ 
\end{thm}

\begin{proof} See Appendix \ref{proof:thm:regret_gmm}.  
\end{proof} 

A similar extension under the audience model is possible, but omitted for brevity.

\subsection{Upper and lower bounds for nonlinear estimators}
\label{subsec:minimax_adaptive}

In this extension we
consider a complementary formulation in which the researcher also chooses the estimator, but
does not restrict it to be linear. Specifically, for a design \((\mathcal E,\Sigma)\), let \(\delta\) denote any measurable function of the
reported statistics \(\tilde{S}_{\mathcal E,\Sigma}\). For a prior \(\pi\in\Pi_B\), the audience incurs an expected loss
$ 
\mathbb E_\pi\left[
\left(
\delta(\tilde{S}_{\mathcal E,\Sigma})-\tau_L(\theta)
\right)^2
\right]$  
where the expectation is taken over the prior predictive distribution induced by \(\pi\) and
the sampling model in Setting~\ref{setting:2}, and the decomposition follows immediately from the definition of the posterior expectation.

As in previous sections, we compare against the oracle that has knowledge about the set of priors $\Pi_B$ and must also choose the estimator $\delta$. 
 
\begin{defn}[Adaptive proportional regret]
\label{defn:adaptive_regret}
For a design \((\mathcal E,\Sigma)\), and given $\Pi$, define
\[
\small 
\begin{aligned} 
\mathcal R^{\mathrm{ad}}(\mathcal E,\Sigma)
=
\inf_{\delta}
\sup_{B \ge 0}
\frac{\sup_{\pi\in\Pi_B}
E_\pi\left[
\left(
\delta(\tilde{S}_{\mathcal E,\Sigma})-\tau_L(\theta)
\right)^2
\right]
}{
\displaystyle
\inf_{(\mathcal E',\Sigma')\in\mathcal D, \delta'}
\sup_{\pi' \in \Pi_B} E_{\pi'}\left[
\left(
\delta'(\tilde{S}_{\mathcal E',\Sigma'})-\tau_L(\theta)
\right)^2
\right]
}.
\end{aligned} 
\]
The infimum is over all measurable estimators based on the report
\(\tilde{S}_{\mathcal E,\Sigma}\). 
\qed
\end{defn}

The problem is typically not quasi-convex and therefore the solution differs from the one in Theorem \ref{thm:1}, see Appendix \ref{app:armstrong}. As we discuss in an example in Appendix \ref{sec:gaussian_priors}, the choice of the design is computationally expensive once we focus on non-linear rules. In the following theorem,  we show that under the class of priors in Assumption \ref{defn:class_priors}, the objective provided in Theorem \ref{thm:1} gives us a valid surrogate objective function for the \textit{design choice}. 
We also provide a lower, easy to compute to assess the quality of the linear regret solution. 

\begin{thm}[Regret comparisons]
\label{cor:adaptive_surrogate} Let Assumptions \ref{ass:1a}, \ref{ass:constraint_set}, \ref{ass:linearity} hold and suppose Equation \eqref{eqn:guassian} hold. Let $\Pi_B$ be the class of priors  in Assumption \ref{defn:class_priors}. 
Then
\[
\small 
\begin{aligned} 
\mathcal R^{\mathrm{Bayes}}(\mathcal E,\Sigma) \le R^{\mathrm{ad}}(\mathcal E,\Sigma)
\le
\min_{a:\ \Lambda_{\mathcal E}^\top a=\omega}
\max\left\{
\frac{a^\top\Sigma a}{\alpha^\star},
\frac{\|a_{1:p_o}\|_*^2}{\beta^\star}
\right\},
\end{aligned} 
\]
where \(\alpha^\star\) and \(\beta^\star\) are defined each as the minimum of $\alpha(\mathcal{E},\Sigma,a), \beta(\mathcal{E},a)$ respectively over feasible $(\mathcal{E},\Sigma)\in \mathcal{D}, a: \Lambda_{\mathcal{E}}^\top a= \omega$. 
\end{thm}

 \begin{proof} See Appendix \ref{proof:adaptive_surrogate}.
 \end{proof}


\subsection{Illustrations of local asymptotics with non linearities}
\label{sec:gmm-extension}

We next provide details of interpretating our linearity assumptions through the lens of local asymptotic approximations. 

\paragraph{Local sequences} Following Example~\ref{exmp:gmm}, let
\(\theta\in\mathbb R^p\) denote the primitive structural parameter, and let
\(\widehat\theta_0^{\mathrm{obs}}\) denote a preliminary estimate; \(\widehat\theta_0^{\mathrm{obs}}\) could be
obtained from the observational study alone, with
$ 
\theta_0^{\mathrm{obs}}
=
\operatorname{plim}
\left(
\widehat\theta_0^{\mathrm{obs}}
\right).
$ 
Let the target estimand be \(\tau(\theta)\). For an experimental choice
\(\mathcal E\), let
$ 
g^{\mathrm{obs}}(\theta)\in\mathbb R^{p_o},
g^{\mathrm{exp}}_{\mathcal E}(\theta)
\in\mathbb R^{|\mathcal E|}
$ 
denote the population counterparts of the observational and experimental
moments, and let
\(\bar G^{\mathrm{obs}}\) and
\(\bar G^{\mathrm{exp}}_{\mathcal E,\Sigma}\)
denote their sample analogues. Let \(n\) index the precision of these sample
moments, so that their sampling standard errors are of order \(n^{-1/2}\).
We allow the underlying parameter to vary along a local sequence, suppressing
its \(n\)-subscript. Consider the setup of Example~\ref{exmp:gmm},  Equation \eqref{eqn:tilde_theta}, with 
$ 
\mathbb E[\bar G^{\mathrm{obs}}]
-
g^{\mathrm{obs}}(\theta)
=
b_n,
\mathbb E[
\bar G^{\mathrm{exp}}_{\mathcal E,\Sigma}
]
-
g^{\mathrm{exp}}_{\mathcal E}(\theta)
=
0,
$ 
where \(b_n\in\mathbb R^{p_o}\) denotes the bias of the observational
moments. Under twice differentiable moment functions with locally bounded
second derivatives, a Taylor expansion gives
\[
\small 
\begin{aligned}
\sqrt n
\left(
\mathbb E[
\tilde{S}_{\mathcal E,\Sigma}
]
-
\Lambda_{\mathcal E}\theta
\right)
=
\begin{pmatrix}
\sqrt n\,b_n\\
0_{|\mathcal E|}
\end{pmatrix}
+
O\left(
\sqrt n
\left\|
\theta-\theta_0^{\mathrm{obs}}
\right\|^2
\right).
\end{aligned} 
\]
We further write, after centering and $\sqrt{n}$-rescaling both the parameter and the statistic to define the local experiment, $ 
\mathbb V
\left(
\sqrt{n} \tilde{S}_{\mathcal E,\Sigma}
\right)
=
\Sigma+o(1)$, so that the representation coincides with the one used in the main analysis and satisfies
Assumption~\ref{ass:constraint_set}.

\paragraph{Second order remainder} For the Taylor remainder to be asymptotically negligible at the sampling-error
scale, it is sufficient that
$ 
\sqrt n
\left\|
\theta-\theta_0^{\mathrm{obs}}
\right\|^2
=
o(1).
$ 
The bias \(b_n\) may
be of the same order as \(n^{-1/2}\), of a faster order, or of a slower order,
provided that the relevant local restrictions hold. In the special
direct-parameter case where
\(g^{\mathrm{obs}}(\theta)=\theta\), and 
$ 
b_n = \theta_0^{\mathrm{obs}} - \theta
$, the preceding condition becomes
$ 
\sqrt n\,\|b_n\|^2=o(1).
$ 
The usual \(b_n=O(n^{-1/2})\) local-misspecification sequence is a special case
\citep[e.g.][]{andrews2020transparency}.

It remains to relate the primitive parameter to the policy target. If
\(\tau\) is twice continuously differentiable in a neighborhood of
\(\theta_0^{\mathrm{obs}}\), with locally bounded Hessian, then
$ 
\sqrt n
\left(
\tau(\theta)-\tau_L(\theta)
\right)
=
O\left(
\sqrt n
\left\|
\theta-\theta_0^{\mathrm{obs}}
\right\|^2
\right).
$ 
Therefore, under
$ 
\sqrt n
\left\|
\theta-\theta_0^{\mathrm{obs}}
\right\|^2
=
o(1),
$ 
$ 
\tau(\theta)-\tau_L(\theta)
=
o(n^{-1/2}).
$ 

Similar reasoning applies when the expansion point is replaced by a
preliminary estimator
$$ 
\small 
\begin{aligned} 
\widehat\Lambda_{\mathcal E}
=
\begin{pmatrix}
\widehat\Lambda^{\mathrm{obs}}\\
\widehat\Lambda^{\mathrm{exp}}_{\mathcal E}
\end{pmatrix}
=
\begin{pmatrix}
\displaystyle
\frac{\partial g^{\mathrm{obs}}(\theta)}
{\partial\theta^\top}
\bigg|_{\theta=\widehat\theta_0^{\mathrm{obs}}}\\[8pt]
\displaystyle
\frac{\partial g^{\mathrm{exp}}_{\mathcal E}(\theta)}
{\partial\theta^\top}
\bigg|_{\theta=\widehat\theta_0^{\mathrm{obs}}}
\end{pmatrix}, \qquad 
\tilde{S}_{\mathcal E,\Sigma}
=
\begin{pmatrix}
\bar G^{\mathrm{obs}}
-
g^{\mathrm{obs}}(\widehat\theta_0^{\mathrm{obs}})
+
\widehat\Lambda^{\mathrm{obs}}
\widehat\theta_0^{\mathrm{obs}}\\[3pt]
\bar G^{\mathrm{exp}}_{\mathcal E,\Sigma}
-
g^{\mathrm{exp}}_{\mathcal E}
(\widehat\theta_0^{\mathrm{obs}})
+
\widehat\Lambda^{\mathrm{exp}}_{\mathcal E}
\widehat\theta_0^{\mathrm{obs}}
\end{pmatrix}.
\end{aligned} 
$$ 
A Taylor expansion under twice differentiability and bounded derivatives gives
\[
\small 
\begin{aligned}
\sqrt n
\left(
\tilde{S}_{\mathcal E,\Sigma}
-
\widehat\Lambda_{\mathcal E}\theta
\right)
={}&
\sqrt n
\begin{pmatrix}
\bar G^{\mathrm{obs}}
-
\mathbb E[\bar G^{\mathrm{obs}}]
\\[3pt]
\bar G^{\mathrm{exp}}_{\mathcal E,\Sigma}
-
\mathbb E[
\bar G^{\mathrm{exp}}_{\mathcal E,\Sigma}
]
\end{pmatrix}
+
\begin{pmatrix}
\sqrt n\,b_n\\
0_{|\mathcal E|}
\end{pmatrix} +
O_p\left(
\sqrt n
\left\|
\theta-\widehat\theta_0^{\mathrm{obs}}
\right\|^2
\right).
\end{aligned}
\]
Thus, the same first-order representation continues to hold if
$ 
\sqrt n
\left\|
\theta-\widehat\theta_0^{\mathrm{obs}}
\right\|^2
=
o_p(1).
$ 
Similar reasoning applies if we replace $\tau_L(\theta)$ with $\omega$ evaluated at the estimated $\widehat{\theta}_0^{\mathrm{obs}}$ instead of its probability limit, provided $\tau$ is twice continuously differentiable with bounded derivatives.

\subsubsection{Local misspecification and estimation of \texorpdfstring{\(\Sigma\)}{Sigma}}
\label{sec:sigma}

Local asymptotics discussed so far also clarify why we treat the observational bias \(b\) and the variance-covariance matrix \(\Sigma\) differently. Under local misspecification, the observational bias generates a shift in the mean of $\tilde{S}$ that need not be asymptotically negligible. By contrast, under continuity of the asymptotic variance, evaluating the variance-covariance matrix at the probability limit of a preliminary estimator changes it only by a negligible amount.

\paragraph{Drifts in asymptotic variance} Suppose that for each experimental choice \(\mathcal E\), the asymptotic variance-covariance matrix can be written as a continuous function
$ 
\Sigma_{\mathcal E}(\theta,b),
$ 
where \(\theta\) can include any nuisance parameters or second moments needed to determine the covariance matrix. The dependence on \(b\) allows local misspecification to affect the variance in finite samples. Under local misspecification, consider a sequence of parameters
$ 
\theta_n-\theta_0^{\mathrm{obs}}=o(1), 
$ 
and write the variance as 
$$  
\begin{aligned} 
\small 
\mathbb V_{\theta_n,b_n}
\left[
\sqrt n
\tilde{S}_{\mathcal E,\Sigma}
\right]
=
\Sigma_{\mathcal E}(\theta_n,b_n)+o(1) \qquad 
\mathbb E_{\theta_n,b_n}
\left[
\tilde{S}_{\mathcal E,\Sigma}
\right]
-
\Lambda_{\mathcal E}\theta_n
=
\begin{pmatrix}
b_n\\
0_{|\mathcal E|}
\end{pmatrix},
\qquad
b_n=\frac{\delta_n}{\sqrt n},
\end{aligned} 
$$ 
where
$ 
\|\delta_n\|=o(n^{1/4}).
$ 
This condition allows \(\delta_n\) to be zero, to converge to a finite limit, or to diverge at any rate slower than \(n^{1/4}\). In every case,
$ 
b_n=o(1),
$ 
so the underlying misspecification remains local to zero. After root-\(n\) rescaling, however, the mean shift becomes
$ 
\sqrt n
\left\{
\mathbb E_{\theta_n,b_n}
\left[
\tilde{S}_{\mathcal E,\Sigma}
\right]
-
\Lambda_{\mathcal E}\theta_n
\right\}
=
\begin{pmatrix}
\delta_n\\
0_{|\mathcal E|}
\end{pmatrix}.
$ 
Thus, observational misspecification remains relevant at the normalized scale and need not be asymptotically negligible.

By contrast, continuity of \(\Sigma_{\mathcal E}(\theta,b)\) at
\((\theta_0^{\mathrm{obs}},0)\), together with
$ 
\theta_n-\theta_0^{\mathrm{obs}}=o(1)
\text{ and } 
b_n=o(1),
$ 
implies
$ 
\Sigma_{\mathcal E}(\theta_n,b_n)
-
\Sigma_{\mathcal E}(\theta_0^{\mathrm{obs}},0)
=
o(1).
$ 
Hence, although the normalized mean may be shifted by \(\delta_n\), the effect of local parameter drift and local misspecification on the asymptotic variance is negligible. The asymptotic variance relevant under the local sequence can therefore be evaluated at \((\theta_0^{\mathrm{obs}},0)\) without affecting the first-order approximation.

Researchers may replace
\(\Sigma_{\mathcal E}(\theta_0^{\mathrm{obs}},0)\) with a plug-in estimate constructed using \(\widehat\theta_0^{\mathrm{obs}}\to_p\theta_0^{\mathrm{obs}}\), satisfying 
$ 
\left\|
\Sigma_{\mathcal E}(\widehat\theta_0^{\mathrm{obs}},0)
-
\Sigma_{\mathcal E}(\theta_0^{\mathrm{obs}},0)
\right\|_{\mathrm{op}}
=
o_p(1)
$ without affecting the approximation.

\section{Empirical illustration} \label{sec:applications}



Because large-scale experimentation is typically infeasible due to cost constraints, researchers often combine experimental and observational data to extrapolate effects from small-scale experiments to learn general-equilibrium effects \citep[e.g.][]{ToddWolpin2006AER,attanasio2012education, meghir2022migration}. The goal of this subsection is to present a step-by-step practitioner's guide with an example, and compare our regret-optimal design to a standard variance-minimizing (Neyman) benchmark. 


Consider a researcher evaluating conditional cash transfers for sending children to school in Siaya, Kenya. Due to budget constraints, the researcher can only randomize at a small scale (partial equilibrium), but ultimately wishes to predict GE effects. Researchers have access to preliminary (external) estimates from the Mexican PROGRESA experiment \citep{ToddWolpin2006AER}, although these preliminary estimates may lack external validity in Kenya.

\subsection{Step 1: model description} 

The first step is to describe how observational/external and experimental estimates in Kenya are jointly used for estimation. The model can use reduced-form sufficient statistics or be a structural model depending on the application. We consider a stylized school choice model following \cite{BonhommeWeidner2021arXiv}, \cite{ToddWolpin2006AER}.

 Let $S\in\{0,1\}$ denote school attendance, $C$ consumption, $Y$ (pre-transfer) household income, $W$ the child's potential wage, and $t$ the stipend when enrolled. Abstracting from covariates (we will introduce covariates in the estimation), utility is
\begin{equation}\label{eq:BW-utility}
\small 
\begin{aligned} 
U(C,S,t,\varepsilon)
= \xi_3\,C + \xi_1\,CS + (\xi_2-\xi_3-\xi_1)\,tS + \xi_4\,S + S\varepsilon,
\qquad \mathbb{E}[\varepsilon] = 0, \quad \mathbb{E}[\varepsilon^2] = 1,
\end{aligned} 
\end{equation}
where $\varepsilon$ is an exogenous individual level utility shock (with variance normalized to one), and 
$C=Y+W(1-S)+tS$ denotes the budget constraint. Let $D(x) = P(\varepsilon \le x)$ and $\delta(x)$ its PDF, with $D$ assumed symmetric. The parametrization $(\xi_2-\xi_3-\xi_1)$ is without loss and simplifies expressions below. Enrollment satisfies
$ 
S=1\big\{U(Y+t,1,t,\varepsilon)>U(Y+W,0,0,0)\big\}.
$ 
Letting
$
Z(Y,W,t)\equiv \xi_3 W-\xi_1 Y-\xi_2 t-\xi_4,
$ 
we have $P(S=1\mid Y,W,t)=D\!\big(-Z(Y,W,t)\big)$.

\paragraph{Model for GE effects}
We are interested in the effect of a marginal increase in the stipend \(t\)
offered to students in all eligible poor households in rural Kenya and paid upon
enrollment. General-equilibrium feedback operates through household income
and wages, which we write as functions of the common stipend level
\(t\).

Specifically, for functions \(y(t)\) and \(w(t)\), and mean-zero
idiosyncratic income and wage shocks
\(\varepsilon_{Yi}\) and \(\varepsilon_{Wi}\), define potential individual
income and wages under the common stipend \(t\) as
$ 
Y_i(t)
=
y(t)+\varepsilon_{Yi},
W_i(t)
=
w(t)+\varepsilon_{Wi}.
$ 
Let
\[
\small 
\begin{aligned} 
y_0
\equiv
\left.
\frac{\partial y(t)}{\partial t}
\right|_{t=0},
\qquad
w_0
\equiv
\left.
\frac{\partial w(t)}{\partial t}
\right|_{t=0}, \qquad 
\delta_0
\equiv
\mathbb E
\left[
\delta
\left(
-Z\bigl(Y(0),W(0),0\bigr)
\right)
\right]. 
\end{aligned} 
\]

Our estimand is the marginal effect on school attendance of increasing the
stipend offered to all eligible households:
\begin{equation}
\label{eq:delta-theta}
\mathbb E
\left[
\left.
\frac{d}{dt}
\Pr
\left(
S=1
\mid
Y(t),W(t),t
\right)
\right|_{t=0}
\right]
=
\delta_0
\left(
\xi_2+\xi_1y_0-\xi_3w_0
\right).
\end{equation}
The expression decomposes into the direct stipend effect
\(\delta_0\xi_2\), the indirect income effect
\(\delta_0\xi_1y_0\), and the indirect wage effect
\(-\delta_0\xi_3w_0\).

\paragraph{Market clearing conditions} Suppose that every potential student who is not
in school supplies a fixed number \(H\) of hours. For a candidate mean wage
\(w\) and transfer \(t\), define labor supply per eligible person as
$ 
L_s(w,t)
=
H\,
\mathbb E\!\left[
1-
D\!\left(
-Z\bigl(Y_i(t),w+\varepsilon_{Wi},t\bigr)
\right)
\right].
$ Let \(L_d(w)\) denote the corresponding labor demand and let
\(W_0\equiv w(0)\) denote the baseline mean opportunity wage. Market
clearing requires
$ 
L_s(w(t),t)=L_d(w(t)).
$ 
Differentiating the market-clearing condition at \(t=0\) gives
$ 
w_0
=
\frac{\delta_0(\xi_2+\xi_1y_0)}
{\delta_0\xi_3+d}, d
\equiv
-
\frac{1}{H}
\left.
\frac{\partial L_d(w)}{\partial w}
\right|_{w=W_0}
>0
$

To relate \(d\) to interpretable economic quantites, let \(S_0\) denote the baseline school
attendance probability. Under the maintained
assumption that every non-enrolled potential student supplies \(H\) hours,
\(L_d(W_0)=H(1-S_0)\), we write 
\begin{equation} \label{eqn:d}
d
=
\zeta \frac{1-S_0}{W_0}, \qquad \zeta
\equiv
-
\left.
\frac{\partial\log L_d(W)}
{\partial\log W}
\right|_{W=W_0}
\end{equation} 
where $\zeta$ denotes the absolute labor-demand elasticity.  

Collecting the main model specifications, our target parameters are
\[
\small 
\begin{aligned} 
\theta_1 \equiv \delta_0\xi_1,\qquad
\theta_2 \equiv \delta_0\xi_2,\qquad
\theta_3 \equiv -\delta_0\xi_3,\qquad
\theta_4 \equiv y_0,\qquad
\theta_5 \equiv \log(\zeta), 
\end{aligned} 
\]
where $\zeta = e^{\theta_5}$ guarantees a positive elasticity. The estimand is
\begin{equation}
\small 
\begin{aligned} 
\label{eq:tau-theta}
\tau(\theta)
=
\underbrace{\theta_2}_{\text{Direct stipend effect}}
+
\underbrace{\theta_4 \theta_1}_{\text{Income effect}}
+
\underbrace{w_0(\theta)\theta_3}_{\text{Wage effect}},
\qquad
w_0(\theta)
=
\frac{\theta_2+\theta_4 \theta_1}{ \frac{1 - S_0}{W_0} e^{\theta_5} -\theta_3}, 
\end{aligned}
\end{equation}
where $\frac{1 - S_0}{W_0}$ will be collected observed from census data/baseline information in Kenya.

\subsection{Step 2: preliminary (possibly misspecified) estimation}

Following the literature, the model is typically estimated separately by gender. We will focus on effects on transfers to female students throughout, while a similar analysis following the same steps can be conducted for male students. 
We set
\(S_0=0.88\) and \(W_0=40 \times 0.64\) purchasing-power-parity U.S. dollars per week,
using the baseline data for females in eligible untreated households in Kenya from
\citet{egger2022general} and assuming forty-hours per week working hours, so that $(1-S_0)/W_0 = 0.0046$.   We treat these quantities as fixed
pre-experimental inputs as they are measured in the target population.\footnote{The calibration of \(S_0\) also closely matches the school-attendance rate
reported for Siaya County in the 2015--2016 Kenya Integrated Household
Budget Survey, where $S_0 = 0.88$. Skeptical researchers about these estimates may also define these as potentially misspecified additional parameters within our framework.} 

We next construct preliminary external estimates of the three marginal
schooling responses
$
\theta_1,
\theta_2,
\theta_3
$
using data from the large-scale Mexican PROGRESA experiment
\citep{ToddWolpin2006AER}. Throughout the remainder of the application, we measure \(Y\),
\(W\), \(t\), and the experimental perturbations in real 2015
purchasing-power-parity U.S. dollars per week.\footnote{The PROGRESA variables used in
our preliminary estimation are recorded
weekly, all in real 1997 Mexican pesos. Let
$ 
\kappa_M
=
\frac{\operatorname{CPI}_{M,2015}}
     {\operatorname{CPI}_{M,1997}}
\frac{1}{\operatorname{PPP}_{M,2015}}
\approx 0.29
$ 
denote the conversion from one real 1997 Mexican peso to a real 2015 PPP U.S.
dollar, where \(\operatorname{PPP}_{M,2015}\) is measured in Mexican pesos per
international dollar. We use the World Bank consumer price index and
private-consumption PPP conversion factor; see
\url{https://data.worldbank.org/indicator/FP.CPI.TOTL} and
\url{https://data.worldbank.org/indicator/PA.NUS.PRVT.PP}. }

We estimate the school-attendance equation by probit, therefore assuming $\varepsilon \sim \mathcal{N}(0,1)$, pooling treated and
control observations and controlling for the individual-level stipend.
The specification also includes the standard covariates used in the
PROGRESA analysis, including age, distance to school, eligibility status,
year, and highest grade completed, estimated separately by
gender. In turn, we
obtain \(S_{1:3}^{\mathrm{obs,raw}}\) that map to the
parameters \(\theta_{1:3}\) after applying the normalization above. Estimates and standard errors are reported below.

\begin{table}[!ht]
\centering
\begin{minipage}[t]{0.55\linewidth}
\vspace{0pt}
\centering
\begin{tabular}{l r r}
\hline
Parameter
&
\(S_{1:3}^{\mathrm{obs}}\)
&
\(\sqrt{\Sigma_{jj}^{\mathrm{obs}}}\)
\\
\hline
\(\theta_1\)
& \(1.83\times10^{-4}\)
& \(2.22\times10^{-4}\)
\\
\(\theta_2\)
& \(6.54\times10^{-3}\)
& \(3.33\times10^{-3}\)
\\
\(\theta_3\)
& \(-6.26\times10^{-3}\)
& \(3.44\times10^{-3}\)
\\
\hline
\end{tabular}

\footnotesize
PROGRESA-based external estimates for female students and corresponding
standard errors, in probability points per real 2015 PPP U.S. dollar per week.
\end{minipage}
\hfill
\begin{minipage}[t]{0.40\linewidth}
\vspace{0pt}
\centering
\[
\Sigma_{1:3,1:3}^{\mathrm{obs}}
=
\begin{bmatrix}
 0.00492 & -0.0129 &  0.00636 \\
 -0.0129 &  1.112 & -0.1441 \\
 0.00636 & -0.1441 & 1.186
\end{bmatrix}
\times 10^{-5}.
\]
\footnotesize
Variance-covariance matrix of the normalized PROGRESA-based external
estimates for female students.
\end{minipage}
\caption{\footnotesize External estimates from the PROGRESA experiment.}
\label{tab:progresa-external-estimates}
\end{table}

We obtain an experimental estimate of the income response in Kenya
\(\theta_4=y_0\) from \citet[Table 5, Panel B]{egger2022general}. The authors'
income-based multiplier is \(2.28\), with a standard error of \(1.73\). We map
this estimate to \(y_0\), so that
$
S_4^{\mathrm{obs}}
=
\widehat y_0
=
2.28,
$
with
\(\sqrt{\Sigma_{44}^{\mathrm{obs}}}=1.73\), which does not depend on the measurement unit.
Similarly, we obtain an external estimate of \(\zeta\) from
\citet{wambugu2017labour}, who studies labor-demand elasticities in the
Kenyan manufacturing sector. Under the author's preferred specification, the
estimated own-wage elasticity is \(-0.1709\), with an implied robust standard
error of \(0.0291\). Because \(\zeta\) denotes the absolute labor-demand
elasticity, we set
$
\widehat\zeta=0.1709.
$
Using the delta method, the corresponding external estimate of
\(\theta_5=\log(\zeta)\) is
$
S_5^{\mathrm{obs}}
=
\log(\widehat\zeta)
=
-1.767,
\sqrt{\Sigma_{55}^{\mathrm{obs}}}
=0.17.
$ Because \(S_4^{\mathrm{obs}}\) and \(S_5^{\mathrm{obs}}\) are obtained from
studies that are independent of the Mexican PROGRESA sample, we set their
covariances with \(S_{1:3}^{\mathrm{obs}}\) equal to zero. We also set the
covariance between \(S_4^{\mathrm{obs}}\) and \(S_5^{\mathrm{obs}}\) equal to
zero because they are estimated using different samples. Thus, 
\[
\small 
\begin{aligned}
\Sigma^{\mathrm{obs}}
&=
\begin{pmatrix}
\Sigma_{1:3,1:3}^{\mathrm{obs}} & 0_{3\times1} & 0_{3\times1}
\\
0_{1\times3} & 1.73^2 & 0
\\
0_{1\times3} & 0 & 0.17^2
\end{pmatrix},
\qquad
S^{\mathrm{obs}}
=
\begin{pmatrix}
1.83\times10^{-4}\\
6.54\times10^{-3}\\
-6.26\times10^{-3}\\
2.28\\
\log(0.1709)
\end{pmatrix}, \qquad 
\omega
&=
\frac{\partial\tau(\theta)}{\partial\theta}
\bigg|_{\theta=S^{\mathrm{obs}}}
=
\begin{pmatrix}
0.2577\\
0.1130\\
0.1115\\
2.071\times10^{-5}\\
6.979\times10^{-4}
\end{pmatrix}.
\end{aligned}
\]

The first-order approximation around
\(\widehat{\theta}_0^{\mathrm{obs}}\equiv S^{\mathrm{obs}}\) gives us
$
\tau(\theta)
=
\tau(\widehat{\theta}_0^{\mathrm{obs}})
+
\omega^\top(\theta-\widehat{\theta}_0^{\mathrm{obs}})
+
O(\lVert\theta-\widehat{\theta}_0^{\mathrm{obs}}\rVert^2),
$
where we treat the second-order remainder as asymptotically negligible under
the local misspecification model described in Section
\ref{sec:gmm-extension}. 

\paragraph{Bias of external estimates} 
A feature of the framework is that it requires the researcher to
specify ex ante which parameters may be biased. We let the PROGRESA estimates be biased. This bias may arise  from several sources, including external validity bias, difference in the utility models across the two countries (e.g., $\varepsilon$ is not Gaussian), or confounding bias for non-experimental coefficients. All of these sources of misspecification can enter as part of the framework without requiring researchers to take a stand on which one is the dominant source.\footnote{Under the local misspecification framework of
Section~\ref{sec:gmm-extension}, for a sequence $r_n: \sqrt{n} r_n^2 = o(1)$, we allow for three sources of local
misspecification. First, the true distribution of the utility shock in the
population used to estimate the probit may satisfy
\(D_n(x)=\Phi(x)+h(x)r_n\), where \(h\) is a smooth function satisfying
the restrictions needed for \(D_n\) to be a valid symmetric CDF with mean zero
and variance one. Second, confounding may arise through
\(\varepsilon=u+q(Y,W)r_n\), where \(u\) has CDF \(D_n\) and is
conditionally independent of \((Y,W,t)\) given the included controls, while
\(q(\cdot)\) captures confounding. Third, the preliminary PROGRESA
estimates may be centered on Mexican population parameters that differ from
the corresponding Kenyan target parameters by \(O(r_n)\), generating
local external-validity bias. These sources may occur separately or jointly,
and their first-order effects are collected in \(b\).} On the other hand, we assume that the two
\textit{Kenyan} estimates correctly identify \(y_0\) and \(\zeta\). This is a restriction researchers can motivate in a pre-analysis plan; when researchers are skeptical about this assumption, they can similarly allow biases in these two parameters as part of our framework. We
implement these restrictions using the misspecification shape
\[
\small 
\begin{aligned} 
b\in\mathcal B(B)=B\mathcal C,
\qquad
\mathcal C
=
\left\{
c\in\mathbb R^5:
\left\lVert c_{1:3}\right\rVert_\infty\leq1,
\quad
c_4=c_5=0
\right\},
\qquad
B\geq0.
\end{aligned}
\]
Here, \(B\) governs the
unknown magnitude of the external-validity bias.  
The $\ell_{\infty}$-ball is natural once each relevant parameter has the same measurement unit. 

\subsection{Step 3: candidate experiments and experimental variances}

In this example, we consider a researcher who can afford only small,
partial-equilibrium experiments in Kenya, and examine three stylized
experiments or combinations thereof:
\begin{itemize}
\item \(j=\{1\}\): \textit{Unconditional transfer (income shock).}
The researcher randomizes a small weekly income shock of size \(\Delta_1\) to a small fraction of households
(implying no general-equilibrium effects). Under a first-order (Taylor)
approximation, the experiment identifies the average marginal effect of income
on school attendance,
$
\theta_1
\;=\;
\mathbb E\left[
\left.
\frac{\partial}{\partial Y}
\Pr\bigl(S=1\mid Y,W,t\bigr)
\right|_{t=0}
\right].
$


\item \(j=\{2\}\): \textit{Conditional cash transfer (stipend).}
The researcher randomizes a small weekly stipend \(t\) of size \(\Delta_2\),
conditional on attending school, to a small fraction of households (with
prices held fixed). Under a first-order approximation, this identifies the
average marginal effect of the stipend on school attendance,
$
\theta_2
\;=\;
\mathbb E\left[
\left.
\frac{\partial}{\partial t}
\Pr\bigl(S=1\mid Y,W,t\bigr)
\right|_{t=0}
\right].
$

\item \(j=\{3\}\): \textit{Randomized outside-option earnings
(wage shock).}
The researcher offers teenagers otherwise identical full-time jobs and
randomly varies weekly compensation around a baseline wage by a small \(\Delta_3\). 
Holding everything else constant, a small randomized earnings differential
identifies
$
\theta_3
=
\mathbb E\left[
\left.
\frac{\partial}{\partial W}
\Pr(S=1\mid Y,W,t)
\right|_{t=0}
\right].
$\footnote{In particular, if otherwise identical jobs offer weekly
compensation \(W-\Delta_3/2\) and \(W+\Delta_3/2\), then
$
\frac{
\mathbb E[S_i\mid W_i^{\mathrm{offer}}=W+\Delta_3/2]
-
\mathbb E[S_i\mid W_i^{\mathrm{offer}}=W-\Delta_3/2]
}{
\Delta_3
}
=
\theta_3+O(\Delta_3^2).
$
Employment programs are often feasible experiments; see, for instance,
\cite{goldberg2016kwacha}.}
\end{itemize}

We consider a scenario where the researcher may either run a cash-transfer
program (unconditional, conditional, or both) and optimally allocate the sample
size between the two experiments, or run a wage experiment only. This implies
that the set of feasible experiments is
\(\{\{1\},\{2\},\{1,2\},\{3\}\}\), with \(\{1,2\}\) also corresponding to
the optimal allocation of sample size between the two cash-transfer
experiments. The job program requires different implementation infrastructure
than a cash-transfer program, motivating this choice.

For each experiment, suppose the researcher constructs a difference-in-means
estimate
\[
\small 
\begin{aligned} 
S_j^{\mathrm{exp}}
=
\frac{
\bar S_{j,\mathrm{treated}}-
\bar S_{j,\mathrm{control}}
}{
\Delta_j
},
\qquad j\in\{1,2,3\},
\end{aligned} 
\]
where \(S\) denotes school attendance in the treated or control group. To
calibrate experimental precision, we use the baseline school-attendance rate
in the target Kenyan population. Because school attendance is binary, we set
the individual-level outcome variance equal to
$
\sigma_S^2
=
S_0(1-S_0)
=
0.105,
$
and assume homoskedasticity across the treatment and control arms
\citep[a common restriction at the stage of experimental design but not
required to conduct ex-post inference; see][]{GerberGreen2012}. For two treatment arms \(j\) and \(j'\) using independent samples,
$\Sigma_{\{j,j'\}}^{\mathrm{exp}}
=
4\sigma_S^2
\begin{pmatrix}
\displaystyle\frac{1}{n_j\Delta_j^2}
&
0
\\[8pt]
0
&
\displaystyle\frac{1}{n_{j'}\Delta_{j'}^2}
\end{pmatrix},
n_j+n_{j'}=n_{\mathrm{tot}}.
$
For simplicity, we use a common monetary perturbation equal to 20 percent of
baseline weekly child earnings,
$ 
\Delta_j
=
0.20W_0
=
5.12$
real 2015 PPP U.S. dollars per week, which matches the minimum transfer under Progresa ($\sim 5$ PPP dollars/week). 
Under this calibration,
$ 
\operatorname{Var}\left(S_j^{\mathrm{exp}}\right) =
\frac{0.0159}{n_j}.
$

\begin{rem}[Shared control group]
In some applications, the control group may be shared between treated arms.
This case corresponds to the same
solution of our optimization program (Appendix \ref{sec:optimization}), where we reparametrize \(\theta\)
to correspond to each treatment arm.
\qed
\end{rem}

\subsection{Step 4: experimental design}

The final step is to select the experiment, allocate the available sample, and, when applicable, choose how to combine the resulting experimental estimate with the external evidence. We compare the design implied by our linear-regret criterion and Bayesian audience criterion (which does not require to commit to one particular estimator)\footnote{The Bayesian solution is computed using a grid of five-hundred $\lambda$ in Corollary \ref{cor:finite_frontier_approximation}. The grid is chosen as follows: we consider 118 values of $\lambda$ corresponding to fifty equally spaced $B$ between zero and 0.03, where \(B=0.03\) is approximately five times the magnitude of the estimated \(\theta_2\) and \(\theta_3\) and approximately ten times the experimental standard error under $n = 3,700$; 128 logarithmic points between $B = 0.03$ and $\lambda = 0.01$, 127 logarithmic point through $\lambda = 0.1$ and 127 logarithmic points through $\lambda = 1$. This grid is sufficiently fine at the relevant values of $B$ that may matter most for misspecification, while it also allows for larger values including $B = \infty$.} 
 with two variance-based benchmarks. The first benchmark jointly chooses the design and the linear estimator to minimize variance, thereby allowing variance-optimal weights on the experimental and external estimates. The second is a classical Neyman design that places weight one on each collected experimental estimate and optimally allocates the sample across the selected experiments.

\begin{figure}[!ht]
    \centering
    \includegraphics[scale = 0.3]{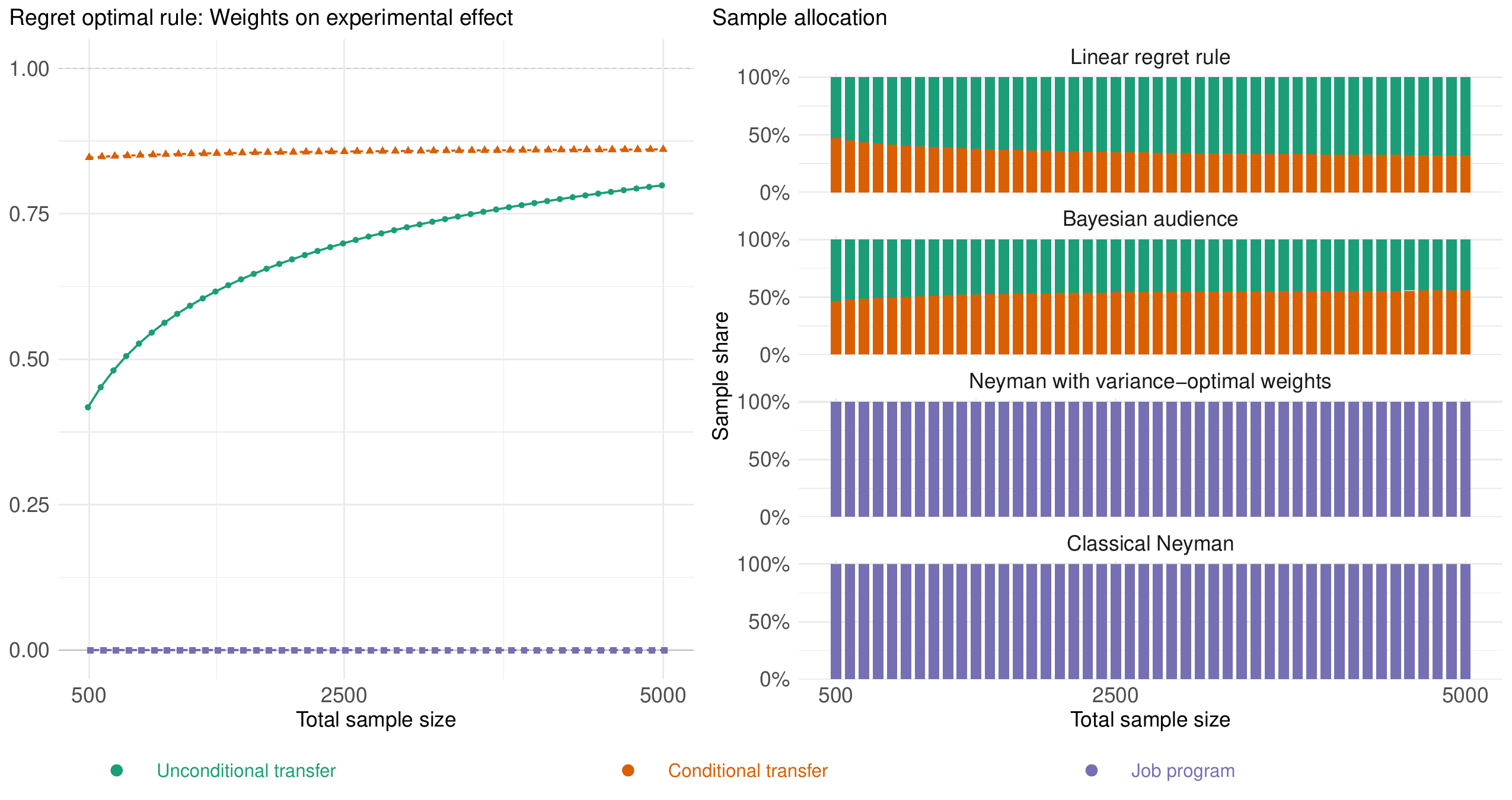}
    \caption{\footnotesize Experimental weights and sample allocation across designs. The left panel reports the regret-optimal linear rule’s weight on the experimental estimate for each feasible experiment as total sample size varies. The right panel reports the corresponding sample allocation shares for the regret-optimal linear rule, the Bayesian audience design, the Neyman allocation with variance-optimal weights, and the classical Neyman allocation.}
    \label{fig:first_fig_application}
\end{figure}

Figure~\ref{fig:first_fig_application} reports the resulting estimators and designs across total sample sizes. The left panel reports the weight placed on the experimental estimate by the linear-regret rule in Theorem~\ref{thm:1}. The right panel reports the corresponding experimental allocation, together with the allocations selected by the two Neyman benchmarks.  The vertical dotted line in the sample-size panels marks $n=3,700$, approximately the average sample size among the meta-analysis of J-PAL funded experimental studies  in \citet{viviano2026model}.

Both the linear-regret criterion and the Bayesian-audience criterion select a combination of the two cash-transfer programs, with the Bayesian criterion placing approximately $50\%$ sample size between the two and the linear regret rule placing between 52 to 65\% for the unconditional cash transfer and the remaining to the conditional one. The linear-regret estimator places approximately 85\% of its weight on the conditional transfer experimental estimate and between about 40 to 80\% to the unconditional transfer experimental estimate, increasing in the experimental sample size. Thus, even around the sample size of a typical study, the optimal rule continues to place a small but meaningful weight on the external estimate. In contrast, both Neyman designs allocate the sample entirely to the job program.  

\vspace{-6mm} 

\subsection{Comparisons and precision gains}

\begin{figure}[!ht]
    \centering
    \includegraphics[scale = 0.3]{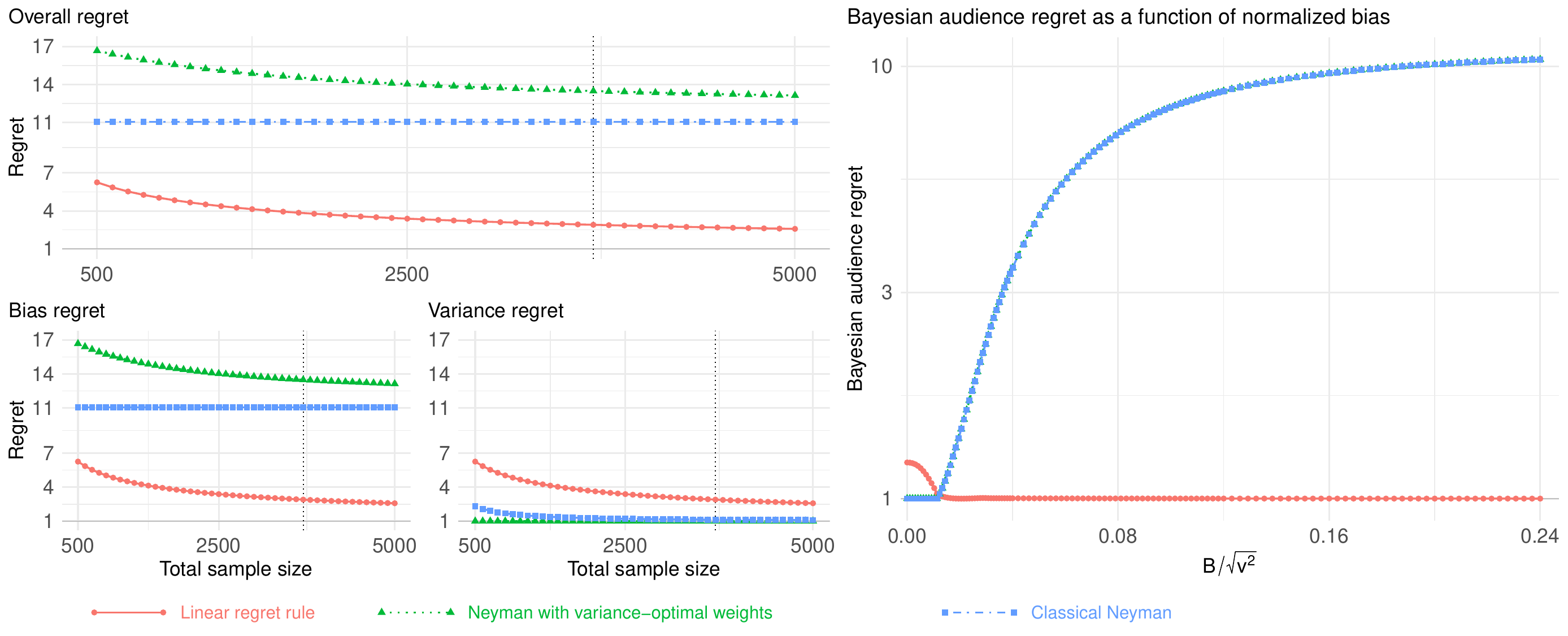}
    \caption{\footnotesize Regret decomposition and Bayesian audience regret. The left panel reports the overall worst-case regret and its bias and variance components across total sample sizes for the regret-optimal linear rule, the Neyman allocation with variance-optimal weights, and classical Neyman allocation. The right panel reports the Bayesian audience regret at each value of the bias bound \(B\) rescaled by the scaled experimental noise parameter $v$ for \(n=3700\) corresponding to the average sample size of J-PAL experiments.}
    \label{fig:second_fig_application}
\end{figure}

Figure~\ref{fig:second_fig_application} illustrates the consequences of these choices. The left panels report overall regret and its bias and variance components, with the dotted line again identifying $n=3,700$. As expected, the Neyman rule attains variance regret equal to one (and the classical Neyman rule without optimizing the weights close to one). Their variance advantage, however, is accompanied by much greater exposure to misspecification. At the empirically calibrated sample size, the Neyman design has bias and overall regret of approximately 13, while the classical Neyman design with variance-optimal weights has regret above 11. 

By comparison, the linear-regret solution is interior and  equalizes its bias and variance regret. It therefore accepts a modest increase in variance relative to the variance-minimizing designs in exchange for a substantial reduction in sensitivity to misspecification. Its overall regret decreases from approximately 6 at $n=500$ to approximately 3 at $n=3,700$, and decreases further as the sample size increases. 

The right panel evaluates the same designs under the Bayesian-audience criterion pointwise as a function of $B$ (divided by the scaled experimental noise parameter $v$). When $B$ is close to zero, the  Neyman designs have regret equal to one, and the design selected by the linear-regret criterion has regret of approximately 1.2. This is the cost of robustness when the external estimates are actually unbiased. The ranking reverses rapidly as the scope for misspecification increases. Once $B/v$ exceeds approximately $2\%$, the regret of both Neyman designs rises sharply, whereas the Bayesian-audience regret of the linear-regret design is close to one and remains there. At $B/v = 8\%$, regret is more than 7 for the Neyman allocation. Thus, at a sample size representative of the J-PAL experiments, the design selected under Theorem \ref{thm:1} remains close to optimal for a Bayesian audience.

\begin{figure}[H]
    \centering
    \includegraphics[width=0.5\linewidth]{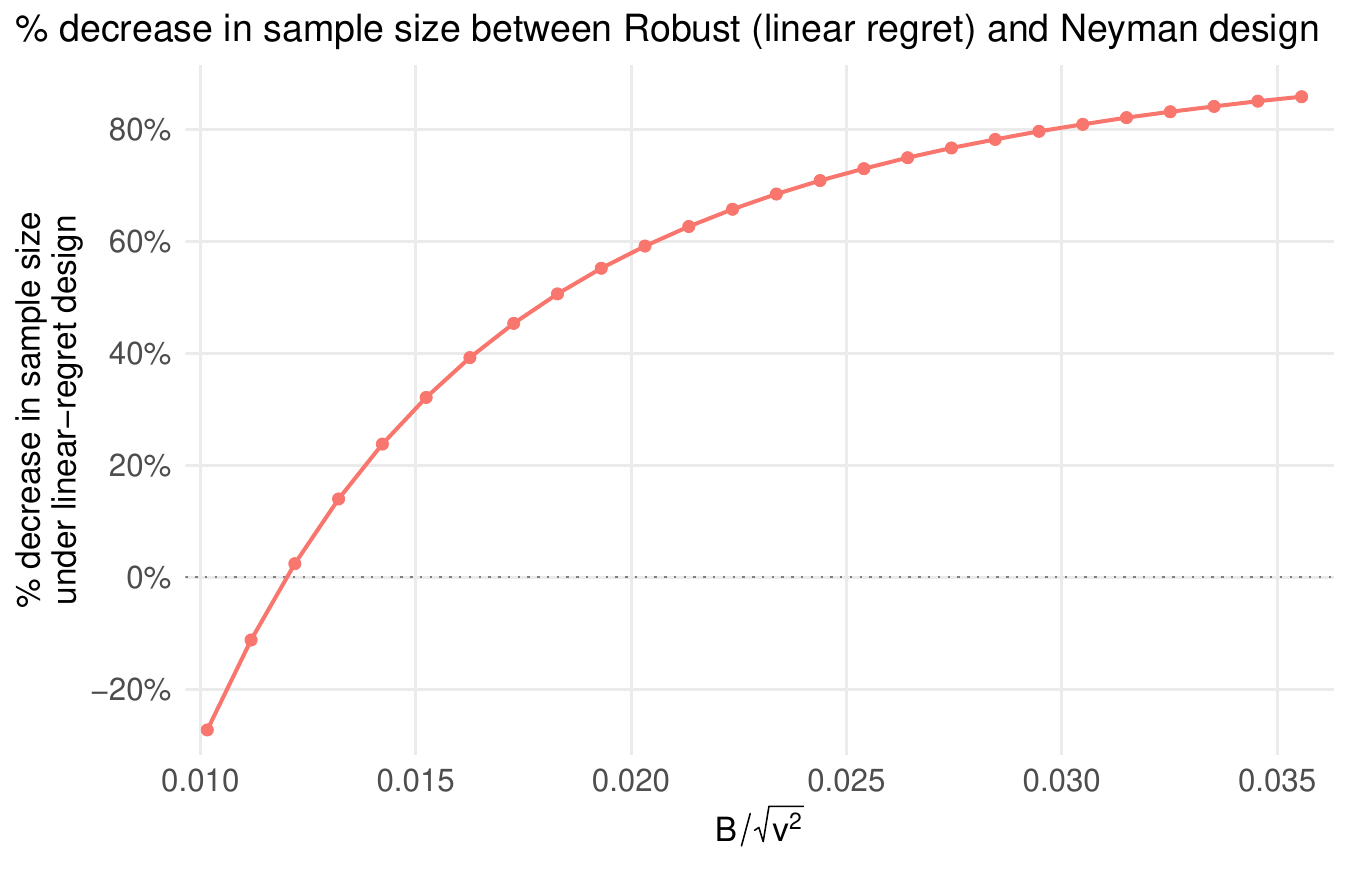}
    \caption{\small Percentage decrease in sample size to achieve the same audience MSE ($\bar{L}_B$ in Equation \eqref{eqn:L_B_audience}) when using the optimal design under the linear regret rule and the one under the Neyman rule at the benchmark sample size $n = 3,700$. The $x$-axis report the bias rescaled by the scaled experimental noise $v$. For $B \approx 0$, the variance-optimal design is preferred since no bias can occur. For $B \ge 2\%$ the scaled experimental noise $v$ our proposed design leads to the same precision as the Neyman design with less than $40\%$ of its sample size.}
    \label{fig:sample_size}
\end{figure}

Figure \ref{fig:sample_size} reports the percentage decrease in $n$ so that, our proposed design under the linear regret rule achieves the same worst-case MSE of the Neyman design at the benchmark $n = 3,700$. We compute the worst-case MSE for a given $B$ under the audience loss (i.e., $\bar{L}_B$ in Equation \eqref{eqn:L_B_audience}), so that this comparison is only driven by the choice of the design. 

For $B \approx 0$, the variance-optimal design is the most precise as the bias equals approximately zero. However, as soon as $B$ is above $2\%$ the outcome standard deviation, the proposed design under the linear-regret criterion reduces sample size by more than $60\%$ compared to the Neyman design. This may translate into substantive monetary gains.

\section{Implications for practice}
\label{sec:guide}

This paper studies experimental design in the presence of informative but
potentially biased external evidence. 
The practical workflow proceeds as follows: 

\begin{itemize}

\item \textbf{Define the estimand(s) of interest.} Specify $\tau(\theta)$ for a known mapping $\tau$ and unknown parameters $\theta \in \mathbb{R}^p$. Some (but not necessarily all) components of $\theta$ can be learned experimentally through e.g., moment restrictions; this clarifies what the experiment can identify, and can be prespecified, often desiderable \citep{miguel2021evidence}.

\item \textbf{Assemble informative observational evidence.} Collect observational estimates and their covariance $\Sigma^{\rm obs}$. These serve as informative (but potentially locally biased) baselines. Such evidence may come from observational designs, structural estimates  with pre-experimental data, or prior experiments conducted in different contexts.
\item \textbf{Compute the sensitivity parameters.} Using the observational baseline, compute the sensitivity weights
$\omega \;=\; \frac{\partial \tau(\theta)}{\partial \theta}$ evaluated at the observational estimates,
which quantify how bias in each coordinate of $\theta$ propagates to $\tau(\theta)$ (through first order local asymptotics in Section \ref{sec:gmm-extension}). In the presence of models implying non-linear smooth moments, evaluate their Jacobian $\Lambda$ at these same preliminary estimates.

\item \textbf{Describe the plausible sources of misspecification.}
A feature of the framework is that it requires researchers to ex-ante describe which external estimates may be biased. For example, 
$ 
\mathcal B_\kappa(B)
=
\left\{
b\in\mathbb R^{p_o}:
|b_\ell|\leq B\kappa_\ell,
\ell=1,\ldots,p_o
\right\}
$ 
allows the researcher to encode greater concern about some components than
others. Setting \(\kappa_\ell=0\) treats the corresponding estimate as
correctly specified.

\item \textbf{Specify feasibility constraints and calibrate experimental variance. Then optimize.} Enumerate the admissible design set, and the budget on sample size allocation. Calibrate per-unit experimental variances using pilot studies or historical data to form the set of feasible designs $\mathcal{D}$ as for standard power-analysis (their estimation error or misspecification is typically of second order relative to $b$, see Section \ref{sec:sigma}).

Finally, optimize over the design (and potentially the estimator) off-the-shelf. 
\end{itemize}

The applicability of the framework spans a wide range of settings. Examples include estimating general-equilibrium
effects or structural models
\citep[e.g.][]{ToddWolpin2006AER,attanasio2012education,meghir2022migration,
de2025decoupling} or choosing among alternative treatment
arms in factorial designs
\citep{muralidharan2019factorial}. In industrial organization,
applications include choosing between demand- and supply-side interventions
\citep{bergquist2020competition}, studying information acquisition in
markets \citep{larroucau2024college}, and deciding
which additional data source to acquire
\citep{allcott2025sources}. Several questions remain for future research, such as sequential or adaptive experimental decisions
\citep[e.g.,][]{cesa2025adaptive}, settings in which experimental
precision is itself learned during the study, or algorithmic procedures for site selection 
\citep[e.g.][]{gechter2024selecting,olea2024externally, abadie2021synthetic}  combined with our approach. The use of our method to inform founding agencies is also an interesting future direction.

\singlespacing
\bibliographystyle{chicago}
\bibliography{bibliography}

\onehalfspacing

\newpage 
\appendix 

\section{Proofs of main results}

\subsection{Proof of Theorems
\ref{thm:1},\ref{thm:restricted_B}, \ref{prop:multivalued_estimands}}
\label{proof:thm:regret_gmm}

Throughout this proof, whenever clear from the context and without loss of generality we will use the notation used in Section \ref{sec:multivalued_estimands} and write with an abuse of notation $\alpha(\mathcal{E},\Sigma,W), \beta(\mathcal{E},W)$ as a function of $W$ (instead of $a_W$).  Theorem~\ref{thm:1} is the special case of
Theorem~\ref{prop:multivalued_estimands} in which the estimand is scalar,
\(\Omega=\omega^\top\). We therefore first prove
Theorem~\ref{prop:multivalued_estimands}. Along the way, we also show how Theorem \ref{thm:restricted_B} follows directly. \\ 
\smallskip
\noindent
\textit{Step 1 (worst-case MSE).}
It follows that we can write 
\[
\begin{aligned}
\tau_L(\hat\theta_W)-\tau_L(\theta)
={}&
\Omega\Gamma_{\mathcal E}(W)
\left(
\tilde{S}_{\mathcal E,\Sigma}
-
\mathbb E[
\tilde{S}_{\mathcal E,\Sigma}
]
\right) +
\Omega\Gamma_{\mathcal E}(W)
\begin{pmatrix}
b\\
0_{|\mathcal E|}
\end{pmatrix},
\end{aligned}
\]
and 
$$ 
\small  
\begin{aligned}
\mathrm{MSE}_b(\mathcal E,\Sigma,W)
={}&
\operatorname{Trace}
\left(
\Omega\Gamma_{\mathcal E}(W)
\Sigma
\Gamma_{\mathcal E}(W)^\top
\Omega^\top
\right)+
\left\|
\left[
\Omega\Gamma_{\mathcal E}(W)
\right]_{.,1:p_o}b
\right\|_2^2
={}&
\alpha(\mathcal E,\Sigma,W)
+
\left\|
\left[
\Omega\Gamma_{\mathcal E}(W)
\right]_{.,1:p_o}b
\right\|_2^2.
\end{aligned} 
$$ 
Therefore, by defining for a matrix $A$, $||A||_*^2 = \sup_{u \in \mathcal{C}}||Au||_2^2$, 
\[
\begin{aligned}
\sup_{b\in\mathcal B(B)}
\left\|
\left[
\Omega\Gamma_{\mathcal E}(W)
\right]_{.,1:p_o}b
\right\|_2^2
&=
B^2
\left\|
\left[
\Omega\Gamma_{\mathcal E}(W)
\right]_{.,1:p_o}
\right\|_{*}^2
=
B^2\beta(\mathcal E,W).
\end{aligned}
\]
Hence,
$ 
\sup_{b\in\mathcal B_l(B)}
\mathrm{MSE}_b(\mathcal E,\Sigma,W)
=
\alpha(\mathcal E,\Sigma,W)
+
B^2\beta(\mathcal E,W).
$ 

Note that $\alpha^\star$ is bounded away from zero by Assumption \ref{ass:constraint_set} and Assumption \ref{ass:linearity}.\footnote{To see this, for any feasible
\((\mathcal E,\Sigma,W)\), define
$ 
A_{\mathcal E,W}
\equiv
\Gamma_{\mathcal E}(W)^\top\Omega^\top.
$ 
Then
$ 
\alpha(\mathcal E,\Sigma,W)
=
\operatorname{tr}\!\left(
\Omega\Gamma_{\mathcal E}(W)\Sigma
\Gamma_{\mathcal E}(W)^\top\Omega^\top
\right) =
\operatorname{tr}\!\left(
A_{\mathcal E,W}^\top\Sigma A_{\mathcal E,W}
\right).
$ 
Moreover, Setting~\ref{setting:1} implies
\(\Gamma_{\mathcal E}(W)\Lambda_{\mathcal E}=I_p\), and hence
$ 
\Lambda_{\mathcal E}^\top A_{\mathcal E,W}
=
\Omega^\top.
$ 
Therefore,
$ 
\|\Omega\|_F
=
\|\Omega^\top\|_F
=
\|\Lambda_{\mathcal E}^\top A_{\mathcal E,W}\|_F
\leq
\|\Lambda_{\mathcal E}\|_{\mathrm{op}}
\|A_{\mathcal E,W}\|_F
\leq
U\|A_{\mathcal E,W}\|_F.
$ 
Assumption~\ref{ass:constraint_set} then gives
$ 
\alpha(\mathcal E,\Sigma,W)
\geq
\underline{\lambda}\|A_{\mathcal E,W}\|_F^2 \geq
\frac{\underline{\lambda}}{U^2}\|\Omega\|_F^2.
$ Since  
\(\Omega\neq 0\), taking the infimum over all feasible designs and
estimators yields
$ 
\alpha^\star
\geq
\frac{\underline{\lambda}}{U^2}\|\Omega\|_F^2
>0.
$}

\smallskip
\noindent
\textit{Step 2 (oracle envelope and basic properties).}
For \(t\ge0\), define
\[
\delta(t)
=
\inf_{\substack{
(\mathcal E,\Sigma)\in\mathcal D\\
W\in\mathcal W(\mathcal E,\Sigma)
}}
\left\{
\alpha(\mathcal E,\Sigma,W)
+
t\beta(\mathcal E,W)
\right\}.
\]
Under Assumption \ref{ass:constraint_set}, \(\delta(t)\) is finite and strictly
positive for $t \ge 0$. Because it is the pointwise infimum of affine and nondecreasing
functions of \(t\), \(\delta\) is concave and nondecreasing for $t \ge 0$. Moreover, it is easy to show that 
$ 
\delta(0)=\alpha^\star, 
$ 
and $ 
\lim_{t\to\infty}\frac{\delta(t)}{t}
=
\beta^\star.
$\footnote{To see this, for every \(t>0\),
$ 
\frac{\delta(t)}{t}
=
\inf_{(\mathcal E,\Sigma,W)}
\left\{
\frac{\alpha(\mathcal E,\Sigma,W)}{t}
+\beta(\mathcal E,W)
\right\}
\geq \beta^\star.
$
Conversely, for every \(\varepsilon>0\), choose a feasible
\((\mathcal E_\varepsilon,\Sigma_\varepsilon,W_\varepsilon)\) such that
\(\beta(\mathcal E_\varepsilon,W_\varepsilon)
\leq\beta^\star+\varepsilon\). Then
$ 
\frac{\delta(t)}{t}
\leq
\frac{\alpha(\mathcal E_\varepsilon,\Sigma_\varepsilon,
W_\varepsilon)}{t}
+\beta^\star+\varepsilon.
$ 
Because this design is fixed and its \(\alpha\) is finite, taking
\(t\to\infty\) gives
\(\limsup_{t\to\infty}\delta(t)/t\leq\beta^\star+\varepsilon\).
Letting \(\varepsilon\downarrow0\) proves that
\(\lim_{t\to\infty}\delta(t)/t=\beta^\star\). When the infimum for $\beta^\star$ is attained, the argument works directly with $\varepsilon = 0$. }

\smallskip
\noindent
\textit{Step 3 (quasi-convexity and boundary maximization).}
Fix a feasible
\((\mathcal E,\Sigma,W)\), and define
\[
\phi(t)
=
\frac{
\alpha(\mathcal E,\Sigma,W)
+
t\beta(\mathcal E,W)
}{
\delta(t)
},
\qquad t\ge0.
\]
The numerator is a nonnegative affine function of \(t\), while the
denominator is positive by Assumption \ref{ass:constraint_set} (since the variance is bounded away from zero) and concave. Hence \(\phi\) is quasi-convex for $t \ge 0$. In particular, for every \(T<\infty\),
\[
\sup_{t\in[0,T]}\phi(t)
=
\max\{\phi(0),\phi(T)\}, 
\]
proving the claim for Theorem \ref{thm:restricted_B}. 

Letting $T\to\infty$ and using the limits in Step~2, by quasi-convexity for any $\beta^\star > 0$, 
\[
\sup_{t\ge 0}\phi(t)=\max\!\left\{\frac{\alpha}{\delta(0)},\ \lim_{t\to\infty}\frac{\alpha+t\beta}{\delta(t)}\right\}
=\max\!\left\{\frac{\alpha}{\alpha^\star},\ \frac{\beta}{\beta^\star}\right\}. 
\]

It remains to consider the boundary case \(\beta^\star=0\). If
\(\beta(\mathcal E,W)>0\), Step~2 implies
$ 
\lim_{t\to\infty}\phi(t)=+\infty,
$ 
which agrees with
\(\beta(\mathcal E,W)/\beta^\star=+\infty\). If instead
\(\beta(\mathcal E,W)=0\), then
$ 
\phi(t)
=
\frac{
\alpha(\mathcal E,\Sigma,W)
}{
\delta(t)
}.
$ 
Because \(\delta(t)\) is nondecreasing,
$ 
\sup_{t\ge0}\phi(t)
=
\phi(0)
=
\frac{
\alpha(\mathcal E,\Sigma,W)
}{
\alpha^\star
}.
$ 
Under the convention \(0/0=1\), this again equals
\[
\max\left\{
\frac{
\alpha(\mathcal E,\Sigma,W)
}{
\alpha^\star
},
\frac{
\beta(\mathcal E,W)
}{
\beta^\star
}
\right\},
\]
since
\(\alpha(\mathcal E,\Sigma,W)/\alpha^\star\ge1\).

This proves Theorem \ref{prop:multivalued_estimands} and Theorem \ref{thm:1} as a special case.

\subsection{Proof of Theorem~\ref{thm:audience_frontier}}
\label{proof:audience_frontier}

Fix a feasible design $(\mathcal E,\Sigma)\in\mathcal D$.

\smallskip
\noindent
\textit{Step 1 (profiled worst-case posterior risk).}
Under Setting~\ref{setting:2}, write
\[
\widetilde S_{\mathcal E,\Sigma}
=
\Lambda_{\mathcal E}\theta
+
\begin{pmatrix}
b\\
0_{|\mathcal E|}
\end{pmatrix}
+\varepsilon,
\qquad
\varepsilon\mid(\theta,b)\sim\mathcal N(0,\Sigma).
\]
For any $a$ satisfying $\Lambda_{\mathcal E}^{\top}a=\omega$, consider
$ 
\delta_a(\widetilde S_{\mathcal E,\Sigma})
=
\omega_0+a^\top\widetilde S_{\mathcal E,\Sigma}.
$ 
Because
$\tau_L(\theta)=\omega_0+\omega^\top\theta$, we have
$ 
\delta_a(\widetilde S_{\mathcal E,\Sigma})
-\tau_L(\theta)
=
a_{1:p_o}^\top b+a^\top\varepsilon.
$ 
Therefore, for every $\pi\in\Pi_B$,
\[
\begin{aligned}
\mathbb E_\pi\!\left[
\left(
\delta_a(\widetilde S_{\mathcal E,\Sigma})
-\tau_L(\theta)
\right)^2
\Bigm|\theta
\right]
&=
a^\top\Sigma a
+
\mathbb E_\pi\!\left[
(a_{1:p_o}^\top b)^2
\mid\theta
\right]                                          
&\leq
a^\top\Sigma a
+
B^2\|a_{1:p_o}\|_*^2.     
\end{aligned}
\]
Since the posterior mean minimizes Bayes risk under squared-error loss,
taking expectations, the supremum over $\pi\in\Pi_B$, and then the
minimum over feasible $a$ gives
\begin{equation}
\label{eq:audience_profile_upper}
\small 
\begin{aligned} 
\bar L_B(\mathcal E,\Sigma)  = \sup_{\pi \in \Pi_B} \inf_\delta \mathbb{E}\left[\Big(\delta(\tilde{S}_{\mathcal{E},\Sigma}) - \tau_L(\theta)\Big)^2\right]
& \leq \inf_\delta \sup_{\pi \in \Pi_B}  \mathbb{E}\left[\Big(\delta(\tilde{S}_{\mathcal{E},\Sigma}) - \tau_L(\theta)\Big)^2\right] \\ & \le \inf_{\delta_a:  \Lambda_{\mathcal{E}}^\top a = \omega} \sup_{\pi \in \Pi_B}  \mathbb{E}\left[\Big(\delta_a(\tilde{S}_{\mathcal{E},\Sigma}) - \tau_L(\theta)\Big)^2\right] \\ & \le 
\min_{a:\Lambda_{\mathcal E}^{\top}a=\omega}
\left\{
\alpha(\mathcal E,\Sigma,a)
+
B^2\beta(\mathcal E,a)
\right\}.
\end{aligned} 
\end{equation}

To prove the reverse inequality, let
$ 
\mathcal K
=
\operatorname{co}
\left\{
uu^\top:u\in\mathcal C
\right\}, 
$ where $\mathrm{co}(\cdot)$ denotes the convex hull. Also recall that $\Lambda_{\mathcal{E}}$ has full column rank and $\Sigma$ is positive definite by Assumption \ref{ass:constraint_set}. 
For any $K\in\mathcal K$, consider priors under which
\[
\theta\sim\mathcal N(0,T^2I_p),
\qquad
b\mid\theta\sim\mathcal N(0,B^2K).
\]
These priors belong to $\Pi_B$ by construction.\footnote{For every \(q\in\mathbb R^{p_o}\),
$ 
\mathbb E_\pi[(q^\top b)^2\mid\theta]
=
B^2q^\top Kq
\leq
B^2\sup_{u\in\mathcal C}(q^\top u)^2
=
\sup_{\widetilde b\in\mathcal B(B)}
(q^\top\widetilde b)^2,
$ 
where the inequality follows from
\(K\in\operatorname{co}\{uu^\top:u\in\mathcal C\}\).}

The posterior covariance of $\theta$ under this prior is
\[
\left[
T^{-2}I_p
+
\Lambda_{\mathcal E}^{\top}
V_{B,K}^{-1}
\Lambda_{\mathcal E}
\right]^{-1}, \qquad V_{B,K}
\equiv 
\Sigma
+
\begin{pmatrix}
B^2K&0\\
0&0_{|\mathcal E|\times|\mathcal E|}
\end{pmatrix}. 
\]
It follows that, as $T\rightarrow\infty$,
$ 
L_\pi(\mathcal E,\Sigma)
\rightarrow
\omega^\top
\left(
\Lambda_{\mathcal E}^{\top}
V_{B,K}^{-1}
\Lambda_{\mathcal E}
\right)^{-1}
\omega.
$ 
In addition, it follows that\footnote{To verify the equality, the first-order conditions of the constrained
quadratic problem give
$ 
a
=
V^{-1}\Lambda_{\mathcal E}
\left(
\Lambda_{\mathcal E}^\top
V^{-1}\Lambda_{\mathcal E}
\right)^{-1}
\omega,
$ 
and substitution yields the stated value using the full rank condition of $\Lambda_{\mathcal{E}}$ and positive definite condition of $\Sigma$ in Assumption \ref{ass:constraint_set}.}
\[
\omega^\top
\left(
\Lambda_{\mathcal E}^{\top}
V_{B,K}^{-1}
\Lambda_{\mathcal E}
\right)^{-1}
\omega
=
\min_{a:\Lambda_{\mathcal E}^{\top}a=\omega}
\left\{
a^\top\Sigma a
+
B^2a_{1:p_o}^{\top}Ka_{1:p_o}
\right\}.
\]
Consequently,
\[
\bar L_B(\mathcal E,\Sigma)
\geq
\sup_{K\in\mathcal K}
\min_{a:\Lambda_{\mathcal E}^{\top}a=\omega}
\left\{
a^\top\Sigma a
+
B^2a_{1:p_o}^{\top}Ka_{1:p_o}
\right\}.
\]
The objective is convex and coercive in $a$, affine in $K$, and
$\mathcal K$ is compact and convex ($\mathcal{C}$ is compact and convex by Assumption \ref{defn:class_priors}). The Sion's minimax theorem therefore implies\footnote{Although $a$ is not in a compact set, it suffices that $\mathcal{K}$ is compact.}
\[
\begin{aligned}
\bar L_B(\mathcal E,\Sigma)
&\geq
\min_{a:\Lambda_{\mathcal E}^{\top}a=\omega}
\sup_{K\in\mathcal K}
\left\{
a^\top\Sigma a
+
B^2a_{1:p_o}^{\top}Ka_{1:p_o}
\right\}                 =
\min_{a:\Lambda_{\mathcal E}^{\top}a=\omega}
\left\{
\alpha(\mathcal E,\Sigma,a)
+
B^2\beta(\mathcal E,a)
\right\}.
\end{aligned}
\]
Combining this inequality with \eqref{eq:audience_profile_upper} yields the first claim 
\begin{equation}
\label{eq:audience_profile_exact}
\bar L_B(\mathcal E,\Sigma)
=
\min_{a:\Lambda_{\mathcal E}^{\top}a=\omega}
\left\{
\alpha(\mathcal E,\Sigma,a)
+
B^2\beta(\mathcal E,a)
\right\}.
\end{equation}

Finally, note that Assumption~\ref{ass:constraint_set} implies
$a^\top\Sigma a\geq c \|a\|_2^2$, for a positive constant $c > 0$ while
$\Lambda_{\mathcal E}^{\top}a=\omega$ and
Assumption~\ref{ass:linearity} imply that $\|a\|_2$ is bounded away from
zero uniformly over $\mathcal E$. This follows from the fact that $||\omega||_2 = ||\Lambda_{\mathcal{E}}^\top a||_2 \le ||\Lambda_{\mathcal{E}}||_{\mathrm{op}} ||a||_2 \le U ||a||_2$ and $\omega \neq 0$. 
Thus 
$ 
\inf_{(\mathcal E',\Sigma')\in\mathcal D}
\bar L_B(\mathcal E',\Sigma')
>0$ $\forall B\geq0.$ 

\smallskip
\noindent

\smallskip
\noindent
\textit{Step 2 (upper set generated by the frontier).}
For a fixed design $(\mathcal E,\Sigma)$, define
\[
\mathcal F^+(\mathcal E,\Sigma)
\equiv
\mathcal F(\mathcal E,\Sigma)+\mathbb R_+^2.
\]
We first show that
\begin{equation}
\label{eq:frontier_upper_set}
\begin{aligned}
\mathcal F^+(\mathcal E,\Sigma)
=
\Big\{
(x,y)\in\mathbb R_+^2:
&\text{ there exists }a
\text{ satisfying }\Lambda_{\mathcal E}^{\top}a=\omega,
\quad \alpha(\mathcal E,\Sigma,a)\leq x,\quad
\beta(\mathcal E,a)\leq y
\Big\}.
\end{aligned}
\end{equation}
It immediately follows that $\mathcal{F}^+$ is a subset of the set on the right-hand side of Equation \eqref{eq:frontier_upper_set}.\footnote{Formally, if $(x,y)\in\mathcal F^+
=\mathcal F+\mathbb R_+^2$, then for some feasible $a$ and
$r_1,r_2\geq0$, $(x,y)=(\alpha(\mathcal E,\Sigma,a),
\beta(\mathcal E,a))+(r_1,r_2)$, which implies
$\alpha(\mathcal E,\Sigma,a)\leq x$ and
$\beta(\mathcal E,a)\leq y$.} We therefore only need to prove the other direction.  Consider $(x,y)$ on the right-hand side
of \eqref{eq:frontier_upper_set}. Among all $\widetilde a$ satisfying
\[
\Lambda_{\mathcal E}^{\top}\widetilde a=\omega,\qquad
\alpha(\mathcal E,\Sigma,\widetilde a)\leq x,\qquad
\beta(\mathcal E,\widetilde a)\leq y,
\]
choose one that minimizes
$ 
\alpha(\mathcal E,\Sigma,\widetilde a)
+
\beta(\mathcal E,\widetilde a).
$ 
Such a minimizer exists, since the feasible set for $\tilde{a}$ is nonempty and closed,
and Assumption~\ref{ass:constraint_set} implies
$ 
c\|\widetilde a\|_2^2
\leq
\alpha(\mathcal E,\Sigma,\widetilde a)
\leq x
$ 
for some $c>0$, so the feasible set is bounded. Let \(\mathcal G(\mathcal E,\Sigma)\) denote the set on the
right-hand side of \eqref{eq:frontier_upper_set}. We first note that
\(\mathcal G(\mathcal E,\Sigma)\) is convex\footnote{To see this, take any
\((x_1,y_1),(x_2,y_2)\in\mathcal G(\mathcal E,\Sigma)\).
By the definition of \(\mathcal G(\mathcal E,\Sigma)\), for each
\(i\in\{1,2\}\), there exists a vector \(a_i\) satisfying
$ 
\Lambda_{\mathcal E}^{\top}a_i=\omega,
\alpha(\mathcal E,\Sigma,a_i)\leq x_i,
\beta(\mathcal E,a_i)\leq y_i.
$ 
For any \(\eta\in[0,1]\), define
$
a_\eta=\eta a_1+(1-\eta)a_2.
$ Then
\(a_\eta=\eta a_1+(1-\eta)a_2\) is feasible and, by convexity of
\(\alpha(\mathcal E,\Sigma,\cdot)\) and
\(\beta(\mathcal E,\cdot)\),
$
\alpha(\mathcal E,\Sigma,a_\eta)
\leq \eta x_1+(1-\eta)x_2,
\beta(\mathcal E,a_\eta)
\leq \eta y_1+(1-\eta)y_2.
$
Therefore,
\(\eta(x_1,y_1)+(1-\eta)(x_2,y_2)
\in\mathcal G(\mathcal E,\Sigma)\).}, and as we define 
$ 
p\equiv
\left(
\alpha(\mathcal E,\Sigma,\widetilde a),
\beta(\mathcal E,\widetilde a)
\right), 
$
 \(p\) lies on the boundary of
\(\mathcal G(\mathcal E,\Sigma)\).\footnote{To show this, suppose not. Then if \(p\) were an interior
point, for some sufficiently small \(\varepsilon>0\),
$ 
p-\varepsilon \cdot (1,1)\in\mathcal G(\mathcal E,\Sigma).
$ 
Consequently, there would exist a feasible \(a_\varepsilon\) such that
$
\alpha(\mathcal E,\Sigma,a_\varepsilon)
\leq
\alpha(\mathcal E,\Sigma,\widetilde a)-\varepsilon,
\beta(\mathcal E,a_\varepsilon)
\leq
\beta(\mathcal E,\widetilde a)-\varepsilon.
$ 
Because \(p\leq(x,y)\) coordinatewise, \(a_\varepsilon\) would satisfy
the constraints defining \(\widetilde a\), while
$ 
\alpha(\mathcal E,\Sigma,a_\varepsilon)
+\beta(\mathcal E,a_\varepsilon)
<
\alpha(\mathcal E,\Sigma,\widetilde a)
+\beta(\mathcal E,\widetilde a),
$ 
contradicting the optimality of \(\widetilde a\).}

As a result, we can apply 
the supporting-hyperplane theorem which implies that for a nonzero vector
\(q=(q_1,q_2)\)
\[
q_1\alpha(\mathcal E,\Sigma,\widetilde a)
+q_2\beta(\mathcal E,\widetilde a)
\leq
q_1x'+q_2y'
\qquad
\text{for every }(x',y')\in\mathcal G(\mathcal E,\Sigma).
\]
Because \(\mathcal G(\mathcal E,\Sigma)\) is also upper comprehensive (if \(z\in\mathcal G(\mathcal E,\Sigma)\), then
\(z+r\in\mathcal G(\mathcal E,\Sigma)\) for every
\(r\in\mathbb R_+^2\)), it must be that \(q_1,q_2\geq0\); otherwise,
increasing the coordinate associated with a negative component of
\(q\) would contradict the supporting inequality (e.g., if $q_1 < 0$, we can make $x'$ arbitrary large and contradict the inequality).

Since \(q\neq0\), \(q_1+q_2>0\). Define
$ 
\lambda\equiv\frac{q_2}{q_1+q_2}\in[0,1].
$ 
Applying the supporting inequality to the point
$ 
\left(
\alpha(\mathcal E,\Sigma,a),
\beta(\mathcal E,a)
\right)\in\mathcal G(\mathcal E,\Sigma)
$ 
for each feasible \(a\) shows that
$ 
\widetilde a\in
\arg\min_{a:\Lambda_{\mathcal E}^{\top}a=\omega}
\left\{
(1-\lambda)\alpha(\mathcal E,\Sigma,a)
+\lambda\beta(\mathcal E,a)
\right\}.
$ 
Therefore,
$ 
p\in\mathcal F(\mathcal E,\Sigma).
$ 
Finally, since \(p\leq(x,y)\) coordinatewise,
$ 
(x,y)\in
\mathcal F(\mathcal E,\Sigma)+\mathbb R_+^2
=
\mathcal F^+(\mathcal E,\Sigma),
$ 
which proves the reverse inclusion in
\eqref{eq:frontier_upper_set}.

\smallskip
\noindent

\textit{Step 3 (properties of the set).} Step 2 shows that 
$\mathcal F^+(\mathcal E,\Sigma)$ is convex and upper comprehensive.
We now claim that it is also a closed set. To see this, consider a convergent sequence
$(x_n,y_n)\in\mathcal F^+(\mathcal E,\Sigma)$ and corresponding feasible
vectors $a_n$. Since
$ 
c\|a_n\|_2^2
\leq
\alpha(\mathcal E,\Sigma,a_n)
\leq x_n,
$  for some finite positive constant $c$, 
the sequence $(a_n)$ is bounded. Any convergent subsequence has a feasible
limit satisfying the inequalities defining the limit $(x,y)$. By continuity of
\(\alpha\) and \(\beta\), and continuity of the affine constraint,
$ 
\Lambda_{\mathcal E}^{\top}a=\omega, 
\alpha(\mathcal E,\Sigma,a)\leq x, \beta(\mathcal E,a)\leq y.
$ 
Equation~\eqref{eq:frontier_upper_set} therefore implies that
\((x,y)\in\mathcal F^+(\mathcal E,\Sigma)\).

It follows from \eqref{eq:audience_profile_exact} and
\eqref{eq:frontier_upper_set} that, for every $q_1>0$ and $q_2\geq0$,
\begin{equation}
\label{eq:frontier_support_risk}
\begin{aligned}
\inf_{(x,y)\in\mathcal F^+(\mathcal E,\Sigma)}
\{q_1x+q_2y\}
&=
\min_{a:\Lambda_{\mathcal E}^{\top}a=\omega}
\left\{
q_1\alpha(\mathcal E,\Sigma,a)
+
q_2\beta(\mathcal E,a)
\right\}               =
q_1\bar L_{\sqrt{q_2/q_1}}(\mathcal E,\Sigma).
\end{aligned}
\end{equation}

\smallskip
\noindent
\textit{Step 4  (frontier domination implies risk domination).} 
Fix $R\geq1$ and suppose that
\begin{equation}
\label{eq:frontier_domination_proof}
\forall(\alpha,\beta)\in\mathcal F^\star,\quad
\exists(\widetilde\alpha,\widetilde\beta)
\in\mathcal F(\mathcal E,\Sigma)
\quad\text{such that}\quad
\widetilde\alpha\leq R\alpha,
\qquad
\widetilde\beta\leq R\beta.
\end{equation}

We want to show that $\mathcal R^{\mathrm{Bayes}}(\mathcal E,\Sigma)\leq R$. 

To show this, by the existence of the solution in the oracle set $\mathcal{F}^\star$ imposed in Definition \ref{defn:pareto},  there exists
$(\alpha_t^\star,\beta_t^\star)\in\mathcal F^\star$ satisfying for given $t \ge 0$,  $\lambda = t/(1 + t)$
\begin{equation}
\label{eq:oracle_frontier_support}
\alpha_t^\star+t\beta_t^\star
=
\inf_{\substack{
(\mathcal E',\Sigma')\in\mathcal D\\
a:\Lambda_{\mathcal E'}^\top a=\omega
}}
\left\{
\alpha(\mathcal E',\Sigma',a)
+t\beta(\mathcal E',a)
\right\}.
\end{equation}
By our working hypothesis in \eqref{eq:frontier_domination_proof}, there exists
$(\widetilde\alpha_t,\widetilde\beta_t)
\in\mathcal F(\mathcal E,\Sigma)$ such that
\[
\widetilde\alpha_t\leq R\alpha_t^\star,
\qquad
\widetilde\beta_t\leq R\beta_t^\star.
\]
Therefore,
$ 
\bar L_{\sqrt t}(\mathcal E,\Sigma)
\leq
\widetilde\alpha_t+t\widetilde\beta_t \leq
R(\alpha_t^\star+t\beta_t^\star) =
R\inf_{(\mathcal E',\Sigma')\in\mathcal D}
\bar L_{\sqrt t}(\mathcal E',\Sigma').
$ 
The oracle risk $\inf_{(\mathcal E',\Sigma')\in\mathcal D}
\bar L_{\sqrt t}(\mathcal E',\Sigma')$ is strictly positive by the last paragraph in Step~1. We therefore divide
by it and take the supremum over $t\geq0$, obtaining
$ 
\mathcal R^{\mathrm{Bayes}}(\mathcal E,\Sigma)\leq R.
$ 

\smallskip
\noindent
\textit{Step 5 (risk domination implies frontier domination by separating hyperplane theorem).}
Conversely, suppose that
\[
\mathcal R^{\mathrm{Bayes}}(\mathcal E,\Sigma)\leq R.
\]
Then we want to show by contradiction that $ 
(R\alpha,R\beta)\in\mathcal F^+(\mathcal E,\Sigma)
$ for every $(\alpha,\beta) \in \mathcal{F}^\star$. 

Fix $(\alpha,\beta)\in\mathcal F^\star$ and 
suppose $(R\alpha,R\beta) \not \in\mathcal F^+(\mathcal E,\Sigma)$. Since $\mathcal F^+(\mathcal E,\Sigma)$ is closed and
convex by Step~3, the separating-hyperplane theorem gives
$(q_1,q_2)\neq(0,0)$ such that
\begin{equation}
\label{eq:frontier_separation}
R(q_1\alpha+q_2\beta)
<
\inf_{(x,y)\in\mathcal F^+(\mathcal E,\Sigma)}
(q_1x+q_2y).
\end{equation}

\smallskip
\noindent
\textit{Step 6 (condition for $q_1 \ge 0, q_2 \ge 0$).}
We next want to show that Equation \eqref{eq:frontier_separation} must hold for 
$q_1,q_2\geq0$. We prove this by contradiction. If $q_1<0$, then for every
$(x,y)\in\mathcal F^+(\mathcal E,\Sigma)$ and every $s\geq0$,
$(x+s,y)\in\mathcal F^+(\mathcal E,\Sigma)$, implying that the infimum
on the right-hand side of \eqref{eq:frontier_separation} is $-\infty$.
This contradicts \eqref{eq:frontier_separation}, because its left-hand side is finite whereas its right-hand side would equal $-\infty$.
The same argument gives $q_2\geq0$.

\smallskip
\noindent
\textit{Step 7 (contradiction for $q_1 > 0, q_2 \ge 0$).}
Suppose first that $q_1>0$ and set $t=q_2/q_1$. Equation
\eqref{eq:frontier_support_risk} and the maintained assumption that $\mathcal{R}^{\mathrm{Bayes}}(\mathcal{E},\Sigma) \le R$ imply for $(\alpha,\beta) \in \mathcal{F}^\star$ selected in Step 5 for the contradiction argument, 
\[
\begin{aligned}
\inf_{(x,y)\in\mathcal F^+(\mathcal E,\Sigma)}
(q_1x+q_2y)
&=
q_1\bar L_{\sqrt t}(\mathcal E,\Sigma) \leq
Rq_1
\inf_{(\mathcal E',\Sigma')\in\mathcal D}
\bar L_{\sqrt t}(\mathcal E',\Sigma')\\
&=
R
\inf_{\substack{
(\mathcal E',\Sigma')\in\mathcal D\\
a:\Lambda_{\mathcal E'}^\top a=\omega
}}
\left\{
q_1\alpha(\mathcal E',\Sigma',a)
+q_2\beta(\mathcal E',a)
\right\} \leq
R(q_1\alpha+q_2\beta).
\end{aligned}
\]
The last inequality uses the existence of a solution in Definition \ref{defn:pareto}.  The resulting inequality
contradicts \eqref{eq:frontier_separation}.

\smallskip
\noindent
\textit{Step 8 (contradiction for $q_1=0$ and conclusion).}
It remains to consider $q_1=0$. Since $(q_1,q_2)\neq(0,0)$ and
$q_1,q_2\geq0$, we must have $q_2>0$.
Note that\footnote{To verify this equality, note first that $\alpha > 0$, so
$ 
\frac{\bar L_{\sqrt t}(\mathcal E,\Sigma)}{t}
=
\inf_{a:\Lambda_{\mathcal E}^\top a=\omega}
\left\{
\frac{\alpha(\mathcal E,\Sigma,a)}{t}
+\beta(\mathcal E,a)
\right\}
\geq
\inf_{a:\Lambda_{\mathcal E}^\top a=\omega}
\beta(\mathcal E,a).
$ 
Conversely, evaluating the infimum at any fixed feasible $a$ and taking
$t\to\infty$ gives an upper limit no larger than $\beta(\mathcal E,a)$.
Taking the infimum over feasible $a$ proves
\eqref{eq:fixed_beta_limit}. A similar argument applies once minimizing over $(\mathcal{E}',\Sigma') \in \mathcal{D}$.} 
\begin{equation}
\label{eq:fixed_beta_limit}
\lim_{t\to\infty}
\frac{\bar L_{\sqrt t}(\mathcal E,\Sigma)}{t}
=
\inf_{a:\Lambda_{\mathcal E}^\top a=\omega}
\beta(\mathcal E,a), \qquad \lim_{t\to\infty}
\frac{
\inf_{(\mathcal E',\Sigma')\in\mathcal D}
\bar L_{\sqrt t}(\mathcal E',\Sigma')
}{t}
=
\inf_{\substack{
(\mathcal E',\Sigma')\in\mathcal D\\
a:\Lambda_{\mathcal E'}^\top a=\omega
}}
\beta(\mathcal E',a). 
\end{equation} 

Consequently, by Step~2, since $\bar{L}_{\sqrt{t}}(\mathcal{E},\Sigma) \le R \inf_{(\mathcal{E}',\Sigma') \in \mathcal{D}} \bar{L}_{\sqrt{t}}(\mathcal{E}',\Sigma')$ for all $t > 0$, 
\[
\begin{aligned}
\inf_{(x,y)\in\mathcal F^+(\mathcal E,\Sigma)}q_2y
&=
q_2\inf_{a:\Lambda_{\mathcal E}^\top a=\omega}
\beta(\mathcal E,a) \leq
Rq_2
\inf_{\substack{
(\mathcal E',\Sigma')\in\mathcal D\\
a:\Lambda_{\mathcal E'}^\top a=\omega
}}
\beta(\mathcal E',a) \leq
Rq_2\beta
\end{aligned}
\]
where the last inequality follows since $(\alpha,\beta) \in \mathcal{F}^\star$ by our starting hypothesis. 
This again contradicts \eqref{eq:frontier_separation}. Hence
$ 
(R\alpha,R\beta)\in\mathcal F^+(\mathcal E,\Sigma).
$ 
By the definition
$\mathcal F^+(\mathcal E,\Sigma)
=\mathcal F(\mathcal E,\Sigma)+\mathbb R_+^2$, there therefore exists
$(\widetilde\alpha,\widetilde\beta)
\in\mathcal F(\mathcal E,\Sigma)$ such that
$ 
\widetilde\alpha\leq R\alpha,
\widetilde\beta\leq R\beta.
$ 
Because $(\alpha,\beta)\in\mathcal F^\star$ was arbitrary, we have proved
that, for every $R\geq1$,
\[
\small 
\begin{aligned} 
\mathcal R^{\mathrm{Bayes}}(\mathcal E,\Sigma)\leq R
\quad\Longleftrightarrow\quad
\forall(\alpha,\beta)\in\mathcal F^\star,\ 
\exists(\widetilde\alpha,\widetilde\beta)
\in\mathcal F(\mathcal E,\Sigma):
\begin{cases}
\widetilde\alpha\leq R\alpha,\\
\widetilde\beta\leq R\beta.
\end{cases}
\end{aligned}
\]
Taking the infimum over $R\geq1$ proves the result.

\subsection{Proof of Corollary~\ref{cor:finite_frontier_approximation}}
\label{proof:cor1}

For concision, for a fixed design write
$ 
F(\mathcal E,\Sigma;\lambda)
\equiv
\min_{a:\Lambda_{\mathcal E}^{\top}a=\omega}
\left\{
(1-\lambda)\alpha(\mathcal E,\Sigma,a)
+\lambda\beta(\mathcal E,a)
\right\}.
 $

\smallskip
\noindent
\textit{Step 1 (uniform Lipschitz continuity).}
We first establish Lipschitz continuity of $F$ in $\lambda$. 
By compactness of $\mathcal C$, there exists
$ 
c_{\mathcal C}
\equiv
\max_{u\in\mathcal C}\|u\|_2^2<\infty
$ 
such that, for every feasible $a$,
$ 
\beta(\mathcal E,a)
=
\sup_{u\in\mathcal C}
(a_{1:p_o}^{\top}u)^2
\leq
c_{\mathcal C}\|a\|_2^2.
$ 
For each feasible
$\mathcal E$ (finite by assumption), choose
\[
s_{\mathcal E}
\in
\arg\min_{a:\Lambda_{\mathcal E}^{\top}a=\omega}
\beta(\mathcal E,a).
\]
Such vector exist because $\Lambda_{\mathcal E}$ has full column rank
and $\beta(\mathcal E,\cdot)^{1/2}$ is a continuous seminorm on a
finite-dimensional space. Hence
$ 
A
\equiv
\max_{\mathcal E}
\|s_{\mathcal E}\|_2^2
<\infty.
$ 

Fix $(\mathcal E,\Sigma)\in\mathcal D$, and let $a^v$ minimize
$\alpha(\mathcal E,\Sigma,a)$ subject to
$\Lambda_{\mathcal E}^{\top}a=\omega$. The uniform eigenvalue bounds in
Assumption~\ref{ass:constraint_set} imply
\[
\begin{aligned}
\alpha(\mathcal E,\Sigma,a^v)
&\leq
\alpha(\mathcal E,\Sigma,s_{\mathcal E})
\leq
\bar\lambda A, \qquad 
\|a^v\|_2^2
&\leq
\frac{\bar\lambda}{\underline\lambda}A, \qquad 
\beta(\mathcal E,a^v)
&\leq
c_{\mathcal C}
\frac{\bar\lambda}{\underline\lambda}A.
\end{aligned}
\]

For $\lambda\in(0,1/2]$, let $a^\lambda$ attain
$F(\mathcal E,\Sigma;\lambda)$ and set
$t=\lambda/(1-\lambda)\leq1$. Optimality gives
\[
\alpha(\mathcal E,\Sigma,a^\lambda)
+t\beta(\mathcal E,a^\lambda)
\leq
\alpha(\mathcal E,\Sigma,a^v)
+t\beta(\mathcal E,a^v).
\]
Because $a^v$ minimizes $\alpha$, it follows that
\[
\begin{aligned}
\beta(\mathcal E,a^\lambda)
&\leq
\beta(\mathcal E,a^v), \qquad 
\alpha(\mathcal E,\Sigma,a^\lambda)
\leq
\alpha(\mathcal E,\Sigma,a^v)
+\beta(\mathcal E,a^v).
\end{aligned}
\]
For $\lambda=0$, choose $a^\lambda=a^v$.

For $\lambda\in(1/2,1)$, set
$t=(1-\lambda)/\lambda<1$. Comparing $a^\lambda$ with
$s_{\mathcal E}$ similarly gives
\[
\beta(\mathcal E,a^\lambda)
+t\alpha(\mathcal E,\Sigma,a^\lambda)
\leq
\beta(\mathcal E,s_{\mathcal E})
+t\alpha(\mathcal E,\Sigma,s_{\mathcal E}).
\]
Because $s_{\mathcal E}$ minimizes $\beta$, it follows that
\[
\begin{aligned}
\alpha(\mathcal E,\Sigma,a^\lambda)
&\leq
\alpha(\mathcal E,\Sigma,s_{\mathcal E})
\leq
\bar\lambda A, \qquad 
\beta(\mathcal E,a^\lambda)
\leq
\beta(\mathcal E,s_{\mathcal E})
+\alpha(\mathcal E,\Sigma,s_{\mathcal E})
\leq
(c_{\mathcal C}+\bar\lambda)A.
\end{aligned}
\]
For $\lambda=1$, choose $a^\lambda=s_{\mathcal E}$.

Consequently, there exists a constant
$ 
M
\equiv
A\max\left\{
\bar\lambda+
c_{\mathcal C}\frac{\bar\lambda}{\underline\lambda},
\,
c_{\mathcal C}+\bar\lambda
\right\}
<\infty
$ 
such that a minimizer $a^\lambda$ can be chosen satisfying
\[
\max\left\{
\alpha(\mathcal E,\Sigma,a^\lambda),
\beta(\mathcal E,a^\lambda)
\right\}
\leq M
\]
uniformly over $(\mathcal E,\Sigma)\in\mathcal D$ and
$\lambda\in[0,1]$.

For any $\lambda,\lambda'\in[0,1]$, optimality of $a^\lambda$ with respect to $F(\mathcal{E},\Sigma,\lambda)$ therefore
implies
\[
\begin{aligned}
F(\mathcal E,\Sigma;\lambda')
&\leq
(1-\lambda')
\alpha(\mathcal E,\Sigma,a^\lambda)
+\lambda'\beta(\mathcal E,a^\lambda) =
F(\mathcal E,\Sigma;\lambda)
+(\lambda'-\lambda)
\left[
\beta(\mathcal E,a^\lambda)
-\alpha(\mathcal E,\Sigma,a^\lambda)
\right]\\
&\leq
F(\mathcal E,\Sigma;\lambda)
+M|\lambda'-\lambda|.
\end{aligned}
\]
Interchanging $\lambda$ and $\lambda'$ gives
\begin{equation}
\label{eq:oracle_lipschitz}
\begin{aligned}
\left|
F(\mathcal E,\Sigma;\lambda)
-
F(\mathcal E,\Sigma;\lambda')
\right|
&\leq
M|\lambda-\lambda'|, \qquad 
\left|
F^\star(\lambda)-F^\star(\lambda')
\right|
&\leq
M|\lambda-\lambda'|.
\end{aligned}
\end{equation}
The second inequality follows because the pointwise minimum of
uniformly $M$-Lipschitz functions is itself $M$-Lipschitz.

\smallskip
\noindent
\textit{Step 2 (scalarization of audience regret).}
Define
$ 
m^\star
\equiv
\min\{F^\star(0),F^\star(1)\}.
$ 
For every feasible $(\mathcal E,\Sigma,a)$,
$ 
\alpha(\mathcal E,\Sigma,a)\geq F^\star(0),
\beta(\mathcal E,a)\geq F^\star(1).
$ 
It follows that, for every $\lambda\in[0,1]$,
\begin{equation}
\label{eq:oracle_uniform_lower_bound}
F^\star(\lambda)
\geq
(1-\lambda)F^\star(0)+\lambda F^\star(1)
\geq m^\star.
\end{equation}
Here
$m^\star>0$ by assumption. By \eqref{eq:audience_profile_exact}, setting
$ 
\lambda=\frac{B^2}{1+B^2}
$ 
gives
$ 
\bar L_B(\mathcal E,\Sigma)
=
(1+B^2)F(\mathcal E,\Sigma;\lambda).
$ 
The common factor $1+B^2$ cancels from the proportional regret.
Therefore,
\begin{equation}
\label{eq:audience_scalarization_ratio}
\mathcal R^{\mathrm{Bayes}}(\mathcal E,\Sigma)
=
\sup_{\lambda\in[0,1]}
\frac{F(\mathcal E,\Sigma;\lambda)}
{F^\star(\lambda)}.
\end{equation}
The endpoint $\lambda=1$ is included by
\eqref{eq:oracle_lipschitz}--\eqref{eq:oracle_uniform_lower_bound} and
continuity.

\smallskip
\noindent
\textit{Step 3 (the grid problem is a lower bound).}
Fix any $(\mathcal E,\Sigma)\in\mathcal D$ and set
$ 
R=\mathcal R^{\mathrm{Bayes}}(\mathcal E,\Sigma).
$ 
Because
$F(\mathcal E,\Sigma;\lambda)\geq F^\star(\lambda)$ for every
$\lambda$, Equation~\eqref{eq:audience_scalarization_ratio} implies
$R\geq1$. For each grid point, choose $a_k$ attaining
$F(\mathcal E,\Sigma;\lambda_k)$. Then
\[
(1-\lambda_k)\alpha(\mathcal E,\Sigma,a_k)
+\lambda_k\beta(\mathcal E,a_k)
\leq
R F^\star(\lambda_k).
\]
Thus $(\mathcal E,\Sigma,R,a_0,\ldots,a_K)$ is feasible for
\eqref{eq:finite_lambda_program}. Since this holds for every feasible
design,
$ 
\underline R_K
\leq
\min_{(\mathcal E,\Sigma)\in\mathcal D}
\mathcal R^{\mathrm{Bayes}}(\mathcal E,\Sigma).
$ 

\smallskip
\noindent
\textit{Step 4 (approximation guarantee).}
Take any feasible
$(\mathcal E,\Sigma,R,a_0,\ldots,a_K)$ in
\eqref{eq:finite_lambda_program}. Since
$F(\mathcal E,\Sigma;\lambda_k)$ is the minimum over feasible weights,
\[
F(\mathcal E,\Sigma;\lambda_k)
\leq
R F^\star(\lambda_k),
\qquad k=0,\ldots,K.
\]

Fix an arbitrary $\lambda\in[0,1]$. If
$\lambda\in[\lambda_j,\lambda_{j+1}]$, choose the nearer of
$\lambda_j$ and $\lambda_{j+1}$ and denote it by $\lambda_k$. Then
\[
|\lambda-\lambda_k|
\leq
\frac{\lambda_{j+1}-\lambda_j}{2}
\leq
\frac{\varepsilon m^\star}{2}.
\]
Using \eqref{eq:oracle_lipschitz},
\[
\begin{aligned}
F(\mathcal E,\Sigma;\lambda)
&\leq
F(\mathcal E,\Sigma;\lambda_k)
+M|\lambda-\lambda_k| \leq
R F^\star(\lambda_k)
+M|\lambda-\lambda_k| \leq
R F^\star(\lambda)
+(R+1)M|\lambda-\lambda_k|.
\end{aligned}
\]
Because $R\geq1$, the grid condition and
\eqref{eq:oracle_uniform_lower_bound} imply
\[
\begin{aligned}
(R+1)M|\lambda-\lambda_k|
&\leq
\frac{\varepsilon(R+1)Mm^\star}{2} \leq
\varepsilon RMm^\star
\leq
\varepsilon RMF^\star(\lambda).
\end{aligned}
\]
Consequently,
\[
F(\mathcal E,\Sigma;\lambda)
\leq
(1+\varepsilon M)R F^\star(\lambda)
\qquad
\text{for every }\lambda\in[0,1].
\]
Equation~\eqref{eq:audience_scalarization_ratio} therefore gives
$ 
\mathcal R^{\mathrm{Bayes}}(\mathcal E,\Sigma)
\leq
(1+\varepsilon M)R.
$ 
Since this holds for every feasible tuple in
\eqref{eq:finite_lambda_program},
$ 
\min_{(\mathcal E,\Sigma)\in\mathcal D}
\mathcal R^{\mathrm{Bayes}}(\mathcal E,\Sigma)
\leq
(1+\varepsilon M)\underline R_K.
$ 
Combining this inequality with Step~3 proves the result.

\subsection{Proof of Theorem \ref{thm:CI_equivalence}}
\label{proof:ci}

Write \(z=z_{1-\eta} > 0\).

\textit{Step 1 (closed form of the worst-case expected distance).}
Fix a feasible design
\((\mathcal E,\Sigma)\in\mathcal D\),
\(W\in\mathcal W(\mathcal E,\Sigma)\), and \(B\geq0\).
By Definition~\ref{defn:variance_regret} and symmetry of $\mathcal{C}$ it follows that 
$ 
\sup_{b\in\mathcal B(B)}
\underline a_{W,1:p_o}b
=
B\sqrt{
\beta (\mathcal E,\underline {a}_W)
}.
$ 
Consequently,
$ 
\iota_B(\mathcal E,\Sigma,W)
= 
\underline{\tau}_L(\hat\theta_W)
-
z\sqrt{
\alpha(\mathcal E,\Sigma,\underline{a}_W)
}
-
B\sqrt{
\beta_{\underline\omega}(\mathcal E,\underline{a}_W)
}.
$ 
Because
\(\Gamma_{\mathcal E}(W)\Lambda_{\mathcal E}=I_p\),
the construction of the estimator in Setting~\ref{setting:1} and the Gaussian property in Equation \ref{eqn:guassian} imply for a bias $b_0$
\[
\small 
\begin{aligned} 
\underline{\tau}_L(\hat\theta_W)
-
\underline{\tau}_L(\theta)
\overset{d}{=}
\underline{a}_{W,1:p_o} b_0
+
\sqrt{
\alpha(\mathcal E,\Sigma,\underline{a}_W)
}\,Z,
\qquad
Z\sim\mathcal N(0,1).
\end{aligned} 
\]
It follows that
\[
\small 
\begin{aligned}
&
\sup_{b_0\in\mathcal B(B)}
\mathbb E_{\mathcal E,\Sigma,b_0}
\left[
\left|
\underline{\tau}_L(\theta)
-
\iota_B(\mathcal E,\Sigma,W)
\right|
\right]
 \\ & =
\sup_{b_0\in\mathcal B(B)}
\mathbb E
\left[
\left|
z\sqrt{
\alpha (\mathcal E,\Sigma,\underline{a}_W)
}
+
B\sqrt{
\beta(\mathcal E,\underline{a}_W)
}
-
\underline{a}_{W,1:p_o}b_0
-
\sqrt{
\alpha(\mathcal E,\Sigma,\underline{a}_W)
}\,Z
\right|
\right].
\end{aligned}
\]
The map
$ 
c\longmapsto
\mathbb E\left[
\left|
c-
\sqrt{
\alpha(\mathcal{E},\Sigma, \underline{a}_W)
}\,Z
\right|
\right]
$ 
is nondecreasing for \(c\geq0\). Moreover,
$ 
\underline a_{W,1:p_o}b_0
\in
\left[
-B\sqrt{\beta(\mathcal E,\underline{a}_W)},
B\sqrt{\beta(\mathcal E,\underline{a}_W)}
\right].
$ 
The worst case is therefore attained when
$ 
\underline a_{W,1:p_o}b_0
=
-B\sqrt{\beta(\mathcal E,\underline{a}_W)}.
$ 
Thus,
\[
\small 
\begin{aligned}
&
\sup_{b_0\in\mathcal B(B)}
\mathbb E_{\mathcal E,\Sigma,b_0}
\left[
\left|
\underline{\tau}_L(\theta)
-
\iota_B(\mathcal E,\Sigma,W)
\right|
\right]
 =
\mathbb E
\left[
\left|
z\sqrt{
\alpha(\mathcal{E},\Sigma, \underline{a}_W)
}
+
2B\sqrt{
\beta(\mathcal E,\underline{a}_W)
}
-
\sqrt{
\alpha(\mathcal{E},\Sigma, \underline{a}_W)
}\,Z
\right|
\right].
\end{aligned}
\]

For \(x,y\geq0\), define
\[
G(x,y)
\equiv
\mathbb E\left[|x(z-Z)+2y|\right].
\]
The function \(G\) is continuous, positively homogeneous, and
nondecreasing in both arguments.\footnote{To see this, note that 
\(G(rx,ry)=rG(x,y)\) for \(r\geq0\). Moreover, for \(x>0\),
direct differentiation gives
\[
\begin{aligned}
\frac{\partial G(x,y)}{\partial x}
&=
2\varphi\left(z+\frac{2y}{x}\right)
+
z\left[
2\Phi\left(z+\frac{2y}{x}\right)-1
\right]
>0, \qquad 
\frac{\partial G(x,y)}{\partial y}
&=
2\left[
2\Phi\left(z+\frac{2y}{x}\right)-1
\right]
>0,
\end{aligned}
\]
where \(\Phi\) and \(\varphi\) are the standard normal distribution
and density. Monotonicity at \(x=0\) follows by continuity.
}  We have just shows that the worst-case expected excess lengh equal
\[
\small 
\begin{aligned} 
G\left(
\sqrt{
\alpha(\mathcal{E},\Sigma, \underline{a}_W)
},
B\sqrt{
\beta(\mathcal E,\underline{a}_W)
}
\right).
\end{aligned} 
\]

\smallskip
\textit{Step 2 (oracle envelope and basic properties).}
For \(t\geq0\), define
\[
\delta(t)
=
\inf_{\substack{
(\mathcal E',\Sigma')\in\mathcal D\\
W'\in\mathcal W(\mathcal E',\Sigma')
}}
G\left(
\sqrt{
\alpha(\mathcal E',\Sigma',\underline{a}_{W'})
},
t\sqrt{
\beta(\mathcal E',\underline{a}_{W'})
}
\right).
\]
The maintained assumptions and the fact that
\(\underline\omega\neq0\) imply
\(\alpha_{\underline\omega}^\star>0\). Hence \(\delta(t)\) is finite
and strictly positive. Because \(G\) is nondecreasing in its second
argument, \(\delta\) is nondecreasing. Positive homogeneity gives
\[
\delta(0)
=
G(1,0)
\sqrt{\alpha_{\underline\omega}^\star}.
\]
Moreover, since \(G(0,y)=2y\),
$ 
\lim_{t\to\infty}\frac{\delta(t)}{t}
=
2\sqrt{\beta_{\underline\omega}^\star}.
$

\smallskip
\textit{Step 3 (homogeneity and boundary maximization).}
Fix a feasible \((\mathcal E,\Sigma,W)\), and define
\[
\phi(t)
=
\frac{
G\left(
\sqrt{
\alpha(\mathcal{E},\Sigma, \underline{a}_W)
},
t\sqrt{
\beta(\mathcal E,\underline{a}_W)
}
\right)
}{
\delta(t)
},
\qquad t\geq0.
\]
By Steps~1-2,
$ 
\mathcal R^{\mathrm{CI}}(\mathcal E,\Sigma,W)
=
\sup_{t\geq0}\phi(t).
$ 

Suppose first that
\(\beta_{\underline\omega}^\star>0\), and let
$ 
R
=
\max\left\{
\sqrt{
\frac{
\alpha(\mathcal{E},\Sigma, \underline{a}_W)
}{
\alpha_{\underline\omega}^\star
}
},
\sqrt{
\frac{
\beta(\mathcal E,\underline{a}_W)
}{
\beta_{\underline\omega}^\star
}
}
\right\}.
$ 
For every feasible
\((\mathcal E',\Sigma',W')\),
\[
\small 
\begin{aligned}
\sqrt{
\alpha(\mathcal{E},\Sigma, \underline{a}_W)
}
&\leq
R\sqrt{
\alpha(\mathcal E',\Sigma',\underline{a}_{W'})
},
\qquad 
\sqrt{
\beta(\mathcal E,\underline{a}_W)
}
&\leq
R\sqrt{
\beta(\mathcal E',\underline{a}_{W'})
}.
\end{aligned}
\]
Monotonicity and positive homogeneity of \(G\) therefore imply
\[
\begin{aligned}
&
G\left(
\sqrt{
\alpha(\mathcal{E},\Sigma, \underline{a}_W)
},
t\sqrt{
\beta(\mathcal E,\underline{a}_W)
}
\right) \leq
R
G\left(
\sqrt{
\alpha(\mathcal E',\Sigma',\underline{a}_{W'})
},
t\sqrt{
\beta(\mathcal E',\underline{a}_{W'})
}
\right).
\end{aligned}
\]
Taking the infimum over feasible
\((\mathcal E',\Sigma',W')\) gives
$ 
\phi(t)\leq R, 
\text{for every }t\geq0.
$ 
On the other hand, Step~2 gives
\[
\phi(0)
=
\sqrt{
\frac{
\alpha(\mathcal{E},\Sigma, \underline{a}_W)
}{
\alpha_{\underline\omega}^\star
}
},
\qquad
\lim_{t\to\infty}\phi(t)
=
\sqrt{
\frac{
\beta(\mathcal E,\underline{a}_W)
}{
\beta_{\underline\omega}^\star
}
}.
\]
Hence,
$ 
\mathcal R^{\mathrm{CI}}(\mathcal E,\Sigma,W)
=
\max\left\{
\sqrt{
\frac{
\alpha(\mathcal{E},\Sigma, \underline{a}_W)
}{
\alpha_{\underline\omega}^\star
}
},
\sqrt{
\frac{
\beta(\mathcal E,\underline{a}_W)
}{
\beta_{\underline\omega}^\star
}
}
\right\}.
$ 

It remains to consider
\(\beta_{\underline\omega}^\star=0\).
If
\(\beta(\mathcal E,\underline{a}_W)>0\), Step~2 implies
$ 
\lim_{t\to\infty}\phi(t)=+\infty.
$ 
If instead
\(\beta(\mathcal E,\underline{a}_W)=0\), the numerator of
\(\phi(t)\) does not depend on \(t\). Since \(\delta(t)\) is
nondecreasing,
\[
\sup_{t\geq0}\phi(t)
=
\phi(0)
=
\sqrt{
\frac{
\alpha(\mathcal{E},\Sigma, \underline{a}_W)
}{
\alpha_{\underline\omega}^\star
}
}.
\]
Under the conventions \(0/0=1\) and \(x/0=+\infty\) for \(x>0\),
the same characterization therefore holds in all cases. Squaring both
sides yields the desired result for the second bullet point.

Finally, suppose that the model is point identified, so that the lower
envelope coincides with the target as a function and
\(\underline\omega=\omega\). Theorem~\ref{thm:1} then gives
$ 
\mathcal R^{\mathrm{CI}}(\mathcal E,\Sigma,W)
=
\sqrt{
\mathcal R^{\mathrm{MSE}}(\mathcal E,\Sigma,W)
}.
$ 
Since the square root is strictly increasing, the two criteria have
the same set of minimizers. This proves the first claim.

\subsection{Proof of Theorem~\ref{cor:adaptive_surrogate}}
\label{proof:adaptive_surrogate}

Fix a feasible design
\((\mathcal E,\Sigma)\in\mathcal D\).

\smallskip
\noindent
\textit{Step 1 (fixed-\(B\) minimax risk).}
Fix \(B\geq0\). For any \(a\) satisfying
\(\Lambda_{\mathcal E}^{\top}a=\omega\), consider the linear rule
\[
\delta_a(\tilde S_{\mathcal E,\Sigma})
=
\omega_0+a^\top\tilde S_{\mathcal E,\Sigma}.
\]
Under Equation~\eqref{eqn:guassian}, write
$ 
\tilde S_{\mathcal E,\Sigma}
=
\Lambda_{\mathcal E}\theta
+
\begin{pmatrix}
b\\
0_{|\mathcal E|}
\end{pmatrix}
+\varepsilon,
\qquad
\varepsilon\mid(\theta,b)\sim\mathcal N(0,\Sigma).
$ 
Because
\(\tau_L(\theta)=\omega_0+\omega^\top\theta\) and
\(\Lambda_{\mathcal E}^{\top}a=\omega\),
$ 
\delta_a(\tilde S_{\mathcal E,\Sigma})
-
\tau_L(\theta)
=
a_{1:p_o}^\top b+a^\top\varepsilon.
$ 
Consequently, for every \(\pi\in\Pi_B\),
\[
\small 
\begin{aligned}
&
\mathbb E_\pi
\left[
\left(
\delta_a(\tilde S_{\mathcal E,\Sigma})
-
\tau_L(\theta)
\right)^2
\Bigm|\theta
\right]
 =
a^\top\Sigma a
+
\mathbb E_\pi
\left[
\left(a_{1:p_o}^\top b\right)^2
\Bigm|\theta
\right]
\leq
a^\top\Sigma a
+
B^2\|a_{1:p_o}\|_*^2,
\end{aligned}
\]
where the inequality follows from Assumption~\ref{defn:class_priors}.
Taking expectations over \(\theta\), the supremum over
\(\pi\in\Pi_B\), and then the minimum over feasible \(a\) gives
\[
\small 
\begin{aligned}
&
\inf_\delta
\sup_{\pi\in\Pi_B}
\mathbb E_\pi
\left[
\left(
\delta(\tilde S_{\mathcal E,\Sigma})
-
\tau_L(\theta)
\right)^2
\right]
 \leq
\min_{a:\Lambda_{\mathcal E}^{\top}a=\omega}
\left\{
a^\top\Sigma a
+
B^2\|a_{1:p_o}\|_*^2
\right\}.
\end{aligned}
\]

Conversely, the minimax inequality and
Equation~\eqref{eq:audience_profile_exact} imply
\[
\small 
\begin{aligned}
\inf_\delta
\sup_{\pi\in\Pi_B}
\mathbb E_\pi
\left[
\left(
\delta(\tilde S_{\mathcal E,\Sigma})
-
\tau_L(\theta)
\right)^2
\right]
& \geq
\sup_{\pi\in\Pi_B}
\inf_\delta
\mathbb E_\pi
\left[
\left(
\delta(\tilde S_{\mathcal E,\Sigma})
-
\tau_L(\theta)
\right)^2
\right]
\\ & =
\bar L_B(\mathcal E,\Sigma)
=
\min_{a:\Lambda_{\mathcal E}^{\top}a=\omega}
\left\{
a^\top\Sigma a
+
B^2\|a_{1:p_o}\|_*^2
\right\}.
\end{aligned}
\]
The preceding two inequalities therefore give
\begin{equation}
\label{eq:adaptive_fixed_B}
\small 
\begin{aligned}
&
\inf_\delta
\sup_{\pi\in\Pi_B}
\mathbb E_\pi
\left[
\left(
\delta(\tilde S_{\mathcal E,\Sigma})
-
\tau_L(\theta)
\right)^2
\right]
 =
\bar L_B(\mathcal E,\Sigma)
=
\min_{a:\Lambda_{\mathcal E}^{\top}a=\omega}
\left\{
a^\top\Sigma a
+
B^2\|a_{1:p_o}\|_*^2
\right\}.
\end{aligned}
\end{equation}
Taking the infimum over feasible designs yields
\begin{equation}
\label{eq:adaptive_oracle_fixed_B}
\small 
\begin{aligned}
&
\inf_{\substack{
(\mathcal E',\Sigma')\in\mathcal D\\
\delta'
}}
\sup_{\pi'\in\Pi_B}
\mathbb E_{\pi'}
\left[
\left(
\delta'(\tilde S_{\mathcal E',\Sigma'})
-
\tau_L(\theta)
\right)^2
\right]
=
\inf_{(\mathcal E',\Sigma')\in\mathcal D}
\min_{a':\Lambda_{\mathcal E'}^\top a'=\omega}
\left\{
(a')^\top\Sigma'a'
+
B^2\|a'_{1:p_o}\|_*^2
\right\}.
\end{aligned}
\end{equation}
These quantities are strictly positive. Indeed, for every feasible
\((\mathcal E',\Sigma',a')\),
\[
\small 
\begin{aligned} 
\|\omega\|_2
=
\|\Lambda_{\mathcal E'}^\top a'\|_2
\leq
U\|a'\|_2,
\qquad
(a')^\top\Sigma'a'
\geq
\underline\lambda\|a'\|_2^2 \qquad \Rightarrow (a')^\top\Sigma'a'
\geq
\frac{\underline\lambda}{U^2}\|\omega\|_2^2
>0
\end{aligned} 
\]

\smallskip
\noindent
\textit{Step 2 (lower bound).}
Applying the minimax inequality to the infimum over \(\delta\) and the
supremum over \(B\), and then using
\eqref{eq:adaptive_fixed_B}--\eqref{eq:adaptive_oracle_fixed_B}, gives
\[
\small 
\begin{aligned}
\mathcal R^{\mathrm{ad}}(\mathcal E,\Sigma)
&\geq
\sup_{B\geq0}
\frac{
\displaystyle
\inf_\delta
\sup_{\pi\in\Pi_B}
\mathbb E_\pi
\left[
\left(
\delta(\tilde S_{\mathcal E,\Sigma})
-
\tau_L(\theta)
\right)^2
\right]
}{
\displaystyle
\inf_{\substack{
(\mathcal E',\Sigma')\in\mathcal D\\
\delta'
}}
\sup_{\pi'\in\Pi_B}
\mathbb E_{\pi'}
\left[
\left(
\delta'(\tilde S_{\mathcal E',\Sigma'})
-
\tau_L(\theta)
\right)^2
\right]
}
 =
\sup_{B\geq0}
\frac{
\bar L_B(\mathcal E,\Sigma)
}{
\displaystyle
\inf_{(\mathcal E',\Sigma')\in\mathcal D}
\bar L_B(\mathcal E',\Sigma')
}
 =
\mathcal R^{\mathrm{Bayes}}(\mathcal E,\Sigma).
\end{aligned}
\]

\smallskip
\noindent
\textit{Step 3 (upper bound).}
Fix again \(a\) satisfying
\(\Lambda_{\mathcal E}^{\top}a=\omega\). Evaluating the infimum over
estimators in Definition~\ref{defn:adaptive_regret} at the linear rule
\(\delta_a\), and using Step~1, gives
\[
\begin{aligned}
\mathcal R^{\mathrm{ad}}(\mathcal E,\Sigma)
&\leq
\sup_{B\geq0}
\frac{
a^\top\Sigma a+B^2\|a_{1:p_o}\|_*^2
}{
\displaystyle
\inf_{(\mathcal E',\Sigma')\in\mathcal D}
\min_{a':\Lambda_{\mathcal E'}^\top a'=\omega}
\left\{
(a')^\top\Sigma'a'
+
B^2\|a'_{1:p_o}\|_*^2
\right\}
}.
\end{aligned}
\]
By the definitions of \(\alpha^\star\) and \(\beta^\star\), every
feasible \((\mathcal E',\Sigma',a')\) satisfies
$ 
(a')^\top\Sigma'a'\geq\alpha^\star,
\|a'_{1:p_o}\|_*^2\geq\beta^\star.
$ 
Therefore,
\[
\small 
\begin{aligned} 
\mathcal R^{\mathrm{ad}}(\mathcal E,\Sigma)
\leq
\sup_{B\geq0}
\frac{
a^\top\Sigma a+B^2\|a_{1:p_o}\|_*^2
}{
\alpha^\star+B^2\beta^\star
}.
\end{aligned}
\]
When \(\beta^\star>0\), setting \(t=B^2\), the ratio on the
right-hand side is the ratio of two affine functions of \(t\), with
derivative of constant sign. Its supremum is therefore attained at
\(t=0\) or as \(t\to\infty\), and hence
\[
\sup_{B\geq0}
\frac{
a^\top\Sigma a+B^2\|a_{1:p_o}\|_*^2
}{
\alpha^\star+B^2\beta^\star
}
=
\max\left\{
\frac{a^\top\Sigma a}{\alpha^\star},
\frac{\|a_{1:p_o}\|_*^2}{\beta^\star}
\right\}.
\]
If \(\beta^\star=0\) and
\(\|a_{1:p_o}\|_*^2>0\), both sides are \(+\infty\). If instead
\(\beta^\star=\|a_{1:p_o}\|_*^2=0\), the supremum equals
\(a^\top\Sigma a/\alpha^\star\), which is at least one. Thus the same
equality holds under the conventions \(0/0=1\) and
\(x/0=+\infty\) for \(x>0\).

Because the resulting inequality holds for every \(a\) satisfying
\(\Lambda_{\mathcal E}^{\top}a=\omega\), taking the minimum over such
weights gives
\[
\mathcal R^{\mathrm{ad}}(\mathcal E,\Sigma)
\leq
\min_{a:\Lambda_{\mathcal E}^{\top}a=\omega}
\max\left\{
\frac{a^\top\Sigma a}{\alpha^\star},
\frac{\|a_{1:p_o}\|_*^2}{\beta^\star}
\right\}.
\]
Combining this inequality with Step~2 proves the result.

\section{Optimization}

\subsection{Linear regret} 
\label{sec:optimization}

We now provide an explicit optimization routine, assuming for simplicity no constraints on the weighting matrix $W$. We assume that observational estimates are independent of the experimental estimates, but may be correlated with each other; experimental estimates are mutually independent. This corresponds to settings in which experiments are conducted on independent samples, separately from the observational sample, as in our applications. For example, the experimental estimates may correspond to separate experiments or to arm-specific means computed on disjoint randomized samples. More general dependence structures can be accommodated by replacing the covariance matrix below.

\paragraph{Notation and decision variables.}

We take \(\Sigma_{\mathrm{obs}}\) to be the variance--covariance matrix of the observational study estimates and \(v_j^2/n_j\) the variance of the experimental estimates, for known \(v_j^2\), where \(n_j\) denotes the sample size allocated to experimental estimate \(j\in\mathcal E\). We optimize over the sample-size allocation throughout. Fix nonnegative weights
$ 
\kappa
=
(\kappa_1,\ldots,\kappa_{p_o})^\top
\in\mathbb R_+^{p_o},
$ 
and consider a general weighted rectangular ambiguity set
\[
\small 
\begin{aligned} 
\mathcal B_\kappa(B)
=
B\mathcal C_\kappa,
\qquad
\mathcal C_\kappa
=
\left\{
c\in\mathbb R^{p_o}:
|c_\ell|\leq\kappa_\ell,\quad
\ell=1,\ldots,p_o
\right\}.
\end{aligned} 
\]
Allowing \(\kappa_\ell=0\) imposes \(b_\ell=0\). The weights \(\kappa\) are fixed ex ante and determine the relative magnitudes of misspecification allowed across components. Additional restrictions encoded in the set \(\mathcal C_\kappa\) can be incorporated provided that the resulting set is convex and compact. For symmetric misspecification sets, we can set $k_\ell = 1$ for all $\ell$, in which case $B \mathcal{C} = \{c \in \mathbb{R}^p: ||c||_{\infty} \le B\}$, our leading example.

Let \(x_j=1\{j\in\mathcal E\}\in\{0,1\}\) indicate whether experiment \(j\) is run, and let
\(x\in\mathcal X\), where \(\mathcal X\) denotes the feasible set of experimental designs. Write
$ 
\mathcal E(x)=\{j:x_j=1\}.
$ 
For a given design \(x\), the researcher observes
$ 
\tilde{S}_{\mathcal E(x),\Sigma},
$ 
which stacks the observational estimates and the experimental estimates available under \(\mathcal E(x)\). By Setting~\ref{setting:1},
$ 
\mathbb E[\tilde{S}_{\mathcal E(x),\Sigma}]
=
\Lambda_{\mathcal E(x)}\theta
+
\begin{pmatrix}
b\\
0
\end{pmatrix},
$ 
where the first block corresponds to observational estimates and the second block corresponds to experimental estimates.

We optimize directly over the linear weights placed on the available estimates. Let \(a\) denote these weights, so that the estimator of the target is
\begin{equation} 
\label{eqn:a_constraint}
\small 
\begin{aligned} 
\hat\tau_a
= \omega_0 + 
a^\top\tilde{S}_{\mathcal E(x),\Sigma},
\qquad
a^\top\Lambda_{\mathcal E(x)}=\omega^\top,
\qquad
a=
\Big(
a^{\mathrm{obs}},
a^{\mathrm{exp}}
\Big)^\top. 
\end{aligned} 
\end{equation} 
This parametrization mantains the same GMM structure but, computationally, avoids optimizing directly over a GMM weighting matrix.\footnote{Indeed, if a minimum-distance estimator with weighting matrix \(W\) is
$ 
\hat\theta_W
=
\left(\Lambda_{\mathcal E(x)}^\top W\Lambda_{\mathcal E(x)}\right)^{-1}
\Lambda_{\mathcal E(x)}^\top W
\tilde{S}_{\mathcal E(x),\Sigma},
$ 
then the induced estimator reads as
$ 
\omega^\top\hat\theta_W
=
a_W^\top\tilde{S}_{\mathcal E(x),\Sigma},
a_W
=
W\Lambda_{\mathcal E(x)}
\left(\Lambda_{\mathcal E(x)}^\top W\Lambda_{\mathcal E(x)}\right)^{-1}
\omega.
$ 
The inverse matrix makes the map \(W\mapsto a_W\) nonlinear. For computation, it is therefore simpler to optimize over the induced linear weights \(a\), subject to the calibration constraint.} All minimizations over \(a\) below are understood to impose the constraint in Equation~\eqref{eqn:a_constraint}, together with any additional restrictions on linear weights.

If experiment \(j\) is run, \(n_j\) denotes the experimental sample size allocated to component \(j\), and \(c_j>0\) denotes the corresponding cost. The budget constraint is
$ 
\sum_{j\in\mathcal E(x)}c_jn_j=n.
$

\paragraph{Variance, bias, and optimal sample size.}

Assuming independent experiments, and independence between the observational and experimental estimates, the variance of the plug-in estimator reads as
$ 
(a^{\mathrm{obs}})^\top
\Sigma_{\mathrm{obs}}
a^{\mathrm{obs}}
+
\sum_{j\in\mathcal E(x)}
(a_j^{\mathrm{exp}})^2
\frac{v_j^2}{n_j},
$ 
where \(\Sigma_{\mathrm{obs}}\) is the covariance matrix of the observational estimates and \(v_j^2/n_j\) is the variance of the experimental estimate for component \(j\). The sample sizes enter only through the experimental component. For fixed \(x\) and \(a\), the optimal sample allocation is\footnote{For simplicity, we consider a continuum of sample size; in practice, researchers may threshold this number to the closest integer. This constraint can also be written explicitly in the optimization program. In addition, researchers may also impose a minimum sample size for the treatment arm to be selected, corresponding to an additional constraints in the optimization program.}
\begin{equation}
\label{eqn:optimal_size_general}
\small 
\begin{aligned} 
n_j^\star(a,x)
=
n\,
\frac{|a_j^{\mathrm{exp}}|v_j/\sqrt{c_j}}
{\displaystyle
\sum_{k\in\mathcal E(x)}
|a_k^{\mathrm{exp}}|v_k\sqrt{c_k}},
\qquad
j\in\mathcal E(x).
\end{aligned} 
\end{equation}
Correspondingly, with a slight abuse of notation, denote the variance and bias sensitivity as
\begin{equation}
\label{eqn:alpha_general}
\small 
\begin{aligned} 
\alpha_x(a)
&=
(a^{\mathrm{obs}})^\top
\Sigma_{\mathrm{obs}}
a^{\mathrm{obs}}
+
\frac{1}{n}
\left(
\sum_{j\in\mathcal E(x)}
|a_j^{\mathrm{exp}}|v_j\sqrt{c_j}
\right)^2, \qquad \beta_\kappa(a)
&=
\left(
\sum_{\ell=1}^{p_o}
\kappa_\ell
|a_\ell^{\mathrm{obs}}|
\right)^2.
\end{aligned}
\end{equation}
We write the oracle solutions as
\begin{equation}
\label{eqn:benchmark1}
\small
\begin{aligned}
\alpha^\star
&=
\min_{
x\in\mathcal X,\;
a:\,
a^\top\Lambda_{\mathcal E(x)}=\omega^\top
}
\alpha_x(a), \qquad \beta_\kappa^\star
&=
\min_{
x\in\mathcal X,\;
a:\,
a^\top\Lambda_{\mathcal E(x)}=\omega^\top
}
\beta_\kappa(a).
\end{aligned}
\end{equation}
For fixed \(x\), both problems are convex. If \(\mathcal X\) has a mixed-integer representation, the same problems can be written as mixed-integer convex programs.

\paragraph{Regret-optimal solution.}
Using the optimal sample allocation in Equation~\eqref{eqn:optimal_size_general}, this problem can be written directly as the following epigraph problem:
\begin{align}
\small 
\min_{x,a,r,z,t}\quad
& t
\label{eq:general_minimax_obj}
\\[3pt]
\text{s.t.}\quad
&
\Lambda_{\mathcal E(x)}^\top a=\omega,
\qquad
t\geq0,
\qquad
x\in\mathcal X
\label{eq:general_calibration}
\\
&
(a^{\mathrm{obs}})^\top
\Sigma_{\mathrm{obs}}
a^{\mathrm{obs}}
+
\frac{1}{n}
\left(
\sum_{j\in\mathcal E(x)}
r_jv_j\sqrt{c_j}
\right)^2
\leq
t\alpha^\star,
\nonumber
\\
&
\left(
\sum_{\ell=1}^{p_o}
\kappa_\ell z_\ell
\right)^2
\leq
t\beta_\kappa^\star
\label{eq:general_alpha}
\\
&
r_j\geq a_j^{\mathrm{exp}},
\qquad
r_j\geq-a_j^{\mathrm{exp}},
\qquad
z_\ell\geq a_\ell^{\mathrm{obs}},
\qquad
z_\ell\geq-a_\ell^{\mathrm{obs}},
\nonumber
\\[-2pt]
&
\hspace{7cm}
j\in\mathcal E(x),
\qquad
\ell=1,\ldots,p_o.
\label{eq:general_abs_exp}
\end{align}
The auxiliary variables \(r_j\) and \(z_\ell\) represent the absolute values of the experimental and observational weights. At any optimum,
$ 
r_j=|a_j^{\mathrm{exp}}|,
z_\ell=|a_\ell^{\mathrm{obs}}|,
$ 
because larger values only tighten the variance and bias constraints. Hence, the constraints in Equation~\eqref{eq:general_alpha} are exactly
$ 
\alpha_x(a)\leq t\alpha^\star,
\beta_\kappa(a)\leq t\beta_\kappa^\star.
$ 
For fixed \(x\), the problem is a convex conic program. With binary \(x\), it becomes a mixed-integer convex conic program and can be solved using standard solvers.

\begin{algorithm}[!ht]
\footnotesize 
\caption{Regret-optimal experimental design}
\label{alg:regret_design}
\begin{algorithmic}[1]
\State \textbf{Inputs:} target weights \(\omega\), loading matrices
\(\Lambda_{\mathcal E(x)}\), observational covariance
\(\Sigma_{\mathrm{obs}}\), experimental variance parameters
\((v_j^2)\), per-unit costs \((c_j)\), bias weights
\(\kappa=(\kappa_1,\ldots,\kappa_{p_o})^\top\), total budget \(n\), and
feasible experiment set \(\mathcal X\).

\State \textbf{Compute oracle quantities.}
Obtain \(\alpha^\star\) and \(\beta_\kappa^\star\) by solving
Equation~\eqref{eqn:benchmark1}.

\State \textbf{Solve the regret problem.}
Solve the optimization problem
\eqref{eq:general_minimax_obj}--\eqref{eq:general_abs_exp}
for each \(x\in\mathcal X\), or solve the corresponding mixed-integer conic
program when \(\mathcal X\) has a mixed-integer representation.

\State \textbf{Outputs.}
Let \((x^\star,a^\star,t^\star)\) denote an optimizer. Report:
\begin{itemize}
\item the regret-optimal experiment set
$ 
\mathcal E^\star
=
\{j:x_j^\star=1\};
$ 

\item the optimal linear weights \(a^\star\);

\item the regret factor
$ 
t^\star
=
\max\left\{
\frac{\alpha_{x^\star}(a^\star)}{\alpha^\star},
\frac{\beta_\kappa(a^\star)}{\beta_\kappa^\star}
\right\};
$ 

\item the optimal sample sizes \(n_j^\star\), obtained from
Equation~\eqref{eqn:optimal_size_general}.
\end{itemize}
\end{algorithmic}
\end{algorithm}

\subsection{Audience regret}
\label{sec:optimization_profiled_posterior}

We now optimize the audience regret in Definition~\ref{defn:audience_regret}. We
maintain the same independence structure and notation as in
Section~\ref{sec:optimization}: observational estimates are independent of
the experimental estimates, but may be correlated with each other;
experimental estimates are mutually independent. We also maintain the
weighted misspecification shape introduced in
Section~\ref{sec:optimization}. In particular, for fixed weights
$ 
\kappa
=
(\kappa_1,\ldots,\kappa_{p_o})^\top
\in\mathbb R_+^{p_o},
$ 
the corresponding dual-norm penalty is
$ 
\left(
\sum_{\ell=1}^{p_o}
\kappa_\ell |a_\ell^{\mathrm{obs}}|
\right)^2.
$ 
Allowing \(\kappa_\ell=0\) imposes no prior uncertainty about the bias in
component \(\ell\).

\paragraph{Grid over prior weights and decision variables.}

Under the audience-regret objective, the solution depends on the prior scale
\(B\), equivalently on the weight
\[
\lambda=\frac{B^2}{1+B^2}\in[0,1].
\] 
To simplify optimization, we consider the finite grid
$ 
\mathcal G
=
\{0=\lambda_0<\lambda_1<\cdots<\lambda_K=1\}.
$ 
The endpoint
\(\lambda_K=1\) incorporates directly the limiting case \(B\to\infty\).
Whenever $B$ is bounded from above by $\bar{B}$, we can instead consider an upper bound for $\lambda \le \bar{B}^2/(1+ \bar{B}^2)$.

As in Section~\ref{sec:optimization}, let
\[
x_j=1\{j\in\mathcal E\}\in\{0,1\},
\qquad
x\in\mathcal X,
\qquad
\mathcal E(x)=\{j:x_j=1\}.
\]
For a given design \(x\), let \(n_j\) denote the sample size allocated to
experimental estimate \(j\in\mathcal E(x)\), and let \(c_j>0\) denote its
per-unit cost. The budget constraint is
$ 
\sum_{j\in\mathcal E(x)}c_jn_j=n.
$ 
We take \(\Sigma_{\mathrm{obs}}\) to be the covariance matrix of the
observational estimates, and \(v_j^2/n_j\) to be the variance of experimental
estimate \(j\).

For fixed \(x\), sample sizes \((n_j)_{j\in\mathcal E(x)}\), and weight
\(\lambda_k\), define the profiled scalarized posterior risk
\begin{equation}
\label{eqn:profiled_risk_x_lambda}
\small 
\begin{aligned}
F_x(\lambda_k,n)
=
\min_{a:\ \Lambda_{\mathcal E(x)}^\top a=\omega}
\Bigg\{
&(1-\lambda_k)
\left[
(a^{\mathrm{obs}})^\top
\Sigma_{\mathrm{obs}}
a^{\mathrm{obs}}
+
\sum_{j\in\mathcal E(x)}
(a_j^{\mathrm{exp}})^2\frac{v_j^2}{n_j}
\right] +
\lambda_k
\left(
\sum_{\ell=1}^{p_o}
\kappa_\ell |a_\ell^{\mathrm{obs}}|
\right)^2
\Bigg\}.
\end{aligned}
\end{equation}

\paragraph{Oracle values at each grid point.}

We first compute the oracle value separately for each weight \(\lambda_k\).
For fixed \(\lambda_k<1\), fixed \(x\), and fixed \(a\), the optimal sample
allocation is
\begin{equation}
\label{eqn:profiled_fixed_lambda_allocation}
n_j^\star(a,x)
=
n\,
\frac{|a_j^{\mathrm{exp}}|v_j/\sqrt{c_j}}
{\displaystyle
\sum_{q\in\mathcal E(x)}
|a_q^{\mathrm{exp}}|v_q\sqrt{c_q}},
\qquad
j\in\mathcal E(x).
\end{equation}
At \(\lambda_K=1\), the sample allocation does not affect the scalarized
objective. Substituting the allocation in
Equation~\eqref{eqn:profiled_fixed_lambda_allocation} gives the
fixed-\(\lambda_k\), fixed-\(x\) objective
\begin{equation}
\label{eqn:profiled_oracle_x_lambda}
\small 
\begin{aligned}
\bar F_x(\lambda_k)
=
\min_{a:\ \Lambda_{\mathcal E(x)}^\top a=\omega}
\Bigg\{
&(1-\lambda_k)
\left[
(a^{\mathrm{obs}})^\top
\Sigma_{\mathrm{obs}}
a^{\mathrm{obs}}
+
\frac{1}{n}
\left(
\sum_{j\in\mathcal E(x)}
|a_j^{\mathrm{exp}}|v_j\sqrt{c_j}
\right)^2
\right]
 +
\lambda_k
\left(
\sum_{\ell=1}^{p_o}
\kappa_\ell |a_\ell^{\mathrm{obs}}|
\right)^2
\Bigg\}.
\end{aligned}
\end{equation}
The oracle scalarized posterior risk at grid point \(\lambda_k\) is
\begin{equation}
\label{eqn:profiled_oracle_lambda}
F^\star(\lambda_k)
=
\min_{x\in\mathcal X}\bar F_x(\lambda_k).
\end{equation}
For fixed \(x\), the problem in
Equation~\eqref{eqn:profiled_oracle_x_lambda} is convex. If \(\mathcal X\)
is finite, one can solve it for each feasible \(x\) and take the minimum. If
\(\mathcal X\) has a mixed-integer representation, the same problem can be
written as a mixed-integer convex conic program. The two endpoint values are
\[
F^\star(0)=\alpha^\star, \qquad 
F^\star(1)
=
\beta_\kappa^\star
=
\min_{
x\in\mathcal X,\;
a:\Lambda_{\mathcal E(x)}^\top a=\omega
}
\left(
\sum_{\ell=1}^{p_o}
\kappa_\ell |a_\ell^{\mathrm{obs}}|
\right)^2,
\]
which we assume to be strictly positive as in
Corollary~\ref{cor:finite_frontier_approximation}. The latter is the same
weighted bias benchmark as in Section~\ref{sec:optimization}.

\paragraph{Audience regret solution on the grid.}

The profiled posterior-regret criterion on the grid is
$ 
\mathcal R_K(x,n)
=
\max_{k=0,\ldots,K}
\frac{
F_x(\lambda_k,n)
}{
F^\star(\lambda_k)
}.
$ 
After computing \(F^\star(\lambda_k)\) for each grid point using
Equation~\eqref{eqn:profiled_oracle_lambda}, the grid approximation can be
written as
\begin{equation}
\label{eq:profiled_grid_obj_lambda}
\small 
\begin{aligned}
\underline R_K
& =
\min_{x,(n_j),r,\{a_k,z_k\}_{k=0}^K}
\quad  r
\\[3pt]
\textnormal{s.t.}\quad
& \Lambda_{\mathcal E(x)}^\top a_k=\omega,
\qquad
k=0,\ldots,K,
\\
&
(1-\lambda_k)
\left[
(a_k^{\mathrm{obs}})^\top
\Sigma_{\mathrm{obs}}
a_k^{\mathrm{obs}}
+
\sum_{j\in\mathcal E(x)}
(a_{k,j}^{\mathrm{exp}})^2\frac{v_j^2}{n_j}
\right]
+
\lambda_k
\left(
\sum_{\ell=1}^{p_o}
\kappa_\ell z_{k,\ell}
\right)^2
\leq
rF^\star(\lambda_k),
\qquad
k=0,\ldots,K,
\\
&
z_{k,\ell}\geq a_{k,\ell}^{\mathrm{obs}},
\qquad
z_{k,\ell}\geq-a_{k,\ell}^{\mathrm{obs}},
\\
&\hspace{6cm}
\ell=1,\ldots,p_o,\quad
k=0,\ldots,K,
\\
&
\sum_{j\in\mathcal E(x)}c_jn_j=n,
\qquad
n_j\geq0,\quad
j\in\mathcal E(x),
\\
&
x\in\mathcal X,
\qquad
r\geq1.
\end{aligned}
\end{equation}

The auxiliary variables \(z_k\) represent the absolute values of the
observational weights. Without loss of generality, they can be taken to
satisfy
$ 
z_{k,\ell}
=
|a_{k,\ell}^{\mathrm{obs}}|,
$ 
because replacing larger values by the corresponding absolute values only
relaxes the constraints. The variables \(a_K,z_K\) at \(\lambda_K=1\)
impose directly the large-\(B\) limit of the regret criterion using the
weighted bias benchmark \(F^\star(1)=\beta_\kappa^\star\).

The closed-form allocation in
Equation~\eqref{eqn:profiled_fixed_lambda_allocation} is used to compute the
oracle values \(F^\star(\lambda_k)\) for each \(\lambda_k<1\). At
\(\lambda_K=1\), the oracle value is the weighted bias benchmark
\(\beta_\kappa^\star\).

For fixed \(x\), the finite-grid problem is convex. The terms
$ 
(1-\lambda_k)
(a_{k,j}^{\mathrm{exp}})^2/n_j
$ 
are convex quadratic-over-linear terms and can be represented by rotated
second-order cone constraints. The weighted absolute-value terms remain
convex because \(\kappa_\ell\geq0\) and \(\lambda_k\geq0\). With binary
design choices, the problem becomes a mixed-integer convex conic program.

\section{Two parameters Example \eqref{subsec:two_param}} \label{proof:cor:gamma_2_param}

By Theorem~\ref{thm:1}, for fixed $(\mathcal E,\Sigma)$ the proportional regret as a function of the shrinkage vector equals
$\max\!\Big\{\alpha(\mathcal E,\Sigma,W)/\alpha^\star,\; \beta(\mathcal E,W)/\beta^\star\Big\},
$ 
with $\alpha^\star,\beta^\star$ constants.  

\emph{Step 1.} With independence and two parameters,
\[
\alpha(\mathcal E,\Sigma,W)=C+\omega_j^2\!\Big((1-W_j)^2v_j^2+W_j^2\sigma_j^2\Big),
\]
where $C$ does not depend on $W_j$. Hence $\alpha(j,W_j)\equiv\alpha(\mathcal E,\Sigma,W)$ is a strictly convex quadratic in $W_j$ with unique minimizer at 
\[
W_j^{\mathrm{var}}  = \frac{v_j^2}{\sigma_j^2+v_j^2}.
\]
Moreover, $\alpha(j,W_j)$ is strictly increasing on $\big(W_j^{\mathrm{var}},1\big]$ and strictly decreasing on $\big[0,W_j^{\mathrm{var}}\big)$.  

Write the worst--case bias component holding the other coordinate fixed as
\[
\beta(j,W_j)\;=\;\Big(A_j + |\omega_j|W_j\Big)^2,
\]
where $A_j\ge 0$ collects all terms not involving $W_j$ (including the contribution from the other coordinate). Thus, $\beta(j,W_j)$ is strictly increasing in $W_j$ and convex with minimum at $W_j = 0$.  
Therefore, any minimizer satisfies
\[
W_j^\star \in \big[0,\,W_j^{\mathrm{var}}\big].
\]

\emph{Step 2.}
On $\big(0,W_j^{\mathrm{var}}\big)$ the map $W_j\mapsto \alpha(j,W_j)/\alpha^\star$ is strictly decreasing, while $W_j\mapsto \beta(j,W_j)/\beta^\star$ is strictly increasing. Hence the function
\[
f(W_j)\equiv\max\!\Big\{\alpha(j,W_j)/\alpha^\star,\ \beta(j,W_j)/\beta^\star\Big\},\qquad 
W_j\in\big[0,W_j^{\mathrm{var}}\big],
\]
is minimized either (i) at a boundary point, or (ii) at the unique interior point where the two arguments are equal (by strict monotonicity, at most one intersection exists).
Therefore, 
\begin{itemize}
\item If $\alpha(j,W_j)/\alpha^\star < \beta(j,W_j)/\beta^\star$ for all $W_j\in\big(0,W_j^{\mathrm{var}}\big)$, then $f(W_j)=\beta(j,W_j)/\beta^\star$ on that interval. Since this term is strictly increasing, the minimizer is the left boundary $W_j^\star=0$.

\item If $\alpha(j,W_j)/\alpha^\star > \beta(j,W_j)/\beta^\star$ for all $W_j\in\big(0,W_j^{\mathrm{var}}\big)$, then $f(W_j)=\alpha(j,W_j)/\alpha^\star$ on that interval. Since this term is strictly decreasing there, the minimizer is the right boundary $W_j^\star=W_j^{\mathrm{var}}=v_j^2/(\sigma_j^2+v_j^2)$.

\item Otherwise, by the intermediate value theorem and strict monotonicity of the two curves, there exists a unique $W_j\in\big(0,W_j^{\mathrm{var}}\big)$ such that
$ 
\frac{\alpha(j,W_j)}{\alpha^\star}
\;=\;
\frac{\beta(j,W_j)}{\beta^\star}.
$ 
\end{itemize}

\section{Non linear estimators}
\label{app:armstrong}

\subsection{Review of non linear estimator with fixed design}
This section reviews the definition of proportional regret with a fixed design in  \cite{armstrong2024adaptingmisspecification} and \cite{tsybakov1998pointwise} and illustrates distinctions from our framework. 

\paragraph{Single parameter with a fixed design}

Different from this paper, 
\cite{armstrong2024adaptingmisspecification} focus on a single parameter with two estimators under a fixed design where one of the two estimators is unbiased for the target estimand. In our notation, this corresponds to fixing a design $(\mathcal{E},\Sigma$ with $\mathcal{E}=\{1\}$, so there is an observational estimate
$S^{\mathrm{obs}}\in\mathbb{R}$ and an experimental estimate $S^{\exp}\in\mathbb{R}$ satisfying
\[
\mathbb{E}[S^{\mathrm{obs}}] = \theta_1 + b,
\qquad
\mathbb{E}[S^{\exp}] = \theta_1, \quad \Delta \tilde{S} = S^{\mathrm{obs}} - S^{\exp}
\]
for an unknown scalar bias $b\in\mathbb{R}$. Without loss, standardize each estimator as well as the estimand $\theta_1$ by the standard deviation of the difference $\Delta \tilde{S}$. With this normalization, the authors further assume  
\[
\begin{pmatrix}
S^{\exp} \\
\Delta \tilde{S}
\end{pmatrix}
\sim
\mathcal{N}\!\left(
\begin{pmatrix}
\theta_1 \\
b
\end{pmatrix},
\,
\Sigma
\right),
\quad
\Sigma
=
\begin{pmatrix}
\sigma_{\exp}^2
&
\rho\,\sigma_{\exp} \\[4pt]
\rho\,\sigma_{\exp}
&
1
\end{pmatrix},
\]
where $\sigma_{\exp}^2$ is the variance of $S^{\exp}$, and $\rho$ is the correlation between the experimental estimate and $\Delta \tilde{S}$.  The question of interest in \cite{armstrong2024adaptingmisspecification} is how to optimally combine $\tilde{S}^{\mathrm{exp}}$ and $\Delta \tilde{S}$ in a single estimator. 

\paragraph{Non linear estimator and corresponding proportional regret} 
Using invariance arguments, \cite{armstrong2024adaptingmisspecification} study estimators of the form
\[
\hat\tau_\delta
=
S^{\exp}
+
\rho \,\sigma_{\exp}\,\big\{\delta(\Delta \tilde{S})-\Delta \tilde{S}\big\},
\]
for some measurable function $\delta: \mathbb{R} \to \mathbb{R}$ to be optimized. The authors show that for this choice, the proportional regret worst case for $|b| \le B$, can be written as 
$$  
\small 
\begin{aligned} 
A(B,\delta)
:=
\frac{\displaystyle \mathrm{MSE}_B(\delta)}
     {R^*(B)}, \quad R^*(B)
=
\sigma_{\exp}^2
\Big[
(1-\rho^2)
+
\rho^2\,r_{\mathrm{BNM}}(B)
\Big]
\end{aligned} 
$$ 
where  $r_{\mathrm{BNM}}(\cdot)$ denotes the B-minimax risk in the bounded normal-mean
problem.

\paragraph{Properties of the objective function and solution} Because of the non-linearity of the estimator (and of the $r_{\mathrm{BNM}}$ function)  $A(B,\delta)$ is not quasi-convex in $B^2$ for the chosen non-linear shrinkage rule. The optimal solution is obtained by solving a minimax problem by using a discrete approximation over the least-favorable prior over $B$. 

This differs from our framework where quasi-convexity arises as we restrict the class of estimators to be linear. A formal argument is provided in the proof of Theorem \ref{thm:1}. Such quasi-convexity simplifies estimation here, since the choice is not only for the estimator, but also for the class of designs and sample size allocations. 

\subsection{An example of non-linear estimators for design choice}
\label{sec:gaussian_priors}

We illustrate an example below for a finite-dimensional approximation to the adaptive
regret problem in Definition~\ref{defn:adaptive_regret} with an example under a class of Gaussian prior, to provide intuition behind the non-linear solution. One challenge is that the solution under Gaussian priors is computationally demanding when optimizing over the design even with restrictions on priors.

Consider a finite grid of prior scales
\[
\mathcal B_G
=
\{B_1,\ldots,B_G\}.
\]
For each \(B_g\), construct an analog of \(\Pi_{B_g}\) by the finite set
\[
\pi_{g,h},
\qquad
h=1,\ldots,H,
\]
where
\[
\theta\sim\mathcal N(0,T^2I_p),
\qquad
b\mid\theta
\sim
\mathcal N(0,B_g^2K_{g,h}),
\qquad
K_{g,h}\in\mathcal K_{\mathcal C}.
\]
Here \(g\) indexes the prior scale \(B_g\), while \(h\) indexes the
possible covariance directions within the class
\(\Pi_{B_g}\).

For a design \((\mathcal E,\Sigma)\), after integrating out the
observational bias,
\[
\tilde{S}_{\mathcal E,\Sigma}\mid\theta
\sim
\mathcal N
\left(
\Lambda_{\mathcal E}\theta,
\Sigma_{g,h}
\right),\qquad \Sigma_{g,h}
=
\Sigma
+
\begin{pmatrix}
B_g^2K_{g,h} & 0\\
0 & 0
\end{pmatrix}. 
\]

Under prior \(\pi_{g,h}\), the posterior covariance of \(\theta\) is
\[
V_{g,h}
=
\left(
T^{-2}I_p
+
\Lambda_{\mathcal E}^{\top}
\Sigma_{g,h}^{-1}
\Lambda_{\mathcal E}
\right)^{-1},
\]
and the posterior mean of the linearized target is
\[
m_{g,h}(y)
=
\omega_0
+
\omega^\top
V_{g,h}
\Lambda_{\mathcal E}^{\top}
\Sigma_{g,h}^{-1}
y.
\]
The corresponding posterior variance is
\[
L_{g,h}(\mathcal E,\Sigma)
=
\omega^\top V_{g,h}\omega.
\]

Let $\phi$ denote the Gaussian PDF and define
\[
p_{g,h}^{\mathcal E,\Sigma}(y)
=
\phi
\left(
y;
0,
\Sigma_{g,h}
+
T^2\Lambda_{\mathcal E}\Lambda_{\mathcal E}^{\top}
\right).
\]
For a measurable estimator \(\delta\), define
\[
M_{g,h}(\mathcal E,\Sigma,\delta)
=
\mathbb E_{\pi_{g,h}}
\left[
\left(
\delta(\tilde{S}_{\mathcal E,\Sigma})
-
\tau_L(\theta)
\right)^2
\right].
\]
The posterior-risk decomposition implies
\[
M_{g,h}(\mathcal E,\Sigma,\delta)
=
\int
\left\{
\left(
\delta(y)-m_{g,h}(y)
\right)^2
+
L_{g,h}(\mathcal E,\Sigma)
\right\}
p_{g,h}^{\mathcal E,\Sigma}(y)\,dy.
\]

\paragraph{Step 1: the adaptive oracle at a known scale.}

At scale \(B_g\), the oracle knows \(g\), but does not know which prior
\(\pi_{g,h}\) in the class is correct. It may choose both the design and
one estimator that performs uniformly over \(h\). Its value is
\begin{equation}
\label{eqn:adaptive_oracle_grid}
V_g^{\mathrm{ad},\star}
\equiv
\inf_{(\mathcal E',\Sigma')\in\mathcal D}
\inf_{\delta'}
\max_{h=1,\ldots,H}
M_{g,h}(\mathcal E',\Sigma',\delta').
\end{equation}
For a fixed design \((\mathcal E',\Sigma')\), let
\(q=(q_1,\ldots,q_H)\in\Delta_H\), where \(\Delta_H\) denotes the
\(H\)-dimensional probability simplex. Under standard minimax regularity
conditions,
\[
\inf_{\delta'}
\max_{h=1,\ldots,H}
M_{g,h}(\mathcal E',\Sigma',\delta')
=
\max_{q\in\Delta_H}
\inf_{\delta'}
\sum_{h=1}^H
q_h
M_{g,h}(\mathcal E',\Sigma',\delta').
\]
For fixed \(q\), the minimizing estimator is
\begin{equation}
\label{eqn:oracle_mixture_rule}
\delta_{g,q}^{\mathcal E',\Sigma'}(y)
=
\frac{
\displaystyle
\sum_{h=1}^H
q_h
p_{g,h}^{\mathcal E',\Sigma'}(y)
m_{g,h}^{\mathcal E',\Sigma'}(y)
}{
\displaystyle
\sum_{h=1}^H
q_h
p_{g,h}^{\mathcal E',\Sigma'}(y)
}.
\end{equation}
Therefore, the oracle value can be written as
\begin{equation}
\label{eqn:adaptive_oracle_mixture}
V_g^{\mathrm{ad},\star}
=
\inf_{(\mathcal E',\Sigma')\in\mathcal D}
\max_{q\in\Delta_H}
\sum_{h=1}^H
q_h
M_{g,h}
\left(
\mathcal E',\Sigma',
\delta_{g,q}^{\mathcal E',\Sigma'}
\right).
\end{equation}
Thus, computing the denominator requires solving a scale-specific minimax
estimation and design problem for each \(g\).

\paragraph{Step 2: adaptation across unknown scales.}
Define the finite-grid adaptive
regret of a fixed design by
\begin{equation}
\label{eqn:adaptive_regret_grid}
\mathcal R_G^{\mathrm{ad}}(\mathcal E,\Sigma)
=
\inf_{\delta}
\max_{\substack{
g=1,\ldots,G,\\
h=1,\ldots,H
}}
\frac{
M_{g,h}(\mathcal E,\Sigma,\delta)
}{
V_g^{\mathrm{ad},\star}
}.
\end{equation}
The researcher must use one estimator \(\delta\) for all \(g\) and \(h\),
whereas the oracle denominator may use a different design and estimator for
each known scale \(B_g\).

Let
\[
\lambda
=
\left(
\lambda_{g,h}:
g=1,\ldots,G,\ h=1,\ldots,H
\right)
\in\Delta_{GH}.
\]
Using
\[
\max_{g,h}x_{g,h}
=
\max_{\lambda\in\Delta_{GH}}
\sum_{g=1}^G\sum_{h=1}^H
\lambda_{g,h}x_{g,h},
\]
and applying the minimax theorem, we obtain
\[
\mathcal R_G^{\mathrm{ad}}(\mathcal E,\Sigma)
=
\max_{\lambda\in\Delta_{GH}}
\inf_{\delta}
\sum_{g=1}^G\sum_{h=1}^H
\lambda_{g,h}
\frac{
M_{g,h}(\mathcal E,\Sigma,\delta)
}{
V_g^{\mathrm{ad},\star}
}.
\]

For fixed \(\lambda\), define
\begin{equation}
\label{eqn:adaptive_mixture_weights}
w_{g,h}
\left(
y;\lambda,\mathcal E,\Sigma
\right)
=
\frac{
\displaystyle
\lambda_{g,h}
\left(
V_g^{\mathrm{ad},\star}
\right)^{-1}
p_{g,h}^{\mathcal E,\Sigma}(y)
}{
\displaystyle
\sum_{g'=1}^G\sum_{h'=1}^H
\lambda_{g',h'}
\left(
V_{g'}^{\mathrm{ad},\star}
\right)^{-1}
p_{g',h'}^{\mathcal E,\Sigma}(y)
}.
\end{equation}
The estimator minimizing the \(\lambda\)-weighted average of normalized
risks is
\begin{equation}
\label{eqn:adaptive_mixture_estimator}
\delta_\lambda^{\mathcal E,\Sigma}(y)
=
\sum_{g=1}^G\sum_{h=1}^H
w_{g,h}
\left(
y;\lambda,\mathcal E,\Sigma
\right)
m_{g,h}^{\mathcal E,\Sigma}(y).
\end{equation}
The weights depend on the realized report \(y\). The estimator therefore
adapts to the prior scale, mean, and covariance direction that are most
consistent with the observed evidence.

Substituting the optimal rule into the objective gives
\begin{equation}
\label{eqn:adaptive_regret_mixture}
\mathcal R_G^{\mathrm{ad}}(\mathcal E,\Sigma)
=
\max_{\lambda\in\Delta_{GH}}
\sum_{g=1}^G\sum_{h=1}^H
\lambda_{g,h}
\frac{
M_{g,h}
\left(
\mathcal E,\Sigma,
\delta_\lambda^{\mathcal E,\Sigma}
\right)
}{
V_g^{\mathrm{ad},\star}
}.
\end{equation}
Hence, for a fixed design, the adaptive regret can be computed by
optimizing over the least-favorable mixture weights \(\lambda\).

Once we optimize over the design, estimation therefore requires a sequence of nested optimization problems and may be computationally demanding for a large design space.



\section{Additional figures and tables}

\begin{figure}[!ht]
    \centering
    \includegraphics[scale = 0.7]{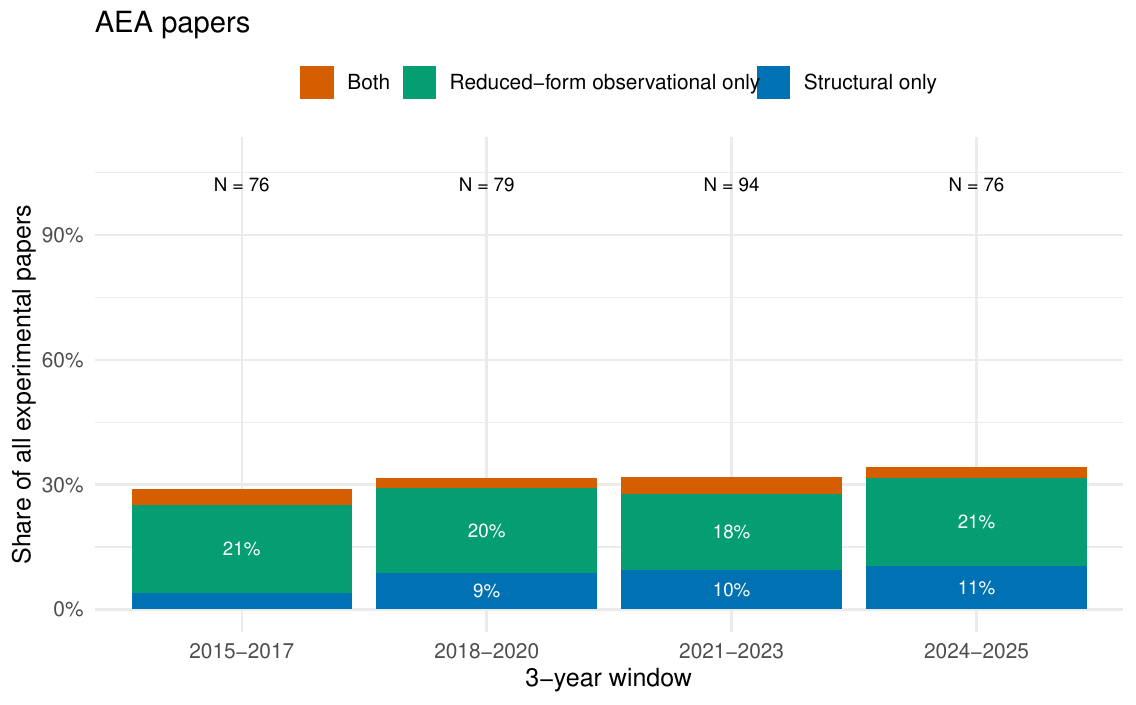}
    \caption{\footnotesize Share of experimental papers published in AEA journals also presenting experimental results in combination with observational estimates (either from a reduced form, or from a structural model or from both).}
    \label{fig:AEA_journals}
\end{figure}

\begin{figure}[!ht]
    \centering
    \includegraphics[scale = 0.7]{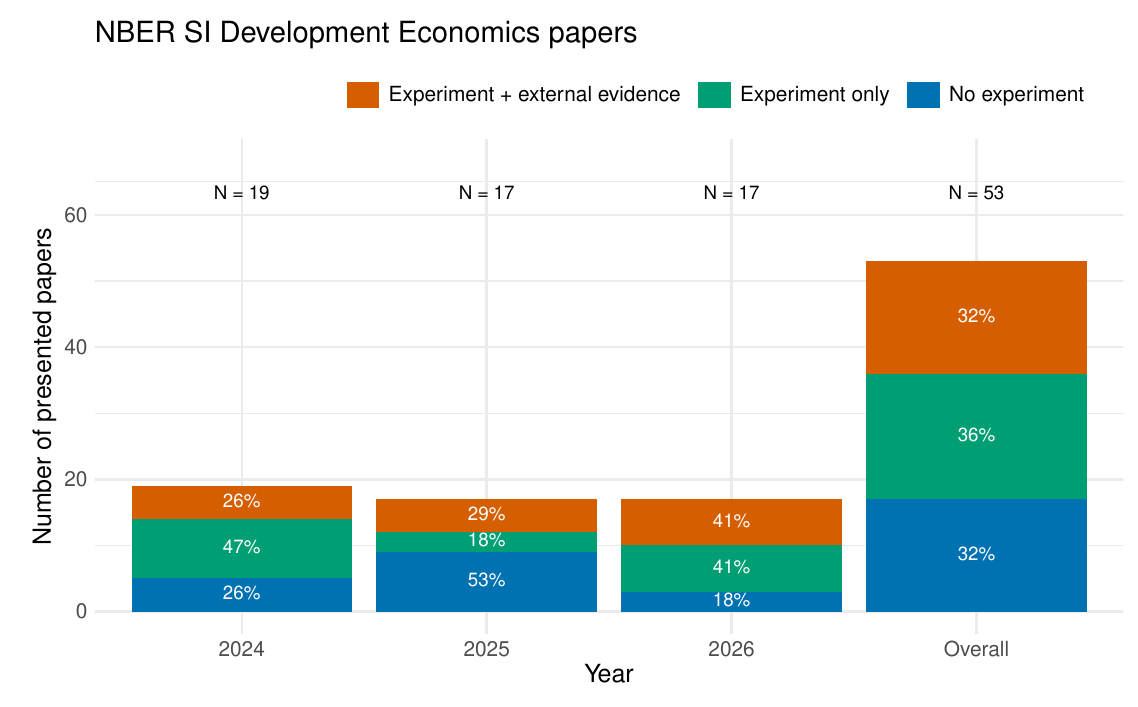}
    \caption{\footnotesize Share of papers at the Development Economics session of the NBER Summer Institute conference from 2024-26 presenting experimental results with and without external evidence. Experiment + external evidence indicates experimental papers that combine experimental evidence with substantive observational or other nonexperimental inputs in their analysis. The figure shows that \textit{within all experimental papers}, the share of experimental evidence + external evidence is close to $50\%$. }
    \label{fig:NBER_SI_dev}
\end{figure}

\end{document}